\documentclass[11pt,a4paper]{article}
\pdfoutput=1

\usepackage{jheppub}
\usepackage{amsmath}
\usepackage[normalem]{ulem}

\usepackage{slashed}
\usepackage{diagbox}
\usepackage{subfigure}
\usepackage{graphics}

\usepackage{simplewick}
\usepackage{multirow}
\usepackage{ulem}
\usepackage{cancel}

\usepackage[dvipsnames]{xcolor}

\DeclareMathOperator{\Tr}{Tr}
\def\II{\hbox{{1}\kern-.25em\hbox{l}}}

\DeclareMathOperator{\Li}{Li}

\newcommand \widebar [1] {\overline{#1}}

\newcommand{\WW}{{\rm\scriptscriptstyle WW}}

\newcommand{\bomega}{\boldsymbol{\omega}}
\newcommand{\btau}{\boldsymbol{\tau}}
\newcommand{\bdeta}{\boldsymbol{\eta}}
\newcommand{\bzeta}{\boldsymbol{\zeta}}

\def \e {\mbox{e}}
\def \qqquad {\qquad\quad}
\def \qqqquad {\qquad\qquad}

\allowdisplaybreaks

\newcounter{MBQ}

\title{\boldmath 
$B\to D \ell  \nu_\ell$ form factors beyond leading power
and extraction of $|V_{cb}|$ and $R(D)$}

\author[a,b,c,d]{Jing Gao,} 

\author[d]{Tobias Huber,}

\author[d,e]{Yao Ji,}

\author[a]{Chao Wang,}

\author[a]{Yu-Ming Wang,}

\author[a,e]{and Yan-Bing Wei}

\subheader{
\begin{flushright}
{\small
TUM-HEP-1331/21\\
SI-HEP-2021-37 \\
P3H-21-103
}
\end{flushright}
}

\affiliation[a]{School of Physics, Nankai University, 
  Weijin Road 94, 300071 Tianjin, China}

\affiliation[b]{Institute of High Energy Physics, CAS, 
   P.O. Box 918(4) Beijing 100049, China}

\affiliation[c]{School of Physics, University of Chinese 
   Academy of Sciences, Beijing 100049, China}

\affiliation[d]{Theoretische Physik 1, Naturwissenschaftlich-Technische 
   Fakult\"at, Universit\"at Siegen, Walter-Flex-Stra{\ss}e 3, 
   D-57068 Siegen, Germany}

\affiliation[e]{
   Physik Department T31, James-Franck-Stra\ss e 1, 
   Technische Universit{\"a}t M{\"u}nchen,\\
   D-85748 Garching, Germany}

\emailAdd{9820210055@nankai.edu.cn}

\emailAdd{huber@physik.uni-siegen.de}

\emailAdd{ yao.ji@tum.de}

\emailAdd{chaowang@nankai.edu.cn}

\emailAdd{wangyuming@nankai.edu.cn}

\emailAdd{yanbing.wei@tum.de}

\abstract{We investigate the subleading-power corrections to the exclusive $B\to D \ell  \nu_\ell$ form factors at ${\cal O} (\alpha_s^0)$ in the light-cone sum rules (LCSR) framework by including the two- and three-particle higher-twist contributions from the $B$-meson light-cone distribution amplitudes up to the twist-six accuracy,
by taking into account the subleading terms in expanding the hard-collinear  charm-quark propagator,
and by  evaluating the hadronic matrix element of the subleading effective current $\bar q \, \gamma_{\mu} \, i \slashed{D}_\perp / (2 \, m_b) \, h_v$.
Employing further the available leading-power results 
for the semileptonic $B \to D$ form factors at the next-to-leading-logarithmic level
and combining our improved LCSR predictions
with the recent lattice determinations,
we then carry out a comprehensive phenomenological analysis on the semi-leptonic $B\to D \ell \nu_\ell$ decay. We extract $|V_{cb}| =
\big( 40.2^{+0.6}_{-0.5} {\big |_{\rm th}}\,\, {}^{+1.4}_{-1.4} {\big |_{\rm exp}} \big)\times 10^{-3}$
($|V_{cb}| = 
\big( 40.9^{+0.6}_{-0.5} {\big |_{\rm th}}\,\, {}^{+1.0}_{-1.0} {\big |_{\rm exp}} \big)\times 10^{-3}$)
using the BaBar (Belle) experimental data, and particularly obtain for the gold-plated ratio $R(D) = 0.302\pm 0.003$.}

\makeatletter
\gdef\@fpheader{}
\makeatother

\begin{document}

\maketitle

%\newpage

%
%%%%%%%%%%%%%%%%%%%%%%%%%%%%%%%%%%%%%%%%%%%%%%%%%%%%%%%%%%%%%%%%%%%%%%%%%%%%
\section{Introduction}
%%%%%%%%%%%%%%%%%%%%%%%%%%%%%%%%%%%%%%%%%%%%%%%%%%%%%%%%%%%%%%%%%%%%%%%%%%%%
%

The precision study of the semileptonic $B\to D \ell \nu_\ell$ decays  is evidently indispensable for determining the CKM matrix element $|V_{cb}|$ from exclusive processes~\cite{ParticleDataGroup:2020ssz}, hence providing a valuable probe of the delicate quark-flavour mixing mechanism 
in the Standard Model (SM) and beyond.
Such heavy-to-heavy $B$-meson decays further offer the unique window into the strong interaction dynamics of the heavy-hadron  system from the first-principles QCD theory. 
The disagreement on the determinations of $|V_{cb}|$ from inclusive and exclusive channels as well as the renowned anomaly~\cite{HFLAV:2019otj} in $R(D)\equiv{\cal B}(B\to D\tau\nu_\tau)/{\cal B}(B\to D\ell\nu_\ell)$
have triggered numerous studies on new-physics interpretations and explanations, both model-independently and in concrete new-physics scenarios, see e.g.~\cite{Faller:2011nj,Crivellin:2015hha,Fajfer:2015ycq,Sahoo:2016pet,Colangelo:2016ymy,Dorsner:2017ufx,Jung:2018lfu,Gambino:2019sif,Iguro:2020cpg}. In~\cite{Crivellin:2014zpa} it was pointed out that it is rather difficult to accommodate the $|V_{cb}|$ discrepancy by new physics, making SM improvements even more important and hence calling for further dedicated investigations on the SM predictions of $B\to D$ form factors with higher accuracy.
While the $B\to D$ form factors have been considered at the \textit{zero}-recoil limit in the framework of heavy quark effective theory (HQET) for some time (see, e.g.,~\cite{Neubert:1993mb} for a pedagogical review,~\cite{Bernlochner:2017jka,Bigi:2017jbd,Bigi:2017njr} for recent studies, and~\cite{Gambino:2020jvv} for a comprehensive report), and calculated with Lattice QCD \cite{MILC:2015uhg,Na:2015kha,Kaneko:2019vkx}, it has only become
possible in recent years to extend the theoretical studies beyond the zero-recoil kinematics.
In particular, it has been demonstrated that the light-cone sum rule (LCSR) approach is particularly helpful in predicting the form factors at the \emph{large}-recoil limit, where a systematic
power-counting scheme can be established, taking the light-cone distribution amplitudes (LCDAs) as the nonperturbative ingredients.
This machinery can be formulated both in the context of QCD as well as in soft-collinear effective theory (SCET), as pioneered in the work of~\cite{Balitsky:1989ry,Belyaev:1993wp,Khodjamirian:2005ea,Khodjamirian:2006st} and~\cite{DeFazio:2005dx,DeFazio:2007hw}, respectively.
To be more specific about the power-counting scheme, which plays a key role in our analysis, we adopt $m_c\sim {\cal O}(\sqrt{\Lambda_{\rm QCD} m_b})$ which is motivated by the mass hierarchy in the Standard Model. 
This scheme was adopted in refs.~\cite{Boos:2005by, Boos:2005qx} to analyze the inclusive semileptonic $B \to X_c\ell\nu_\ell$ decay, in contrast to the commonly used scheme $m_c \sim {\cal O}(m_b)$ in the study of heavy decays (see, for instance~\cite{Beneke:2000ry, Gambino:2012rd}). 
The power counting of $m_c\sim {\cal O}(\sqrt{\Lambda_{\rm QCD} m_b})$ then validates the identification of the on-shell bottom-, charm-, and light-quark field respectively as hard, hard-collinear and soft modes in QCD following the convention established in~\cite{Wang:2015vgv}.

Taking advantage of the parton-hadron duality and the light-cone operator product expansion (OPE), the LCSR approach has yielded fruitful theoretical predictions for the exclusive $B \to D$ transition form factors with an energetic charmed meson.
Furnished by the explicit power-counting scheme introduced above, QCD factorization for the leading-power contribution to the corresponding vacuum-to-$B$-meson correlation function with an interpolating current for the $D$-meson has been established at tree level~\cite{Faller:2008tr} and at one loop~\cite{Wang:2017jow}, respectively.
The subleading power corrections from the three-particle $B$-meson LCDAs were further calculated at ${\cal O}(\alpha_s^0)$~\cite{Faller:2008tr,Wang:2017jow} based upon the  then-available parametrization of the three-body light-ray matrix element~\cite{Kawamura:2001jm,Kawamura:2001bp}  and subsequently revisited  in~\cite{Gubernari:2018wyi} at the twist-four accuracy by employing the general decomposition of the non-local HQET matrix element~\cite{Braun:2017liq}.
The updated version of the higher-twist LCDAs given in~\cite{Braun:2017liq} were further adopted in~\cite{Gubernari:2018wyi,Bordone:2019vic,Bordone:2019guc} for phenomenological studies where the two-particle contributions were found
to dominate over their three-particle counterparts as previously observed in the context of $B\rightarrow \pi,K$ form factors~\cite{Lu:2018cfc}.
It is also worth mentioning that alternative approaches to the LCSR have been attempted in recent decades.
For example, in ref.~\cite{Li:2009wq, Fu:2013wqa,Zhong:2018exo, Zhang:2021wnv}, the $B \rightarrow D \ell \nu_\ell$ form factors were computed from QCD sum rules via the vacuum-to-$D$-meson correlation function interpolated by a $B$-meson state where the nonperturbative strong interaction dynamics are encapsulated in the $D$-meson LCDAs instead.
The main issue of such considerations lies in the yet poor constraints on the LCDAs of the $D$ meson.
Another QCD-motivated framework relies on the transverse-moment-dependent (TMD) factorization to evaluate the $B\to D$ form factors~\cite{Li:1994zm, Kurimoto:2002sb} in which case a systematic power-counting $m_b\gg m_c \gg\Lambda_{\rm QCD}$ was proposed. 
This route has been further explored in ref.~\cite{Fan:2013qz,Hu:2019bdf} and a wealth of perturbative matching coefficient functions has been obtained~\cite{Li:2010nn, Li:2012nk, Li:2012md, Li:2013xna} supplementing this endeavor.   
The TMD factorization for hard exclusive processes is yet to be fully understood as it requires a definite power-counting scheme for the intrinsic transverse momentum~\cite{Wang:2015qqr} as well as a proper construction of the Wilson-line for the TMD wave functions which is necessary to circumvent the rapidity and pinch singularities in the infrared subtraction procedure~\cite{Li:2014xda}. 
As an active on-going field of research, we are looking forward to seeing all these matters settled in the near future.

The main goal of the present work is to establish a systematic evaluation of the next-to-leading-power (NLP) corrections of the $B\to D$ process at the tree-level employing up to twist-six $B$-meson LCDAs as the nonperturbative inputs. 
From a phenomenological point of view, this allows us to quantify the size of such corrections which is of paramount importance given the potential BSM signal the $B\to D\ell\nu_\ell$ process may carry. 
Besides, a better estimate of the subleading power corrections is essential in providing evidence for a clear hierarchy of power expansion, which is crucial for any theory prediction to be trustworthy.
From the theory side, it is also an interesting subject as it represents a nontrivial example for exploring the factorization properties of $B$-meson decays at subleading power, which has drawn much attention in recent years, and has been explicitly shown to hold at tree level for certain decay channels, see e.g.~\cite{Beneke:2018wjp,Shen:2020hfq}. Besides the more theoretically motivated insights, it is also of phenomenological importance to achieve better determinations of $|V_{cb}|$ and $R(D)$ by bridging the gap between recent lattice results at the low hadronic recoil and our advanced LCSR computations at large recoil. Improved  unitarity bounds~\cite{Bigi:2016mdz,DiCarlo:2021dzg,Martinelli:2021onb} provide another handle to lower uncertainties in our predictions for $|V_{cb}|$ and $R(D)$.

The present article is organized as follows. In section~\ref{sec:lp}, we establish the relevant definitions, kinematics and notations and present the $\alpha_s$ corrections to the leading-power contribution. 
Section~\ref{sec:subpower} is then devoted to the derivation of the various contributions at subleading power. %
Section~\ref{sec:num} is reserved for numerical studies where we extract the Boyd-Grinstein-Lebed (BGL) parameters using our theoretical predictions complemented by lattice data~\cite{MILC:2015uhg, Na:2015kha,Kaneko:2019vkx}, which is currently still restricted to the small-recoil region, in order to survey the entire kinematic range of the decay process.
In this section, we also provide our prediction for the decay rates of $B\to D\ell\nu_\ell$ from which we are able to extract $|V_{cb}|$ by fitting to the BaBar and Belle data. 
The celebrated ratio $R(D)$ predicted in this work can be found in~\eqref{eq:RD}. Finally in section~\ref{sec:summary}, we discuss the implications of our results as well as potential avenues for future investigations. Most of the technical details are collected in the appendices.

%
%%%%%%%%%%%%%%%%%%%%%%%%%%%%%%%%%%%%%%%%%%%%%%%%%%%%%%%%%%%%%%%%%%%%%%%%%%%%
\section{Leading-power contributions}
\label{sec:lp}

\subsection {General framework}
\label{sec:framework}
%%%%%%%%%%%%%%%%%%%%%%%%%%%%%%%%%%%%%%%%%%%%%%%%%%%%%%%%%%%%%%%%%%%%%%%%%%%%
%
The $B\to D$ decay amplitude is parametrized by two form factors $f_{BD}^{+}$ and $f_{BD}^{0}$ through Lorentz decomposition of the $B\to D$ matrix element as follows
\begin{align}
\langle{D}(p)\left|\bar c\gamma_\mu b\right| \bar B(p_B) \rangle = f_{BD}^{+}(q^2)\left[2p_\mu
+\left(1-\frac{m_B^2-m_D^2}{q^2}\right)q_\mu\right]+f_{BD}^{0}(q^2)\frac{m_B^2-m_D^2}{q^2}q_\mu\, ,
\end{align}
where $p$ and $p_B$ respectively denote the momenta of the $D$ meson and the $B$ meson and $q\equiv p_B-p=m_Bv-p$ is the momentum transfer between the two meson states.
It is convenient to consider the decay process in the rest frame of the $B$ meson fixing $v^\mu=(1,0,0,0)$.
The $B\to D$ transition is induced by the flavor changing current $\bar c\gamma_\mu b$ which effectively converts the bottom quark into the charm one in accordance with the weak interaction.
The main objective of this work is to offer theoretical predictions for $f_{BD}^+(q^2)$ and $f_{BD}^0(q^2)$. 

The LCSR approach to the study of the $B\to D$ decay takes advantage of  vacuum-to-$B$-meson correlation function 
\begin{align}
\Pi_\mu(n\cdot p, \bar n\cdot p)=&~
i\int d^4x\,\e^{ip\cdot x}\langle{0} \left | T\left\{\bar q(x)\slashed{n}\gamma_5 c(x)\,,\bar c(0)\gamma_\mu b(0)\right\} \right|\bar B(p_B)\rangle \, ,
\label{eq:Pi-correlation}
\end{align}
with the constituents of the $D$-meson state acting as an interpolating operator.
This correlation function will constitute the main object of study in section~\ref{sec:subpower}.
In the above formula $q$ ($c$) denotes the light (charm) quark field and we have introduced two light-like vectors $n$ and $\bar n$ satisfying\footnote{
To fix the normalization, we explicitly take $n_\mu=(1,0,0,1)$ and $\bar n_\mu=(1,0,0,-1)$,
which means $v_\mu=(n_\mu+\bar n_\mu)/2$ in the $B$-meson rest frame.} $\bar n^2=n^2=0$ and $n\cdot\bar n=2$.
Together, they span the two-dimensional plane where the $B\to D$ decay takes place, 
allowing to decompose any vectors of this process in terms of these two light-like vectors.
In order to facilitate the power expansion, we adopt the following power counting scheme in the kinematic region of large hardonic recoil
\begin{align}\label{eq:power-counting}
\left|\bar n\cdot p\right| \sim {\cal O}(\lambda^2 m_b)\, ,\qquad n\cdot p=\frac{m_B^2+m_D^2-q^2}{m_B}\sim{\cal O}(m_b)\, ,\qquad  m_c\sim {\cal O}(\lambda m_b)\,,
\end{align}
where $\lambda\sim {\cal O}(\sqrt{\Lambda_{\rm QCD}/m_b})$.
The hadronic representation of the correlation function~\eqref{eq:Pi-correlation} is
\begin{align}
\Pi^{\rm had}_{\mu}(n\cdot p,\bar{n}\cdot p) 
=&~ \frac{if_{D} \, m_B}{2 \, (m_{D}^2/ n \cdot p - \bar n \cdot p)}
\bigg \{ \left [ \frac{n \cdot p}{m_B} \, f_{BD}^{+} (q^2) + f_{BD}^{0} (q^2)  \right ]\,\bar n_{\mu}
\nonumber \\
& + \frac{m_B}{n \cdot p-m_B}\,
\left [ \frac{n \cdot p}{m_B} \, f_{BD}^{+} (q^2) -  f_{BD}^{0} (q^2)  \right ] \,n_{\mu} \bigg \} \, \nonumber \\
&+ \int_{\omega_s}^{\infty}  \frac{d \omega^{\prime}}{\omega^{\prime} - \bar n \cdot p-i0} \,
\left [\rho_{\bar n}(n \cdot p,\omega^{\prime})  \, \bar{n}_{\mu}  
+\rho_{n}(n \cdot p,\omega^{\prime})   \, n_{\mu} \right ]\,, 
\label{eq:hadronic}
\end{align}
where $\omega_s$ is the threshold parameter for which we will explain in detail later. 
The $D$-meson decay constant $f_D$ is defined as
\begin{align}
\langle{0} \left| \bar q \slashed{n}\gamma_5 c \right| D(p)\rangle = in\cdot p\,f_D\, .
\end{align}
The general rationale of the LCSR approach is that in certain Euclidean kinematic regions ($|p^2|\sim{\cal O}(\lambda^2 m^2_b)$), the integral in~\eqref{eq:Pi-correlation} 
is dominated by light-like separations where the light-cone expansion can be properly constructed.
In this perspective, the partons account for the fundamental degrees of freedom subject to the QCD perturbative corrections forming the foundation for our calculations.
The partonic-level results of the correlation function can be parameterized as
\begin{align}
\Pi^{\rm par}_\mu(n\cdot p, \bar n\cdot p)
= i\frac{\widetilde{f}_B(\mu)m_B}{n\cdot p}\,
\Big[\Pi_{\bar n}(n\cdot p, \bar n\cdot p)\, \bar{n}_\mu
+ \Pi_{n} (n\cdot p, \bar n\cdot p)\, n_\mu\Big]\, .
\label{eq:partonic}
\end{align}
To extract the form factors $f_{BD}^{+}(q^2)$ and $f_{BD}^{0}(q^2)$, we make extensively use of dispersion relation and express the partonic level representation of the correlation function as
\begin{align}
\Pi_{n,\bar n}(n\cdot p,\bar{n}\cdot p) = \int^{\infty}_{0} \frac{d\omega'}{\omega'-\bar n\cdot p-i0}\,
\frac{1}{\pi} {\rm Im_{\omega'}} \Pi_{n,\bar n}(n \cdot p,\omega') \,.
\end{align}

The LCSR procedure matches two distinct descriptions of the correlator --- the QCD representation $\Pi^{\rm par}_\mu$
derived with the perturbative factorization approach,
and the physical one $\Pi^{\rm had}_\mu$ obtained with the hadronic dispersion relation.  
In practice, certain duality assumptions have to be drawn to make the matching possible. 
The general interpretation of the quark-hadron duality is that above a certain (continuum) threshold $\omega_s$ the hadronic, in our current case corresponding to the hadronic states with the same quantum number of the $D$ meson, spectral density of
 physical observables coincides with that of the QCD one averaged by a weight function when the space-like $p^2$ is large
\begin{align}
\int_{\omega_s}^{\infty}  \frac{d \omega^{\prime}}{\omega^{\prime} - \bar n \cdot p-i0} \,
\rho_{n,\bar n}(n \cdot p,\omega^{\prime})
=\frac{i\widetilde{f}_B(\mu)m_B}{n\cdot p}
\int_{\omega_s}^{\infty}  \frac{d \omega^{\prime}}{\omega^{\prime} - \bar n \cdot p-i0} \,
\frac{1}{\pi}
{\rm Im}_{\omega'} 
\Pi_{n,\bar n}(n \cdot p,\omega^{\prime}) \,.
\end{align}
After the continuum subtraction, the resulting expression from the matching then in principle enables one to make theory predictions in the large-recoil region of the physical process.
In reality, however, caution has to be taken in assessing the LCSR results as systematic uncertainties from the parton-hadron duality must be under control. 
The final trick of this technique is to introduce the Borel transform for the purpose of suppressing the contributions from the continuum states.
Then the form factors can be written as
\begin{align}
&~f_D 
\exp\left[-\frac{m_D^2}{n\cdot p\, \omega_M}\right]\left\{\frac{n\cdot p}{m_B}f_{BD}^{+} (q^2),f_{BD}^{0} (q^2)\right\}
\nonumber \\
=& ~ 
\frac{\widetilde{f}_B(\mu)}{n\cdot p} \int^{\omega_s}_0 d\omega' \exp\left[-\frac{\omega'}{ \omega_M}\right]\frac{1}{\pi}
\left[{\rm Im_{\omega'}} \left( \Pi_{\bar n}(n \cdot p,\omega')
\pm\frac{n\cdot p-m_B}{m_B} \Pi_n(n \cdot p,\omega')\right)\right]\,,
\label{eq:lcsr}
\end{align}
where the ``$+$" and ``$-$" signs contribute to $f_{BD}^{+}$ and $f_{BD}^{0}$ separately.
$\omega_M$ and $\omega_s$ denote, respectively, the Borel and threshold parameter with mass dimension one, constituting the two inputs in the framework of LCSR.

\subsection{Leading-power contribution at tree level}
\label{subsection: LP at tree level}

%%%%%%%%%%%%%%%%%%%%%%%%%%%%%%%%%%%%%%%%%%%%%%%%%%%%%%%%%%%%%%%
%%%%%%%%%%%%%%%%%%%%%%%%%%%%%%%%%%%%%%%%%%%%%%%%%%%%%%%%%%%%%%%
\begin{figure}[!ht] 
\begin{center} 
\includegraphics[width=0.25\textwidth]{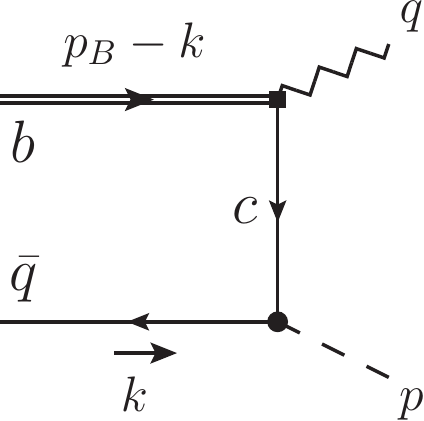} 
\end{center}
\vspace{-0.5cm}
\caption{Two-particle contribution to the correlation function $\Pi_\mu(n\cdot p,\bar n\cdot p)$ at tree level. 
}
\label{fig:2ptree}
\end{figure}

The leading power contribution at tree level to the correlation function~\eqref{eq:Pi-correlation} can be obtained in 
the $p^2<0$ region by evaluating the Feynman diagram in figure~\ref{fig:2ptree}. 
Following the power counting scheme~\eqref{eq:power-counting} and applying the definition of the $B$-meson LCDAs given in~\eqref{def:two}, 
one obtains the leading-power (LP) contribution at tree level to the correlation function\footnote{After integrating out hard/hard-collinear degrees of freedom the matrix element is written solely in terms of the soft (non-perturbative) dynamics in HQET. At the LP approximation this amounts to replacing $b \to h_v$ and $\bar{B}(p_B) \to \bar B_v$.},
\begin{align}
\contraction{\Pi_{\mu,\rm LP}(n\cdot p,\bar n\cdot p)=i\int d^4x\,\e^{ip\cdot x}\langle{0} |\bar q(x)\slashed{n}\gamma_5\,}{c}{(x)}{c}
\Pi_{\mu,\rm LP}^{\rm tree}(n\cdot p,\bar n\cdot p)&=i\int d^4x\,\e^{ip\cdot x}\langle{0} \left |\bar q(x)\slashed{n}\gamma_5 c(x)\bar c(0)\gamma_\mu h_v \right| {\bar B_v}\rangle\notag\\
&=-i\widetilde{f}_B(\mu)\,m_B \,\bar n_\mu \int^\infty_0d\omega\,\frac{\phi_B^{-}(\omega,\mu)}{\bar n\cdot p-\omega-\omega_c+i0} \,\notag\\
&=-i\widetilde{f}_B(\mu)\,m_B \, \bar n_\mu \int^\infty_{\omega_c} d\omega^\prime\,\frac{\phi_B^{-}(\omega^\prime-\omega_c,\mu)}{\bar n\cdot p-\omega^\prime+i0}\, ,\label{eq:2P-tree}
\end{align}
where $\omega_c\equiv m_c^2/n\cdot p$, and we have applied the heavy quark equation of motion (EOM) $\slashed{v}h_v=h_v$ and performed the Fourier transform according to~\eqref{def:LCDAFourier} in the intermediate step. 
The scale-dependent $B$-meson decay constant $\widetilde{f}_B(\mu)$ in HQET can be further expressed in terms of its QCD counterpart $f_B$
via a perturbatively calculable matching coefficient
\begin{align}
\widetilde{f}_B(\mu)  & 
= f_B\left[1-\frac{\alpha_s{(\mu)} C_F}{4\pi}\!
\left(3\ln\frac{m_b}{\mu}-2\right)+ {\cal O}(\alpha_s^2) \right].\label{defF}
\end{align}
Here we have dropped the subleading power terms in the hard-collinear charm-quark propagator, whose factorization properties  will be explored at length with the aid of the HQET equations of motion in  section~\ref{subsec:QPE}.
The third equality in~\eqref{eq:2P-tree} presents the spectral representation of the correlation function $\Pi_\mu$.
Taking the massless limit $m_c\to 0$, the obtained tree-level factorization formula \eqref{eq:2P-tree} reduces to the $B\to\pi$ case \cite{Wang:2015ndk} as expected.
We mention in passing that the achieved LP contribution to the vacuum-to-$B$-meson correlation function under discussion is governed by the \emph{subleading}-twist LCDA  $\phi_B^-(\omega)$,   thus obstructing the extraction of the fundamental shape parameters  $\lambda_B$ and $\widehat{\sigma}_n$ as defined in~\eqref{eq:log-moms}.
It proves more convenient to access such non-perturbative quantities by investigating the non-hadronic $B$-meson decay processes \cite{Beneke:2018wjp,Beneke:2011nf,Braun:2012kp,Wang:2016qii,Wang:2018wfj,Beneke:2020fot,Beneke:2021rjf,Wang:2021yrr}.

%
%%%%%%%%%%%%%%%%%%%%%%%%%%%%%%%%%%%%%%%%%%%%%%%%%%%%%%%%%%%%%%%%%%%%%%%%%%%%
\subsection{Leading-power contribution at \texorpdfstring{${\cal O}(\alpha_s)$}{Lg}}
\label{sec:NLL}
%%%%%%%%%%%%%%%%%%%%%%%%%%%%%%%%%%%%%%%%%%%%%%%%%%%%%%%%%%%%%%%%%%%%%%%%%%%%
%

As it was already demonstrated in ref.~\cite{Wang:2017jow}, the (partial) next-to-leading-logarithmic (NLL) corrections of the leading power are in general numerically more prominent compared to the contributions of the three-particle LCDAs.
Therefore, in order to make a sensible prediction for the $B \to D$ form factors with an inclusion of the subleading power contributions, it is necessary to take into account the NLL corrections of the leading power as well. 
The factorization formula for the correlation function~\eqref{eq:Pi-correlation} at the one-loop level in $\alpha_s$, with the hard function and the jet function encoding separately contributions of the hard scale and the hard-collinear scale, can be obtained with the help of the method of regions~\cite{Beneke:1997zp,Smirnov:2002pj}, see ref.~\cite{Wang:2017jow} for the detail.

The NLL resummation improved LCSR prediction for the leading-power contribution to the form factors was derived in~\cite{Wang:2017jow} which we quote here,
\begin{align}
& f_D \, \exp\left[-\frac{m_D^2}{n\cdot p\, \omega_M}\right] \,
\left \{ \frac{n \cdot p}{m_B} \, f_{B D}^{+,\rm LP}(q^2) \,, \,  f_{B D}^{0,\rm LP}(q^2)  \right  \}  \nonumber \\
= &  -  U_2(\mu_{h_2}, \mu) \,
\widetilde{f}_B(\mu_{h_2}) \,
\int_{0}^{\omega_s} \, d \omega^{\prime} \,\exp\left[-\frac{\omega'}{ \omega_M}\right]
\left\{{\cal F}^{\rm LP}_{+}(\omega^\prime,\mu)\, ,\, {\cal F}^{\rm LP}_{-} (\omega^\prime,\mu)\right\}\, ,
\label{eq:NLL-master}
 \end{align}
 where $\omega_s$ denotes the effective continuum threshold in the $D$-meson channel, and
 \begin{align}
 {\cal F}^{\rm LP}_{\pm} &= C_{+, \bar n}(n \cdot p, \mu) \,   \Phi_{+, \bar n}^{\rm eff}(\omega^{\prime}, \mu) \,
+ U_1(n \cdot p, \mu_{h_1}, \mu) \, C_{-,\bar n}(n \cdot p, \mu_{h_1}) \,
\Phi_{-, \bar n}^{\rm eff}(\omega^{\prime}, \mu)   \nonumber \\
& \pm \frac{n \cdot p - m_B}{m_B} \, \left [  C_{+, n}(n \cdot p, \mu) \,   \Phi_{+, n}^{\rm eff}(\omega^{\prime}, \mu)
 + C_{-, n}(n \cdot p, \mu) \,   \Phi_{-, n}^{\rm eff}(\omega^{\prime}, \mu) \right ]\,.
 \label{eq:calFLP}
\end{align}
The simplified $\Phi_{\pm,\bar n}^{\rm eff}$ and $\Phi_{\pm,n}^{\rm eff}$ are collected in~\eqref{eq:Phi-eff-1} -- \eqref{eq:Phi-eff-4}. 
The hard-matching coefficient functions $C_{\pm,\bar n}(n\cdot p,\mu) $ and $C_{\pm,n}(n\cdot p,\mu)$ can be found in~\eqref{eq:hard-coeff}. 
$U_1(n\cdot p, \mu_{h_1},\mu)$ and $U_2(\mu_{h_2},\mu)$ are the evolution factor for $C_{-,\bar n}(n\cdot p,\mu)$ and $\widetilde{f}_B(\mu)$, respectively 
which effectively resum large logarithms of the form $\ln(\mu/\mu_{h_1})$ and $\ln(\mu/\mu_{h_2})$ allowing the matching calculation to
be carried out at the hard scales $\mu_{h_1}$ and $\mu_{h_2}$, significantly above the factorization scale $\mu$ so that the perturbative calculation is under good control.
The explicit form of these functions corresponding to the NLL resummation can be found in ref.~\cite{Beneke:2011nf}.
The large logarithms $\ln(\mu/\mu_0)$ in the LCDAs $\phi^\pm_B(\omega,\mu)$ with $\mu_0$ of ${\cal O}(\Lambda_{\rm QCD})$ are also resummed to the leading-logarithmic accuracy~\eqref{eq:LL_lcda}.

The LP form factors at the tree level, which correspond to the contribution of the effective heavy-to-light current $J^{\rm (A0)}$ in SCET~\cite{Beneke:2002ph}, preserve the large-recoil symmetry relation $f^0_{BD}(q^2)=n\cdot p/m_B f^+_{BD}(q^2)$.
The symmetry is broken by both the power corrections in $\Lambda_{\rm QCD}/m_b$ and the perturbative corrections in $\alpha_s$~\cite{Beneke:2000wa}.
The breaking of the symmetry relation by ${\cal O}(\alpha_s)$ effects can be seen directly from~\eqref{eq:NLL-master} noticing the last line of~\eqref{eq:calFLP} is proportional to $\alpha_s$.
The culprit responsible for the symmetry breaking comes from integrating out the hard and hard-collinear fluctuations. We will show in the following section that the NLP contributions can also break the large-recoil symmetry.
It is also worth noting that the power-enhanced contribution (of ${\cal O}(\lambda^2)$), which is proportional to the charm-quark mass $m_c$, is only present in $\Phi_{+, \bar n}^{\rm eff}(\omega^{\prime}, \mu)$ and therefore preserves the large-recoil symmetry relation. 
Here we draw an analogy to the SCET analysis of $B\to V$ form factors \cite{Gao:2019lta} to understand the power enhancement of $\Phi_{+, \bar n}^{\rm eff}(\omega^{\prime}, \mu)$. Upon the replacement of $m_q\to m_c$, with $m_q$($m_c$)  maintaining the power counting of ${\cal O}(\lambda^2 m_b)$(${\cal O}(\lambda m_b)$),
the form factor receives the power enhancement $ {\cal O}(\lambda^{3}) \to {\cal O}(\lambda^2)$.
%

%
%%%%%%%%%%%%%%%%%%%%%%%%%%%%%%%%%%%%%%%%%%%%%%%%%%%%%%%%%%%%%%%%%%%%%%%%%%%%
\section{Subleading-power contribution at tree level}
\label{sec:subpower}
%%%%%%%%%%%%%%%%%%%%%%%%%%%%%%%%%%%%%%%%%%%%%%%%%%%%%%%%%%%%%%%%%%%%%%%%%%%%
%

The main motivation of this paper is to systematically investigate the subleading power corrections 
to the $B\to D$ form factors at the tree level. 
These corrections can be organized into three classes according to their origins.
\begin{itemize}
\item Subleading-power corrections generated by higher-twist LCDAs including both two-particle and three-particle contributions. 
\item An expansion of the charm quark propagator in figure~\ref{fig:2ptree} in which case the leading-twist $B$-meson LCDA contributes with a power suppression $\Lambda_{\rm QCD}/(n\cdot p)\sim {\cal O}(\lambda^2)$. 
\item Heavy quark expansion which takes into account the $\Lambda_{\rm QCD}/m_b$ corrections in the HQET framework.
\end{itemize}
%

%
%%%%%%%%%%%%%%%%%%%%%%%%%%%%%%%%%%%%%%%%%%%%%%%%%%%%%%%%%%%%%%%%%%%%%%%%%%%%
\subsection{Higher-twist contribution}
\label{subsec:HT}
%%%%%%%%%%%%%%%%%%%%%%%%%%%%%%%%%%%%%%%%%%%%%%%%%%%%%%%%%%%%%%%%%%%%%%%%%%%%
%

Generally speaking, the higher-twist $B$-meson LCDAs can give rise to the higher-power contributions as already observed in, e.g.~\cite{Beneke:2018wjp,Wang:2017jow}, which can be evaluated presumably
by applying the perturbative factorization technique for the correlation function $\Pi_\mu$ defined in~\eqref{eq:Pi-correlation}.

One source for such contributions comes from the interactions between the hard-collinear charm quark and the soft gluon(s) emitted from the $B$-meson state.
To properly motivate this particular type of contribution, 
it is necessary to consider the scenario where the massive $c$-quark propagates in the vicinity of the light-cone with a  soft-gluon background,  which acquires an expansion in terms of  
the strong coupling~\cite{Balitsky:1987bk,Belyaev:1994zk} (see also \cite{Khodjamirian:2010vf, Khodjamirian:2012rm} for an alternative but symmetric form)
\begin{align}
\contraction{}{q}{(x)}{\,\,q}
c_i (x) \bar c_j(0)
&=\int \!\! \frac{d^4k}{(2\pi)^4}\frac{i\,\e^{-ik\cdot x}}{\slashed{k}-m_c}\bigg\{\delta_{ij}
+g_s\int^1_0 \!\! du\left[\frac{ux^\mu\gamma^\nu}{\slashed{k}+m_c}-\frac{\sigma^{\mu\nu}}{2(k^2-m_c^2)}\right]G^a_{\mu\nu}(ux)(T^a)_{ij}\bigg\} +\ldots
\label{eq:prop-expand}
\end{align}
with $\sigma^{\mu\nu}=\frac{i}{2}[\gamma^\mu,\gamma^\nu]$, $G^a_{\mu\nu}$ the gluon field strength tensor, and $T^a$ denotes the generators in the fundamental representation. 
It is clear that the second term in the curly bracket generates the subleading-power corrections involving the three-particle $B$-meson  LCDAs (i.e.\ LCDA with particle content as antiquark-gluon-quark).
Substituting this term into the correlation function~\eqref{eq:Pi-correlation}
and evaluating the resulting matrix element using~\eqref{def:three} leads to the three-particle higher-twist (3PHT) contribution
\begin{align}
\Pi^{\rm 3PHT}_{\mu}(n\cdot p,\bar n\cdot p) &=
-\frac{i \widetilde{f}_B(\mu) m_B}{n\cdot p}\,
\int^\infty_0 d\omega\,d\xi\int_0^1  
\frac{d u}{(\bar n\cdot p-\omega-u\xi - \omega_c)^2}
\notag \\
&\times \bigg\{
\Big[(1-2u)\,\psi_5(\omega,\xi)+\widetilde\psi_5(\omega,\xi)
+\frac{2m_c}{n\cdot p}\,\phi_6(\omega,\xi)\,\Big]\bar n_\mu \notag\\
&\quad+\Big[2(u-1)\,\phi_4(\omega,\xi)
+\frac{m_c}{n\cdot p}\,\left(\widetilde\psi_5(\omega,\xi) -\psi_5(\omega,\xi)\right)\Big]n_\mu
\bigg\}\, ,\label{eq:3pht}
\end{align}
where we have neglected terms that generate power corrections beyond the subleading order. We would like to note that the linear $m_c$ term is of ${\cal O}(\lambda^3)$ suppressed compared to the LP contribution of the correlation function with $m_c\sim {\cal O}(\lambda m_b)$. We will keep this term because we also want to explore the NLP corrections to the linear $m_c$ terms which will be more relevant with the power counting $m_c\sim  {\cal O}(m_b)$~\cite{Beneke:2000ry,Gambino:2012rd}.

In addition, the first term in the curly bracket of~\eqref{eq:prop-expand} also generates subleading 
power contributions from the higher-twist two-particle LCDAs, e.g., $g_B^\pm(\omega)$ 
in~\eqref{def:two}, after being sandwiched between the $B$-meson state and the vacuum.
In the partonic language, such contributions are attributed to the nonvanishing transverse momenta 
of the light quark travelling in close proximity to, but not exactly on the light-cone.
We therefore obtain the subleading-power corrections to the correlator due to the higher-twist two-particle LCDAs
\begin{align}
\Pi^{\rm 2PHT}_{\mu} &=
i\int\frac{d^4k}{(2\pi)^4}\int d^4 x \, e^{i(p-k)\cdot x}
\langle 0 | 
\bar q(x) \slashed{n}\gamma_5\frac{i(\slashed{k}+m_c)}{k^2-m^2_c} 
\gamma_\mu   h_v(0) 
 |  \bar B_v \, \rangle
\nonumber \\
&= \frac{i \widetilde{f}_B(\mu)\, m_B}{4}\,
\int_0^{\infty} d \omega \int\frac{d^4k}{(2\pi)^4} \int d^4 x \,
\e^{i(p-k-\omega v)\cdot x}\bigg\{
\frac{x^2}{k^2-m^2_c}
\nonumber \\
& \times \Tr\bigg [  \frac{1+\slashed{v}}{2}
\bigg( 2 g_B^{+}(\omega)
- \frac{\slashed{x}}{ v \cdot x}   \Big[ g_{B}^{+}(\omega) - g_{B}^{-}(\omega)  \Big]   \bigg)
\gamma_5\slashed{n}\gamma_5(\slashed{k}+m_c)\gamma_\mu \bigg]+ {\cal O}(x^4)\bigg\}\notag\\
&=  i \widetilde{f}_B(\mu) \, m_B
\int_0^{\infty}  d \omega \, g_B^{-}(\omega)\, 
\partial^2_p\left[
\frac{n\cdot p\,\bar n_\mu+m_c\,n_\mu}{(p-\omega v)^2-m^2_c}\right] \notag\\
&=4\frac{i \widetilde{f}_B(\mu) \,m_B}{-n\cdot p}
\int_0^{\infty} 
\frac{d \omega\, g_B^{-}(\omega)}{(\bar n\cdot p-\omega-\omega_c)^3}\,
\bigg[ \big(\bar n\cdot p-\omega-3\omega_c\big)\,\bar n_\mu
-2\frac{\omega_c\,m_c}{n\cdot p}\,n_\mu\,\bigg]\, ,\label{eq:2pht}
\end{align}
where we have used $\slashed{x}/(v\cdot x)=\slashed{\bar n} + \cdots$ with the ellipses representing terms that generate power contributions beyond the subleading order and are therefore ignored together with ${\cal O}(x^4)$ contributions at the third equal sign.
Note that this expression is in agreement with~(3.12) in ref.~\cite{Lu:2018cfc} after taking the limit of $m_c \to 0$, serving as a consistency check.

Summing up~\eqref{eq:3pht} and \eqref{eq:2pht} therefore yields the tree-level higher-twist (HT) contribution
to the correlator at the subleading power
\begin{align}
\Pi^{\rm HT}_{\mu}(n\cdot p,\bar n\cdot p) &=
-\frac{i \widetilde{f}_B(\mu) \,m_B}{n\cdot p}\int^\infty_0 d\omega
\bigg\{
 \frac{\tau^{\bar n}_{2}(\omega)\,\bar n_\mu}{\bomega_2^2}
+\frac{m_c}{n\cdot p}\,\frac{\widetilde\tau^{\,n}_{2}(\omega)\,n_\mu}{\bomega_2^2}
\nonumber \\
&+\int^\infty_0 d\xi\int_0^1
\frac{d u}{\bomega_3^2}
\Big[
\Big( \tau^{\bar n}_{\rm 3}(u,\omega,\xi)
+\frac{m_c}{n\cdot p}\,\widetilde\tau^{\,\bar n}_{3}(\omega,\xi)\,\Big)\bar n_\mu
\nonumber \\
& \hspace{2.5cm}
+\Big(\tau^{n}_{3}(u,\omega,\xi)
+\frac{m_c}{n\cdot p}\,\widetilde\tau^{\,n}_{3}(u,\omega,\xi)\Big)\, n_\mu
\Big]
\bigg\}\,,\label{eq:HT-master}
\end{align}
where 
\begin{align}
{\boldsymbol{\omega}}_2=\bar n\cdot p-\omega-\omega_c\, ,\qquad  {\boldsymbol{\omega}}_3=\bar n\cdot p-\omega-u\xi-\omega_c\, ,\label{eq:bomega12}
\end{align}
with $\bomega_i$ modulating the contribution of the $i$-particle LCDA and
\begin{align}
\tau^{\bar n}_{2}(\omega)
&=4\widehat{g}^{\, -}_B(\omega)\Big[1
-2\frac{\omega_c}
{\bomega_2}\Big]\,,
& \widetilde\tau^{\,n}_{2}(\omega)
&=
-\frac{8\omega_c\,\widehat{g}^{\,-}_B(\omega)}
{\bomega_2}\,,
\nonumber \\
\tau^{\bar n}_{3}(u,\omega,\xi)
&=\widetilde\psi_5(\omega,\xi)-\psi_5(\omega,\xi)
+\frac{4\bar{u}\,\omega_c\,\psi_5(\omega,\xi)}
{\bomega_3}\,,
& \tau^{n}_{3}(u,\omega,\xi)
&=-2\bar{u}\,\phi_4(\omega,\xi)\,,
\nonumber \\
\widetilde\tau^{\,\bar n}_{3}(\omega,\xi)
&=2\phi_6(\omega,\xi)\,,
& \widetilde\tau^{\, n}_{3}(u,\omega,\xi)
&=\tau^{\bar n}_{3}(u,\omega,\xi)\, ,
\end{align}
where $\bar u =1-u$.
Taking the massless limit $\omega_c\to0$ reproduces~(3.17) of \cite{Lu:2018cfc} as expected.
The explicit definitions of the higher-twist LCDAs are collected in~\eqref{eq:3pt3} --- \eqref{eq:3pt6}, and~\eqref{def:two}.
The object $\widehat{g}^{\,-}_B(\omega)$ defined in~\eqref{def:G-} corresponds to the ``Wandzura-Wilczek" (WW) component of the two-particle twist-5 LCDA $g^-_B(\omega)$. Namely, $\widehat{g}^{\,-}_B(\omega)$ is entirely determined by the lower-twist LCDAs as shown in~\eqref{eq:G-}.

%
%%%%%%%%%%%%%%%%%%%%%%%%%%%%%%%%%%%%%%%%%%%%%%%%%%%%%%%%%%%%%%%%%%%%%%%%%%%%
\subsection{Charm-quark expansion}
\label{subsec:QPE}
%%%%%%%%%%%%%%%%%%%%%%%%%%%%%%%%%%%%%%%%%%%%%%%%%%%%%%%%%%%%%%%%%%%%%%%%%%%%
%

The correlation function $\Pi_{\mu}(n\cdot p,\bar n\cdot p)$ receives corrections from the charm quark propagator expansion (CE). 
The tree-level CE contribution to the correlation function is derived from
\begin{align}
\Pi_{\mu}(n\cdot p,\bar n\cdot p) 
\supset -\int d^4x\int \frac{d^4k}{(2\pi)^4}
e^{i k\cdot x}
\langle 0 | \bar q(x)\slashed{n}\gamma_5
\frac{(\slashed{p}-\slashed{k})+m_c}{(p-k)^2-m^2_c+i0} \gamma_\mu h_v(0) | \bar B_v \rangle \,,
\end{align}
with the expansion of the charm-quark propagator
\begin{align}
\frac{(\slashed{p}-\slashed{k})+m_c}{(p-k)^2-m^2_c}
=&~ \mathop{\underbrace{
\frac{1}{\bar n \cdot \hat{p}}
\frac{\slashed{\bar n}}{2}}}\limits_{\rm LP}
+
\mathop{\underbrace{
\frac{1}{n\cdot p\bar n \cdot \hat{p}}
\Big[\bar n\cdot p
\frac{\slashed{n}}{2}-\slashed{k}+
\frac{n\cdot k \bar{n}\cdot p}{\bar n \cdot \hat{p}}\frac{\slashed{\bar n}}{2}\Big]}}\limits_{\rm NLP}
+
\mathop{\underbrace{\frac{m_c}{n\cdot p}
\frac{1}{\bar n \cdot \hat{p}}
\Big[1+
\frac{n\cdot k\bar{n}\cdot p}{n\cdot p\bar n \cdot \hat{p}}\Big]}}\limits_{m_c~\rm NLP},
\label{eq:charm_expansion}
\end{align}
where $\bar n \cdot \hat{p}=\bar{n}\cdot p-\bar{n}\cdot k-\omega_c$.
Note that although the second term in the last brackets is of ${\cal O}(\lambda^3)$ suppressed compared to the LP term with $m_c\sim {\cal O}(\lambda m_b)$, 
we will keep this term as explained above.
The NLP of this type reads,
\begin{align}
&\Pi^{\rm CE}_{\mu,\rm NLP}(n\cdot p,\bar n\cdot p) 
= \int d^4x\int \frac{d^4k}{(2\pi)^4}
e^{i k\cdot x}\,\frac{n\cdot k\,}{n\cdot p\,\bar n \cdot \hat{p}}
\nonumber \\
&\times\langle  0 |  \bar q(x) 
\left[
-\frac{m_c}{n\cdot k}\slashed{n} n_\mu+\left(\slashed{n}\bar n_\mu+\slashed{\bar n}n_\mu\right)
+\frac{n_\mu \slashed{k}\slashed{n}}{n\cdot k}
-\frac{\bar{n}\cdot p}
{\bar n \cdot \hat{p}}
\Big(\bar{n}_{\mu}+\frac{m_{c}}{n\cdot p}n_{\mu}\Big)\slashed{n}\right]
\gamma_5  h_v(0)  |   \bar B_v \rangle\, ,
\label{eq:CE-matrix-element}
\end{align}
where we have applied the heavy quark EOM $\slashed{v}h_v=h_v$ in the second step.

We now proceed to relate the matrix elements in \eqref{eq:CE-matrix-element} to the $B$-meson LCDAs in a factorized form.
For the first term in the square bracket, we write $n \cdot k=2v\cdot k-\bar n\cdot k$.
 While $\bar n\cdot k$ is readily equal to $\omega$ after integrating over $k^\mu$ and evaluating the yielding matrix element using the definition of the two-particle higher-twist LCDAs given in eq.~\eqref{def:two}, 
 the $2v\cdot k$ part demands more effort. 
 We thus detail the derivation of the $2v\cdot k$ contribution in the following, we have,
 \begin{align}
\Pi^{\rm CE,(1)}_{\mu,\rm NLP} (n\cdot p,\bar n\cdot p) &\equiv \int d^4x\int \frac{d^4k}{(2\pi)^4}\, \e^{ik\cdot x}\,
\frac{2v\cdot k\,\langle  0 |  \bar q(x) \Gamma_\mu  h_v(0) |  \bar B_v \rangle}{n\cdot p\,(\bar{n}\cdot p-\bar n\cdot k-\omega_{c})}
\label{eq:CE1}\\
&=2i\int \frac{d^4k}{(2\pi)^4}\int d^4x\, \e^{ik\cdot x}\, \frac{v\cdot\partial_x \langle  0 |  \bar q(x) \Gamma_\mu  h_v(0) |  \bar B_v \rangle}{n\cdot p\,(\bar{n}\cdot p-\bar n\cdot k-\omega_{c})}
\notag\\
&=2\int \frac{d^4k}{(2\pi)^4}\int d^4x\, \frac{\e^{ik\cdot x}}{n\cdot p}\, \Big[\frac{\bar\Lambda \langle  0 |  \bar q(x) \Gamma_\mu  h_v(0) |  \bar B_v \rangle}{\bar{n}\cdot p-\bar n\cdot k-\omega_{c}}
\notag\\
&\qquad\qquad-\int^1_0du\,\bar u\, \frac{\langle  0 |\bar q(x)[x,ux]g_s\,x_\sigma\,v_\rho\, G^{\sigma\rho}(ux)[ux,0]\Gamma_\mu h_v(0)|  \bar B_v \rangle}{\bar{n}\cdot p-\bar n\cdot k-\omega_{c}}\Big]\, ,\nonumber
 \end{align}
 where $\Gamma_\mu = (\slashed{n}\bar n_\mu+\slashed{\bar n}n_\mu)\gamma_5$, $\partial_x^\rho\equiv\partial/\partial x_\rho$, and $G^{\sigma \rho}\equiv G^{\sigma\rho}_{a} T^a$. 
Notice that 
 the gauge link $[x,0]$ is restored in the second equality after integration-by-parts to ensure gauge invariance.
 We have also applied the HQET EOM $v\cdot Dh_v=0$, and the operator level relation with arbitrary Dirac structure $\Gamma$~\cite{Balitsky:1990ck, Kawamura:2001jm, Braun:2017liq}
 \begin{align}
 \partial_x^\rho\bar q(x)\Gamma[x,0]h_v(0)&=\partial^\rho[\bar q(x)\Gamma[x,0]h_v(0)]-\bar q(x)\Gamma[x,0]D^\rho h_v(0)
 \notag\\
 &+i\int^1_0du\,\bar u\, \bar q(x)[x,ux]g_s\,x_\sigma\, 
 G^{\sigma \rho}(ux)[ux,0]\Gamma h_v(0)\, .\label{eq:op-id1}
 \end{align}
 Here $\partial^\rho$ is the total translation operator, namely, for an arbitrary operator ${\cal O}(x_1,x_2,\ldots,x_n)$ with $n$ spacetime arguments,
 \begin{align}
 \partial^\rho {\cal O}(x_1,x_2,\ldots,x_n)\equiv\frac{\partial}{\partial\zeta_\rho}{\cal O}(x_1+\zeta,x_2+\zeta,\ldots,x_n+\zeta)\Big|_{\zeta=0}\, .
 \end{align}
 The matrix element identity in HQET 
 \begin{align}
 i v\cdot \partial \langle 0 | \bar q(x)\Gamma h_v(0) | \bar B_v \rangle = \bar\Lambda \langle 0 | \bar q(x) \Gamma h_v(0) | \bar B_v\rangle \, ,\label{eq:MI-translation-id}
 \end{align}
due to the translational covariance, is implemented subsequently, with the effective mass $\bar\Lambda= m_B-m_b^{\rm pole}+{\cal O}(\Lambda^2_{\rm QCD}/m_b)$, allowing us to make the replacement $iv\cdot\partial\mapsto\bar\Lambda$~\cite{Neubert:1993mb} to finally reach eq.~\eqref{eq:CE1}.

It is evident that the three-particle LCDAs are non-negligible in the charm-quark-propagator expansion at the subleading order of the power correction. 
After some algebra and with the help of eqs.~\eqref{def:three} and~\eqref{def:two}, one obtains,
\begin{align}
\Pi^{\rm CE,(1)}_{\mu,\rm NLP} 
=\frac{2i\widetilde{f}_B(\mu)\,m_B}{n\cdot p}\int^\infty_0 &d\omega\bigg\{\bigg[\frac{\bar\Lambda\,\phi_B^+(\omega)}{\bomega_2}+\int^\infty_0d\xi\int^1_0du\,\frac{\bar u\,\psi_4(\omega,\xi)}{\bomega_3^2}\bigg]n_\mu\, \\
&\qquad+\bigg[\frac{\bar\Lambda\,\phi_B^-(\omega)}{\bomega_2}+\int^\infty_0d\xi\int^1_0du\,\frac{\bar u\,\psi_5(\omega,\xi)}{\bomega_3^2}\bigg]\bar n_\mu
\bigg\}\, .\notag
\end{align}
The rest of the contributions in~\eqref{eq:CE-matrix-element} are calculable in the same fashion allowing us to write down      
the total subleading power correction to the correlator at the tree level due to charm-quark propagation as follows,
\begin{align}
& \Pi^{\rm CE}_{\mu,\rm NLP}(n\cdot p,\bar n\cdot p) 
\nonumber \\
= &
i\,\frac{\widetilde{f}_B(\mu)\,m_B}{n\cdot p}\int^\infty_0 d\omega\bigg\{
\bigg[\phi^-_B(\omega)\,
\frac{(\omega+\omega_c)\,(\omega -2\bar\Lambda)}{\bomega_2^2}
+2\,\int^\infty_0 d\xi\int^1_0 du\,\bar u\,
\psi_5(\omega,\xi)\,\frac{\widebar\bomega_3}{\bomega_3^3}
\bigg]\bar n_\mu
\nonumber\\
&\hspace*{0.5cm}+\bigg[
-
\frac{m_c}{\bomega_2}\,\phi^-_B(\omega)
+\frac{2\,\bar\Lambda-\omega}{\bomega_2}\,
\Big(\phi^+_B(\omega)\,
-\frac{m_c}{n\cdot p}\,
\frac{\bar n\cdot p\,\phi^-_B(\omega)}{\bomega_2}
\Big)
\nonumber \\
&\hspace*{1cm}+2\int^\infty_0 d\xi\int^1_0
\frac{du}{\bomega_3^2}\,
\Big(\psi_4(\omega,\xi)+u \, \phi_4(\omega,\xi)
-\frac{m_c}{n\cdot p}\,\bar u\,
\psi_5(\omega,\xi)\,\frac{2\,\bar n\cdot p}{\bomega_3}\Big)
\bigg]n_\mu\bigg\}\,,
\label{eq:CE-master}
\end{align}
where $\widebar\bomega_3= \bomega_3 -2\bar n\cdot p =-\bar n\cdot p-\omega-u\xi-\omega_c$.

%
%%%%%%%%%%%%%%%%%%%%%%%%%%%%%%%%%%%%%%%%%%%%%%%%%%%%%%%%%%%%%%%%%%%%%%%%%%%%
\subsection{Heavy-quark expansion}
\label{subsec:HQE}
%%%%%%%%%%%%%%%%%%%%%%%%%%%%%%%%%%%%%%%%%%%%%%%%%%%%%%%%%%%%%%%%%%%%%%%%%%%%
%

We proceed to take into account the subleading term in the HQET representation
of the bottom-quark field in order to  reach our goal of accuracy~\cite{Beneke:2018wjp}. 
At the tree level, it is straightforward to express the heavy-to-light QCD current in the following form
\begin{align}
\bar q\gamma_\mu b=\bar q\gamma_\mu h_v+\frac{1}{2m_b}\bar q\gamma_\mu i \slashed{D}_\perp h_v+\cdots
\label{eq:heavy-expand}
\end{align}
where $\slashed{D}_\perp h_v=[\slashed{D}-(v\cdot D)\slashed{v}]h_v=\slashed{D}h_v$ due to the HQET EOM.
The ellipses denote ${\cal O}(1/m_b^2)$ terms 
which are beyond the scope of our current study.
The second term in the expansion also appears in the effective heavy-to-light current $J^{\rm (A2)}$ in SCET~\cite{Beneke:2002ni}, which gives rise to ${\cal O}(\lambda^2)$ suppressed correction compared to the leading $J^{\rm (A0)}$ current.

Focusing on the subleading-power correction to the correlator, we write,
\begin{align}
\contraction{\Pi_{\mu,\rm NLP}^{\rm HQE}(n\cdot p,\bar n\cdot p)=\frac{i}{2m_b}\int d^4x\,\e^{ip\cdot x}\langle{0} |\bar q(x)\slashed{n}\gamma_5}{c}{(x)}{c}
\Pi_{\mu,\rm NLP}^{\rm HQE} (n\cdot p,\bar n\cdot p)&=
\frac{i}{2m_b}\int d^4x\,\e^{ip\cdot x}\langle{0}   |\bar q(x)\slashed{n}\gamma_5 c(x)\bar c(0)\gamma_\mu i\slashed{D} h_v | {\bar B_v}\rangle\notag\\
&=\frac{-i}{2\,m_b\,n\cdot p}\int\frac{d^4k}{(2\pi)^4}
\langle  0 | \bar q(k) \slashed{n}\gamma_5\Big(\frac{n\cdot p}{\bomega_2}\bar n_\mu+\frac{m_c\,n_\mu\slashed{\bar n}}{2\bomega_2}\Big) \slashed{D} h_v
| \bar B_v \rangle\, ,
\end{align}
where one only keeps the leading term from the expansion of the charm quark propagator.
Employing eqs.~\eqref{eq:op-id1}, \eqref{eq:MI-translation-id} and the EOMs together with integration-by-parts 
to move derivatives to the desired position, and then projecting out the resulting matrix elements to the appropriate  
LCDAs following eqs.~\eqref{def:three} and~\eqref{def:two}, one finds the contribution of the heavy-quark expansion 
to the correlator at the subleading power takes the following form,
\begin{align}
\Pi^{\rm HQE}_{\mu,\rm NLP}(n\cdot p,\bar n\cdot p) &= 
\frac{i \widetilde{f}_B(\mu)\,m_B}{2\,m_b}\,\Big[\bar n_\mu-\frac{m_c}{n\cdot p}\, n_\mu\Big]\int^\infty_0 d\omega
\bigg\{
\Big[\frac{2\bar\Lambda-\omega}{\bomega_2}\,\phi^+_B(\omega)
+\frac{\bar\Lambda-\omega}{\bomega_2}\,\phi^-_B(\omega)\Big]
\nonumber \\
&\hspace*{24mm} +2\int^\infty_0 d\xi
\int_0^1\, \frac{du}{\bomega_3^2} \,
\Big[\phi_4(\omega,\xi)+\psi_4(\omega,\xi)\Big] \bigg\}\,.\label{eq:HQE-master}
\end{align}

%
%%%%%%%%%%%%%%%%%%%%%%%%%%%%%%%%%%%%%%%%%%%%%%%%%%%%%%%%%%%%%%%%%%%%%%%%%%%%
\subsection{LCSR for the form factors}
\label{subsec:LCSR}
%%%%%%%%%%%%%%%%%%%%%%%%%%%%%%%%%%%%%%%%%%%%%%%%%%%%%%%%%%%%%%%%%%%%%%%%%%%%
%

Following the procedure described above, we present 
the LCSR for the form factors at the subleading order of power corrections
\begin{align}
f_{BD}^{\ell,\rm NLP}(q^2)=\sum_{{\rm C}={\rm HT,CE}}f^{\ell,\rm C}_{BD}(q^2)+f_{BD}^{\ell,\rm HQE}(q^2)\,,
\qquad \ell=+, 0.
\label{master formula of the NLP LCSR}
\end{align}
From the correlation functions of the higher-twist LCDAs~\eqref{eq:HT-master}, the charm-quark-propagator expansion~\eqref{eq:CE-master} and the heavy-quark expansion~\eqref{eq:HQE-master} contributions,
the corresponding $B \to D$ form factors can be extracted employing~\eqref{eq:lcsr} 
\begin{align}
f_{B D}^{\ell\rm,C} (q^2)=& 
\frac{\widetilde{f}_B(\mu)\,m_B^{\ell}}{f_D\,(n\cdot p)^{\ell+1}}\,
\exp\left[\frac{m^2_D}{n\cdot p\,\omega_M}\right]
\Big[\mathcal{F}^{\,\bar n}_{\rm C}(q^2)
-\frac{n\cdot p-m_B}{(-1)^{\ell} \, m_B}\,\mathcal{F}^{\,n}_{\rm C}(q^2)\Big]\,,\quad 
{\rm C}={\rm HT, CE}
\nonumber \\
f_{B D}^{\ell \rm,HQE} (q^2)=&
\frac{\widetilde{f}_B(\mu)\,m_B^{\ell}}{f_D\,(n\cdot p)^\ell\,m_b}\,
\exp\left[\frac{m^2_D}{n\cdot p\,\omega_M}\right]
\Big[\mathcal{F}^{\,\bar n}_{\rm HQE}(q^2)
-
\frac{n\cdot p-m_B}{(-1)^{\ell}\, m_B}\,\mathcal{F}^{\,n}_{\rm HQE}(q^2)\Big],
\label{eq:FFs-general}
\end{align}
by taking $\ell=\{+,0\} \mapsto \{+1,0\}$ in the mathematical expressions. 
The explicit expressions for ${\cal F}_{\rm C, HQE}^{n,\bar n}$ will become clear later.
Here we have made the power-counting of $1/(n\cdot p)^k$ and $1/[m_b(n\cdot p)^k]$ manifest and included the exponential factor
$\e^{m_D^2/(n\cdot p\,\omega_M)}$ from the Borel transform with $\omega_M$ commonly known as the Borel mass.

The task for us now is to find ${\cal F}_{\rm C}^{n} (q^2)\, , {\cal F}_{\rm C}^{\bar n}(q^2)\, ,{\cal F}_{\rm HQE}^{n} (q^2)$, and ${\cal F}_{\rm HQE}^{\bar n}(q^2)$. 
Applying the dispersion relation to eq.~\eqref{eq:HT-master} yields,
\begin{align}
\mathcal{F}^{\,\bar n}_{\rm HT}(q^2)&=
\mathcal{F}_{2,2}(\btau^{\bar n,2}_{2})
+\mathcal{F}_{2,3}(\btau^{\bar n,3}_{2})
+\mathcal{F}_{3,2}(\btau^{\bar n,2}_{3})
+\mathcal{F}_{3,3}(\btau^{\bar n,3}_{3})
+\frac{m_c}{n\cdot p}\,\mathcal{F}_{3,2}(\widetilde\btau^{\,\bar n,2}_{3})\,,
\nonumber\\
\mathcal{F}^{\,n}_{\rm HT}(q^2)&=
\mathcal{F}_{3,2}(\btau^{n,2}_{3})
+\frac{m_c}{n\cdot p}\,\Big[
\mathcal{F}_{2,3}(\widetilde\btau^{\,n,3}_{2})
+\mathcal{F}_{3,2}(\widetilde\btau^{\, n,2}_{3})
+\mathcal{F}_{3,3}(\widetilde\btau^{\, n,3}_{3})\Big]\,,\label{cal-F-HT}
\end{align}
where ${\cal F}_{i,k}({\btau}_i^{N,k})$ with $N=\bar n, n$ denotes the $i$-particle LCDA contribution generated by the tree level partonic corrections of the form ${\tau_i^{N}(\underline{\omega})}/{(\omega+\cdots)^k}$.  
The ellipses indicate terms that depend on the nature of the LCDAs, (see~\eqref{eq:HT-master}). 
Notice that we employ the bold-font letters for the ``modified DA'' in the LCSR formulas with 
$\widetilde\btau$ specifically reserved for contributions generated by the charm quark mass $m_c$. 

The explicit expressions for the functional ${\cal F}_{i,j}(\phi)$ are collected in eqs.~\eqref{eq:cal-F21}~--~\eqref{eq:cal-F33}, 
whereas $\btau_{i}^{N,j}\, , \widetilde\btau_{i}^{N,j}$ are as follows,
\begin{align}
&\btau^{\bar n,2}_{2}(\omega)= -4 \widehat{g}^{\,-}_B(\omega)\,,
&&
\btau^{n,2}_{3}(u,\omega,\xi)= 2\bar u\,\phi_4(\omega,\xi)\,,
&&
\btau^{\bar n,2}_{3}(u,\omega,\xi)= \big[\psi_5-\tilde \psi_5\big](\omega,\xi)\,,
\nonumber \\
&\btau^{\bar n,3}_{2}(\omega)=8\omega_c\, \widehat{g}^{\,-}_B(\omega)\,,
&& \btau^{\bar n,3}_{3}(u,\omega,\xi)=-4\bar u\, \omega_c\,\psi_5(\omega,\xi)\,,
&& \widetilde\btau^{\,\bar n,2}_{3}(u,\omega,\xi)=-2\phi_6(\omega,\xi)\,,
\nonumber \\
&\widetilde\btau^{\, n,3}_{2}(\omega)= \btau^{\bar n,3}_{2}(\omega)\,,
&& \widetilde\btau^{\,n,2}_{3}(u,\omega,\xi)= \btau^{\bar n,2}_{3}(u,\omega,\xi)\,,
&& \widetilde\btau^{\,n,3}_{3}(u,\omega,\xi)= \btau^{\bar n,3}_{3}(u,\omega,\xi)\,.
\end{align}
Taking the massless quark limit $\omega_c = m_c^2/n\cdot p \to0$ reproduces the corresponding $B\to\pi$ contribution given in eq. (3.17) of \cite{Lu:2018cfc} as expected.

Similarly from~\eqref{eq:CE-master}, we obtain the subleading power corrections due to charm quark propagator expansion,
\begin{align}
\mathcal{F}^{\,\bar n}_{\rm CE}(q^2)=&~
\mathcal{F}_{2,2}(\bdeta^{\bar n,2}_{2})
+\mathcal{F}_{3,2}(\bdeta^{\bar n,2}_{3})
+\mathcal{F}_{3,3}(\bdeta^{\bar n,3}_{3})\,,
\nonumber \\
\mathcal{F}^{\,n}_{\rm CE}(q^2)=&~
\mathcal{F}_{2,1}(\bdeta^{n,1}_{2})
+\mathcal{F}_{3,2}(\bdeta^{n,2}_{3})
\nonumber \\
&+\frac{m_c}{n\cdot p}\,\Big[
\mathcal{F}_{2,1}(\widetilde\bdeta^{\, n,1}_{2})
+\mathcal{F}_{2,2}(\widetilde\bdeta^{\, n,2}_{2})
+\mathcal{F}_{3,2}(\widetilde\bdeta^{\, n,2}_{3})
+\mathcal{F}_{3,3}(\widetilde\bdeta^{\, n,3}_{3})
\Big]\,,\label{eq:cal-F-CE}
\end{align}
following the convention for the super and subscripts of the bold-font letters established before, and,  
\begin{align}
& \bdeta^{\bar n,2}_{2}(\omega)= (\omega+\omega_c)\,(\omega-2\,\bar\Lambda)\,
\phi^-_B(\omega)\,,
&& \bdeta^{n,1}_{2}(\omega)= (2\bar\Lambda-\omega)\, \phi^+_B(\omega) \,,
\nonumber \\
& \widetilde\bdeta^{\,n,1}_{2}(\omega)= (-n\cdot p+\omega-2\bar\Lambda)\, \phi^-_B(\omega)\,,
&& \widetilde\bdeta^{\, n,2}_{2}(\omega)= \bdeta^{\bar n,2}_{2}(\omega)\,,
\nonumber \\
& \bdeta^{\bar n,2}_{3}(u,\omega,\xi)=-2\bar u\,
\psi_5(\omega, \xi) \,,
&& \bdeta^{\bar n,3}_{3}(u,\omega,\xi)=-4\bar u\,(\omega+u\xi+\omega_c)\,
\psi_5(\omega,\xi) \,,
\nonumber \\
& 
\bdeta^{n,2}_{3}(u,\omega,\xi)=2\,
\big[\psi_4+u\,\phi_4\big](\omega,\xi)\,,
&&\widetilde\bdeta^{\,n,2}_{3}(u,\omega,\xi)=-4\bar u\,\psi_5(\omega,\xi)\,,
\nonumber \\
& \widetilde\bdeta^{\,n,3}_{3}(u,\omega,\xi)=
\bdeta^{\bar n,3}_{3}(u,\omega,\xi)\, .&&
\end{align}

The subleading power corrections from the heavy-quark expansion can be obtained along the 
same line from eq.~\eqref{eq:HQE-master} which takes the form, 
\begin{align}
& \mathcal{F}^{\,\bar n}_{\rm HQE}(q^2)=
\mathcal{F}_{2,1}(\bzeta^{\bar n,1}_{2})
+\mathcal{F}_{3,2}(\bzeta^{\bar n,2}_{3})\,,
&&
\mathcal{F}^{\,n}_{\rm HQE}(q^2)=
-\frac{m_c}{n\cdot p}\,\mathcal{F}^{\,\bar n}_{\rm HQE}(q^2)\,,
\end{align}
where the explicit expressions for the functionals ${\cal F}_{ij}(\phi)$ can be found in appendix~\ref{App:NLP}, and
\begin{align}
& \bzeta^{\bar n,1}_{2}(\omega)=
\frac{1}{2}\,\Big[(2\bar\Lambda-\omega)\,\phi^+_B(\omega)
+(\bar\Lambda-\omega)\,\phi^-_B(\omega)\Big]\,,
&& \bzeta^{\bar n,2}_{3}(u,\omega,\xi)=\big[\phi_4+\psi_4\big](\omega,\xi)\, .
\label{eq:pow.coun_NLP}
\end{align}

Before we proceed to the numerical study, let us have a look at the power counting of the various quantities, adopting the asymptotic behaviours of the LCDAs.  We identify another small parameter $\lambda_{sc}\equiv\widebar\omega_{sc}/\Lambda_{\rm QCD}$ with $\widebar\omega_{sc}=\omega_s-\omega_c\sim {\cal O}(\lambda \, \Lambda_{\rm QCD})\ll\Lambda_{\rm QCD}$  entering ${\cal F}_{i,k}$. This introduces a non-homogeneous scaling of the form factors \textit{after} LCSR. While we are interested here in the scaling behaviour of the NLP form factors in the parameter $\lambda_{sc}$, we note that, following ref.~\cite{Beneke:2018wjp}\footnote{For a detailed discussion, see text below eq.~(4.9) in ref.~\cite{Beneke:2018wjp}.}, we will not perform the expansion with respect to $\lambda_{sc}$ in our subsequent numerical study, but keep the full expressions there. It is useful to apply the following scaling property of ${\cal F}_{i,k}(\phi)$
\begin{align}
{\cal F}_{2,k}(\phi)& 
\sim {\cal O} \left(\widebar\omega_{sc}%
^{2-k}\,\phi(
\widebar\omega_{sc}%
)
\right)\,,
&& k=1,2,3\,,
\nonumber\\
{\cal F}_{3,k}(\phi)& 
\sim {\cal O} \left(
\widebar\omega_{sc}^{3-k}\,\phi\big(\lambda_{sc},\widebar\omega_{sc},\Lambda_{\rm QCD}\big)\right)\,,
&& k=2,3\,,
\end{align}
where we have employed $\omega_M \sim \omega_s \gg \widebar\omega_{sc}$, and the RHS displays the dominant behaviour only.
Then one obtains directly
\begin{align}
& {\cal O}(\lambda^0 \lambda^{1}_{sc})&&
\mathcal{F}_{2,1}(\widetilde\bdeta^{n,1}_{2}),
\nonumber \\
& {\cal O}(\lambda^2 \lambda^{-1}_{sc})&&
\mathcal{F}_{2,3}(\btau^{\bar n,3}_{2}),
\nonumber \\
& {\cal O}(\lambda^2 \lambda^{0}_{sc})&&
\mathcal{F}_{2,2}(\btau^{\bar n,2}_{2}),~ 
\mathcal{F}_{3,3}(\btau^{\bar n,3}_{3}),~
\mathcal{F}_{2,2}(\bdeta^{\bar n,2}_{2}),~
\mathcal{F}_{3,3}(\bdeta^{\bar n,3}_{3}),
\nonumber \\
& {\cal O}(\lambda^2 \lambda^{1}_{sc})&& 
\mathcal{F}_{3,2}(\btau^{n,2}_{3}),~ 
\mathcal{F}_{3,2}(\btau^{\bar n,2}_{3}),~
\mathcal{F}_{3,2}(\widetilde\btau^{\,\bar n,2}_{3}),~
\mathcal{F}_{3,2}(\bdeta^{\bar n,2}_{3}),\notag\\
& &&\mathcal{F}_{3,2}(\widetilde\bdeta^{n,2}_{3}),~
\mathcal{F}_{2,1}(\bzeta^{\bar n,1}_{2}),~
\mathcal{F}_{3,2}(\bzeta^{\bar n,2}_{3}),
\nonumber \\
& {\cal O}(\lambda^2 \lambda^{2}_{sc})&&
\mathcal{F}_{2,1}(\bdeta^{n,1}_{2}),~
\mathcal{F}_{3,2}(\bdeta^{n,2}_{3}),
\end{align}
and for the form factors
\begin{align}
f^{(+,0),\rm HT}_{BD} \sim 
{\cal O}(\lambda^4\, \lambda^{-1}_{sc}),
\quad f^{(+,0),\rm CE}_{BD} \sim 
{\cal O}(\lambda^3\, \lambda^{1}_{sc}),
\quad f^{(+,0),\rm HQE}_{BD} \sim 
{\cal O}(\lambda^4\, \lambda^{1}_{sc}).
\end{align}
The difference in the power counting compared to~\cite{Lu:2018cfc} can be attributed to the asymptotic behaviour of $\widehat{g}^{\,-}_B(\omega)$, which we updated in the present work according to constraints from the equations of motion.

%
%%%%%%%%%%%%%%%%%%%%%%%%%%%%%%%%%%%%%%%%%%%%%%%%%%%%%%%%%%%%%%%%%%%%%%%%%%%%
\section{Numerical study}
\label{sec:num}
%%%%%%%%%%%%%%%%%%%%%%%%%%%%%%%%%%%%%%%%%%%%%%%%%%%%%%%%%%%%%%%%%%%%%%%%%%%%
%

\subsection{Input parameters}

We collect the input parameters for our numerical study in table~\ref{tab:numerics} along with the references from where the numbers are quoted.
The central values and ranges of all parameters used in the study are also explicitly given therein.
$m_{B^{*-}}$ ($m_{D^{*0}}$) is the mass of the first excited state of the $B$($D$)-meson.
The heavy quark masses $ m_b$ and $  m_c$ are given in the $\widebar{\text{MS}}$-scheme at the scale of their respective (${\text{MS}}$) masses.
The scale dependence of the strong coupling $\alpha_s(\mu)$ as well as the heavy quark masses $ m_b(\mu)$ and $ m_c(\mu)$ is evaluated numerically with the help of \texttt{RunDec}~\cite{Chetyrkin:2000yt}, with the former performed at the four-loop order with five-flavor $\Lambda_{\rm QCD}$ corresponding to $\alpha_s(m_Z)=0.1181$.
The $B$-meson decay constant $f_B$ is taken from the $N_f=2+1+1$ Lattice-QCD simulation~\cite{Aoki:2021kgd} which takes into account results from the HPQCD Collaboration~\cite{Dowdall:2013tga,Hughes:2017spc}, the ETM Collaboration~\cite{ETM:2016nbo} and the FNAL/MILC Collaboration~\cite{Bazavov:2017lyh} (see \cite{Carrasco:2015xwa,Beneke:2017vpq,Beneke:2019slt,Frezzotti:2020bfa} for more discussions about the QED corrections to the determination of the decay constant).
The two hard scales $\mu_{h_1}$ and $\mu_{h_2}$ introduced in the matching procedure of the hard coefficient function and the $B$-meson decay constant, respectively, are varied independently. 
The effects of their variations, independent or not, turn out to be minuscule.
The factorization scale $\mu$ is taken to be the same as the hard-collinear scale in the range of $1.5\pm 0.5~{\rm GeV}$.

\begin{table}[t] 
\centering \setlength\tabcolsep{8pt} \def\arraystretch{1.5}
\begin{tabular}{| l  l  l  | l  l  l |} 
\hline 
Parameters & values & ref. & Parameters & values & ref.  \\
\hline 
$m_{B^-}$ & 5.27933 GeV & \cite{ParticleDataGroup:2020ssz} & 
$m_{D^0}$ & 1.86483 GeV & \cite{ParticleDataGroup:2020ssz} \\
\hline 
$m_{B^{\ast-}}$ & 5.32470 GeV & \cite{ParticleDataGroup:2020ssz} & 
$m_{D^{\ast0}}$ & 2.00685 GeV & \cite{ParticleDataGroup:2020ssz} \\
\hline 
${m}_b({m}_b)$ & $4.193^{+0.022}_{-0.035}$ GeV & \cite{Beneke:2014pta} & 
${m}_c({m}_c)$ & $1.288(20)$ GeV & \cite{Dehnadi:2015fra} \\
\hline 
$f_B$ & $(190.0 \pm 1.3)$ MeV & \cite{Aoki:2021kgd} & 
$f_D$ & $212.0 (7) $ MeV & \cite{Aoki:2021kgd} \\
\hline 
$\mu_{h_1}$ & $[{m}_b/2,\, 2\,{m}_b]$ &  
& 
$\mu_{h_2}$ & $[{m}_b/2,\, 2\,{m}_b]$ &  \\
\hline 
$\mu=\mu_{hc}$ & $1.5(5)$ GeV &  
& 
$\mu_{0}$ & $1.0$ GeV &  \\
\hline 
$M^2$ & $(4.5 \pm 1.0)$ GeV$^2$ & \cite{Wang:2017jow} & 
$s_0$ & $6.0(5)$ GeV$^2$ & \cite{Wang:2017jow} \\
\hline 
$\lambda_B$ & $0.35(15)$ GeV&  \cite{Beneke:2018wjp}
& \multirow{3}{*}{$\{\widehat{\sigma}_1,\widehat{\sigma}_2\}$} 
& $\{0.7\,,6.0\}$ 
& \multirow{3}{*}{\cite{Beneke:2018wjp}} \\
\cline{1-3} 
$(\lambda^2_E/\lambda^2_H)$ & $0.5(1)$ & \cite{Beneke:2018wjp} 
&~  & $\{0.0\,,\pi^2/6\}$  &~  \\
\cline{1-3} 
$(2\lambda^2_E+\lambda^2_H)$ & $0.25(15)$ GeV$^2$ 
&  \cite{Beneke:2018wjp}
&~  &$\{-0.7\,,-6.0\}$   &~ \\
\hline 
\end{tabular} 
\caption{Parameters used for the numerical analysis. $\widehat\sigma_1$ and $\widehat\sigma_2$ are the first two (modified) inverse logarithmic moments of the leading twist DA with explicit definitions given in \eqref{eq:log-moms}. $\lambda_E$ and $\lambda_H$ parameterize the matrix element of the local antiquark-gluon-quark operator \eqref{def:lambdaEH}. The quark masses are in the $\widebar{\rm MS}$-scheme evaluated at the scale of their $\widebar{\rm MS}$ masses. }
\label{tab:numerics}
\end{table}
The numerical values of the hadronic parameters $\lambda_B\, , \widehat\sigma_{1,2}\, , \lambda_{E,H}$ are customarily given at the reference scale $\mu_0=1~\rm{GeV}$.
They are then evolved to the common factorization scale $\mu$ as all other scale-dependent quantities that enter in the final predictions for the form factors\footnote{In practice, however, the evolution of $\lambda_B$ as well as $\widehat\sigma_{1,2}$ are achieved by evolving the leading-twist DA $\phi_+^B(\omega)$ whilst the evolution $\lambda_E$ and $\lambda_H$ are neglected numerically.}.
In order to make numerical predictions for the $B\to D$ form factors, it is necessary to have explicit expressions for all the relevant $B$-meson LCDAs.  
On the one hand, the nonperturbative determination of the $B$-meson LCDA from the method of QCD sum rules~\cite{Braun:2003wx} becomes more sensitive to the parton-hadron duality ansatz and can be validated only for the light-cone separation between the light quark field and the HQET $b$-quark field of order $1$~--~$3$~$\rm{GeV}^{-1}$.
On the other hand, the experimental and lattice constraints on the LCDAs are still far from adequate even in the most innocent case of $\lambda_B$. 
We therefore resort to constructing the LCDAs via modeling, the details of which are collected in appendix~\ref{App:Model} of this work.
It is, however, necessary to highlight several guidelines available through perturbative calculations that the LCDA models must follow. 
First and foremost, the models must satisfy the EOMs which relate LCDAs of different twists by differential equations~\cite{Braun:2017liq}.
For our current accuracy, it is sufficient to take the EOMs at the tree level.
Second, the LCDAs should resemble their asymptotics at sufficiently large scales dictated by their RGEs. 
We point out that the second constraint is not as serious as the first one in the sense that it is not entirely clear what counts as ``sufficiently large'' scales.
This in principle gives rise to a reasonable amount of model variations allowed at the reference scale $\mu_0$ as suggested with tantalizing evidence from the study of $\gamma^{\ast} \gamma \to\pi$ transition~\cite{Agaev:2010aq, Agaev:2012tm,Cloet:2013tta, Stefanis:2015qha,Gao:2021iqq}.

 We should mention that the key hadronic input $\lambda_B (\mu_0)$ has been also determined from an updated QCD sum rule analysis \cite{Khodjamirian:2020hob} invoking further non-local quark condensates in analogy to Ref. \cite{Braun:2003wx} and the yielding prediction $\lambda_B (\mu_0) = 383 \pm 153 {\rm MeV}$  is compatible with the adopted value shown in table \ref{tab:numerics}. Moreover, this fundamental quantity was also recently extracted 
by matching the NLO computation of twist-one and twist-two contributions to $\bar B_u \to \gamma \ell \nu_{\ell}$ from the light-cone sum rule method  \cite{Janowski:2021yvz} and an alternative  SCET analysis \cite{Beneke:2018wjp} including the subleading power correction estimated with the dispersion technique. The obtained result  $\lambda_B (\mu_0) = 360 \pm 110 {\rm MeV}$ \cite{Janowski:2021yvz} is again in excellent agreement with what we have employed in our numerical analysis, but with somewhat smaller theory uncertainty than the corresponding number quoted in table \ref{tab:numerics}.  Additional determinations of the inverse moment $\lambda_B$ have been also pursued \cite{Wang:2015vgv,Lu:2018cfc,Wang:2017jow,Gao:2019lta}  by matching two distinct versions of the light-cone QCD  sum rules.
Taking advantage of the estimated intervals of  $\lambda_B$ with reduced uncertainties  will indeed be beneficial for pinning down the theory predictions for the partial decay rates of  $B \to D \ell \nu_{\ell}$
as well as for the celebrated CKM matrix element $|V_{cb}|$. However, such update is not expected to bring about the  notable impact on the extracted result of the gold-plated ratio $R(D)$ due to the substantial cancellation of the very non-perturbative uncertainty.
Since we aim at performing the conservative evaluation of the semileptonic $B \to D \ell \nu_{\ell}$ decay observables, we prefer to take the valus of $\lambda_B (\mu_0)$ with larger uncertainty as specified in table \ref{tab:numerics} (see also \cite{Lu:2022fgz}), also keeping in mind that the robust determination of this essential parameter is  not available at present.

The central values and ranges of the LCSR parameters $M^2$ and $s_0$ are obtained following the standard procedure outlined in~\cite{Wang:2015vgv} which yields
\begin{align}
M^2\equiv n\cdot p\,\omega_M=(4.5\pm 1.0)~{\rm GeV}^2\, ,\qqquad s_0\equiv n\cdot p\,\omega_s = (6.0\pm 0.5)~{\rm GeV}^2\, ,\label{eq:LCSRnums}
\end{align}
in agreement with the choice in refs.~\cite{Faller:2008tr, Khodjamirian:2009ys}.

\subsection{Numerical results from LCSR}

By making use of the numerical inputs listed in table~\ref{tab:numerics} and the LCDA models in appendix~\ref{App:Model}, we are able to produce theory predictions for the $B\to D$ form factors in the large-recoil limit.
 In this subsection, we adopt the simple-exponential model corresponding to taking  $\{\widehat{\sigma}_1,\widehat{\sigma}_2\}=\{0.0,\pi^2/6\}$ as our default choice for the leading-twist LCDAs, which is sufficient for our subsequent analysis.
 The explicit expressions for the higher-twist LCDAs, which are necessary for the evaluation of the higher-power corrections, are then generated accordingly from the general ansatz presented in appendix~\ref{App:Model}.
The uncertainty induced by varying the values of $\{\widehat{\sigma}_1,\widehat{\sigma}_2\}$ will be included for estimating the total uncertainties of the form factors.

\begin{figure}[t] 
\begin{center} 
\includegraphics[width=\textwidth]{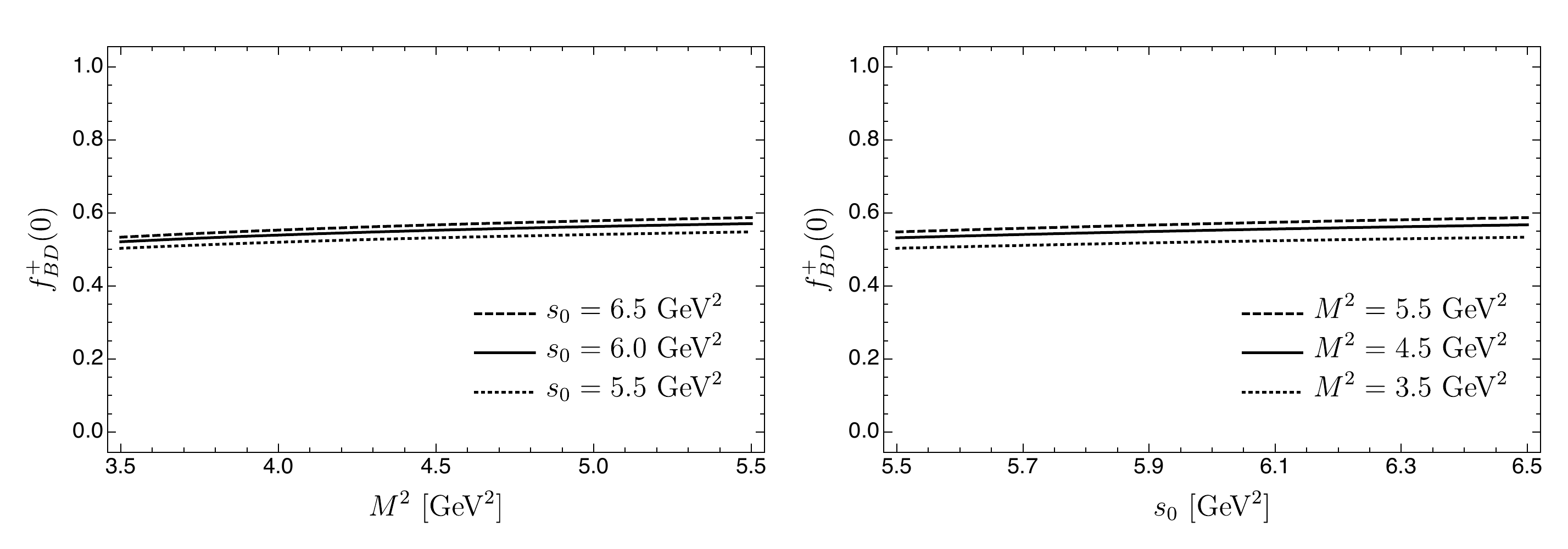} 
\end{center} 
\vspace{-0.5cm}
\caption{Form factor $f^+_{BD}(q^2=0)$ dependence on the Borel parameters $M^2$ and $s_0$.}
\label{fig:s0mm}
\end{figure}

As it was elucidated earlier, the application of LCSR requires the introduction of two (auxiliary) parameters $M^2$ and $s_0$ commonly known as the Borel mass and threshold, respectively.
Ideally, the final predictions of the physical observables should be independent of the actual values of these parameters, at least within a reasonable interval.
In practice, however, this is normally not the case.
Yet, from varying these parameters in a certain range, one is able to get a handle on the reliability of the LCSR formalism itself.
This is exactly what we would like to demonstrate in figure~\ref{fig:s0mm} where it is clear  
that the effects of the Borel mass and the threshold parameter on the prediction of $B\to D$ form factor $f^+_{BD}(0)$ are rather small, validating the application of LCSR near $q^2=0$.

\begin{figure}[t] 
\begin{center} 
\includegraphics[width=0.618\textwidth]{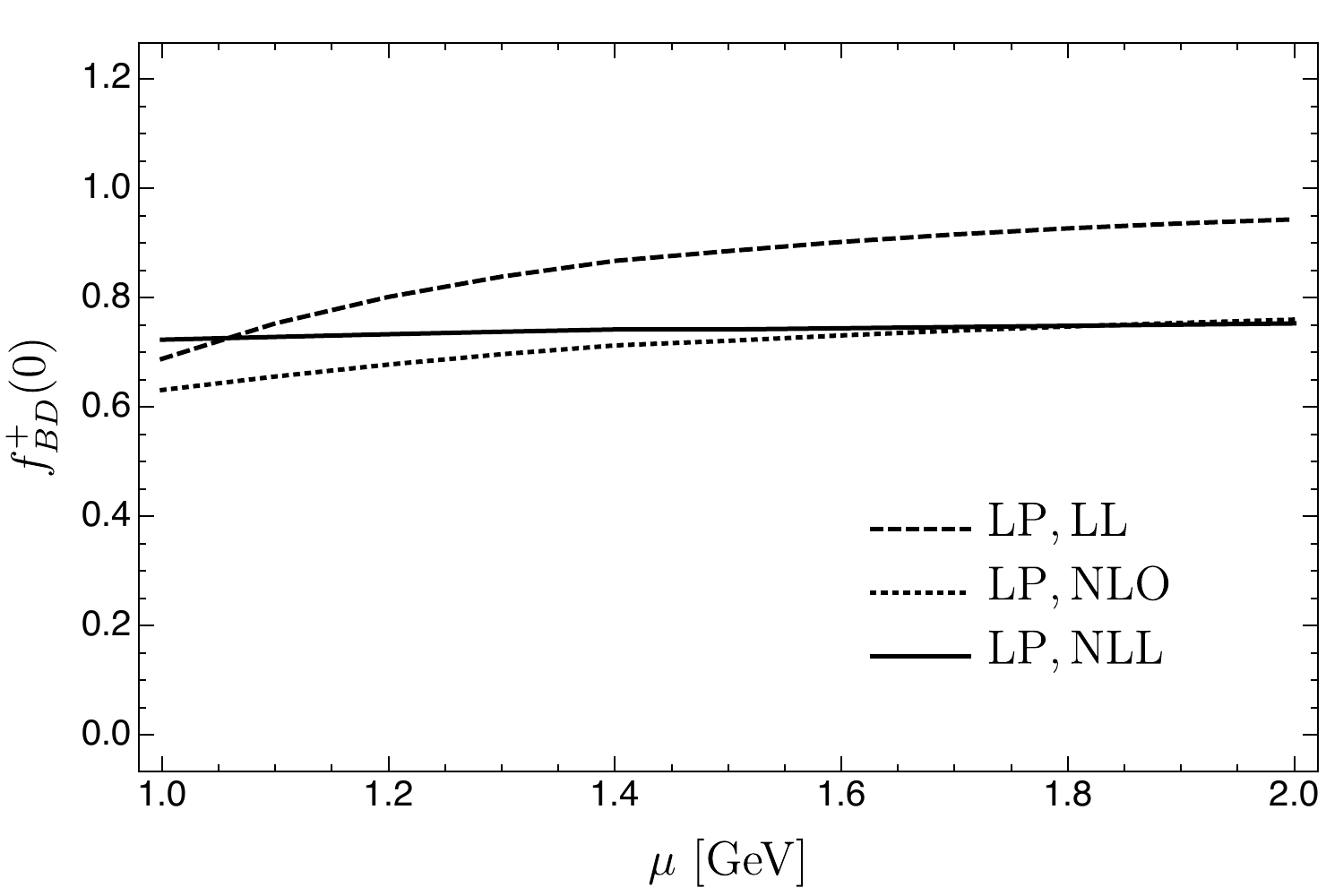} 
\end{center} 
\vspace{-0.5cm}
\caption{Factorization scale $\mu$ dependence of the form factor $f_{BD}^+(q^2=0)$ at leading power approximation. An explanation for the nomenclature is in order: ``LL", ``NLO", and ``NLL" correspond to $C^{0}_{1}\otimes\Phi_1,\, C^{1}_1\otimes\Phi_0$, and $C^1_2\otimes\Phi_1$, respectively, where $C^i_j$ is the $i$-loop matching coefficient including up to $j$-loop anomalous dimension (scale dependence) whereas $\Phi_i$ denotes the twist-3 LCDA with $i$-loop evolution improvement. The symbol $\otimes$ stands for a general convolution integral.}
\label{fig:mu}
\end{figure}

\begin{figure}[t] 
\begin{center} 
\includegraphics[width=0.618\textwidth]{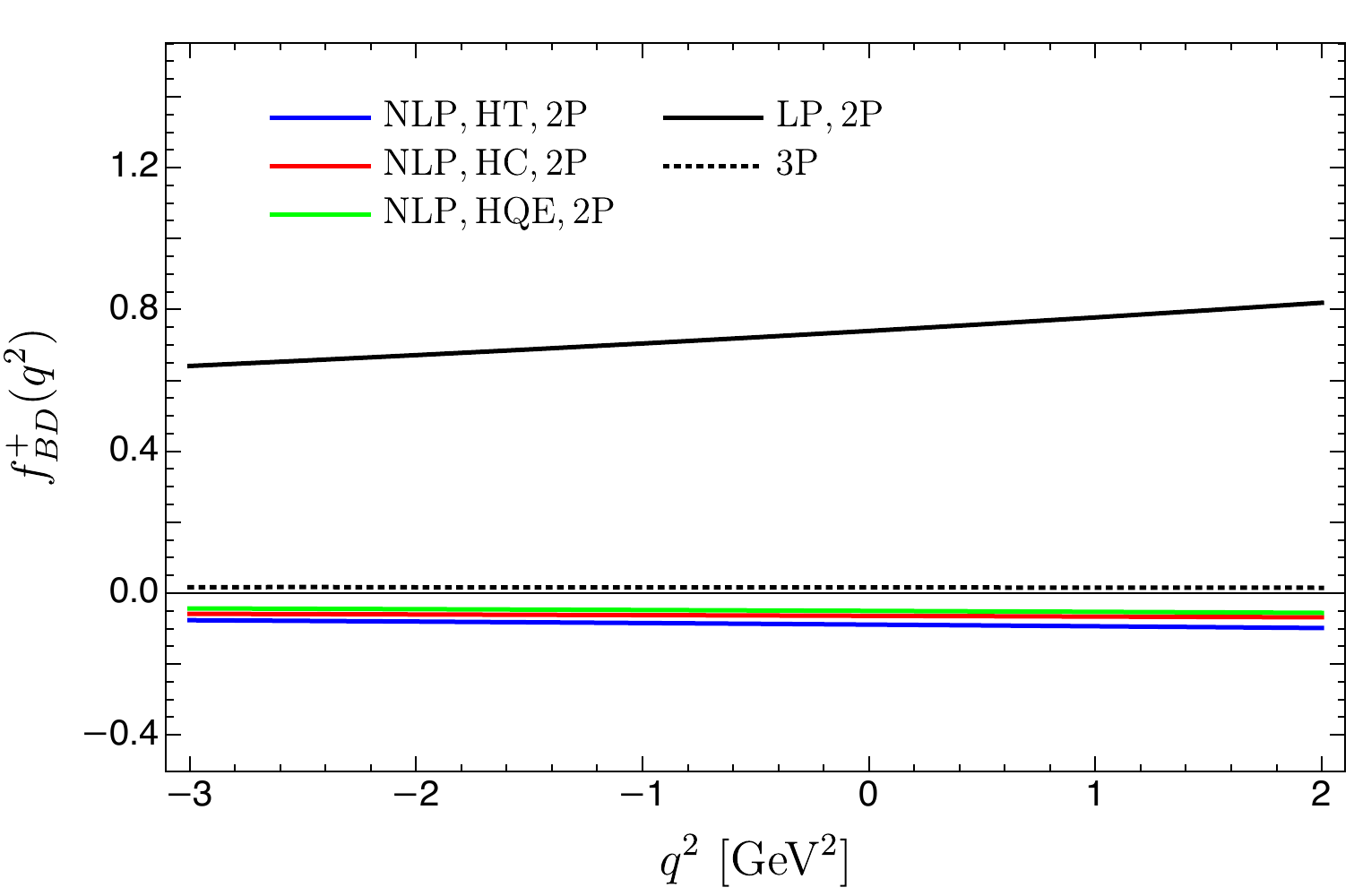} 
\end{center} 
\vspace{-0.5cm}
\caption{A comparison of different contributors to $f_{BD}^+(q^2)$ at different $q^2$. The leading power contribution is clearly dominant across the full range of $q^2$ depicted in the figure. The explicit formulas for each contribution $\{{\rm HT,HC,HQE,3P,LP}\}$ are given in eqs.~\eqref{eq:2pht}, \eqref{eq:CE-master}, \eqref{eq:HQE-master}, ~\eqref{eq:3pht}, and~\eqref{eq:NLL-master}, respectively. Similar hierarchies are also observed for $f_{BD}^0(q^2)$.}
\label{fig:break}
\end{figure}

The dependence of the form factor $f_{BD}^+(q^2=0)$ on the factorization scale $\mu$ provides us with another error estimate for our theory predictions.
Since, as we will see below (figure~\ref{fig:break}), the subleading-power contributions are significantly smaller compared to their leading-power counterpart, it is sufficient for us to look into the scale dependence of the leading-power contribution only as illustrated in  figure~\ref{fig:mu}.
Here we can readily observe that the scale dependence of the leading power contribution becomes almost invisible after taking into account  the NLL resummation improvement. 
This is in agreement with the recent study of the two-loop evolution of the leading-twist $B$-meson LCDA~\cite{Braun:2019wyx, Braun:2019zhp} where it was demonstrated that the two-loop effects are minuscule for a variety of $B$-meson LCDA models\footnote{Technically speaking, the leading-power contribution to the $B\to D$ form factors is induced by the twist-3 LCDA $\phi_B^-$, whose two-loop evolution is still unknown. However, it is expected that the Wandzura-Wilczek contribution is dominating and hence $\phi_B^-$ can be well approximated by its twist-2 part.}.

Numerically, we find roughly a $20\%$ decrease in the leading-power contribution to $f^+_{BD}(0)$ at $\mu=1.5~{\rm GeV}$ with the inclusion of $\alpha_s$ correction to the perturbative matching coefficients, namely, from LL to NLO. 
With the (partial) NLL resummation improvement, the leading-power contribution then increase by $\sim3\%$ from the NLO result at the same factorization scale.
While at a small scale $\mu=1.0~{\rm GeV}$, the NLL resummation effects can generate a $\sim15\%$ enhancement to the NLO form factor. From this observation we conclude that the NLL resummation completely stabilizes the $\mu$-dependence of the form factor $f^+_{BD}(0)$, see figure~\ref{fig:mu}. The same holds true for $f^0_{BD}(0)$.

We now proceed to investigate the numerical predictions for each contributing factors of $f_{BD}^+(q^2)$ shown in figure~\ref{fig:break}. 
As it was already pointed out in~\cite{Faller:2008tr}, the light-cone OPE for the correlation function~\eqref{eq:Pi-correlation} is justifiable only at the vicinity of maximal recoil where the power counting $n\cdot p\gg m_c\gg\Lambda_{\rm QCD}$ and $m_c\sim{\cal O}(\sqrt{n\cdot p\,\Lambda_{\rm QCD}})$ is valid. 
Furthermore, the power-counting analysis implies that the QCD factorization is fully applicable to the correlation function~\eqref{eq:Pi-correlation} in the entire space-like region of $q^2$. 
Therefore, we conclude that the LCSR predictions for the $B\to D$ form factors are trustworthy in a modest range of $q^2$ for which we take $q^2\in [-3,2]~{\rm GeV}^2$~\cite{Khodjamirian:2009ys} where the lower bound is set to match the sum rule parameters $M^2$ and $s_0$ adopted in table~\ref{tab:numerics}, albeit we have the freedom to go much deeper into the space-like region. 
The upper bound, which is in principle only subject to constraints much higher than $2~{\rm GeV}^2$ due to issues of parameterization (see subsection~\ref{eq:BGL} for detailed discussions), is somewhat arbitrary.
Such a choice is based on our intention of preserving the small uncertainties of the lattice results which is the determining factor for the uncertainties in our final predictions of the form factors as well as the ratio $R(D)$.
In other words, we choose to take the upper bound of $q^2$ close to the large-recoil limit so that the relatively large uncertainties of the LCSR predictions are still competitive compared to the high precision lattice calculation which still struggles to go beyond the zero-recoil limit, hence having the two approaches complementing each other.

On the one hand, it is immediately evident from figure~\ref{fig:break} that the resummation-improved leading-power contribution is dominant across the kinematic region $q^2\in[-3,2]$ GeV$^2$.
This is reassuring as the foundation of any sensible theory predictions requires a hierarchy of clear power separation. 
On the other hand, all the subleading-power contributions generated by the two-particle LCDAs are of the same order with an opposite sign to the leading-power contribution (black curve). 
Whereas the three-particle contribution is significantly smaller ($\sim10\%$) compared to its two-particle counterpart which can be attributed to its higher Fock-state nature.
This strongly suggests that from a phenomenological point of view, a complete two-loop matching coefficient function for the leading-power contribution is far more desirable than going beyond the tree-level for the subleading part.

\subsection{BGL fitting with strong unitarity bound}
\label{eq:BGL}

As our LCSR predictions are effective only at the \emph{large}-recoil limit, we have to rely on additional information in order to properly access the entire kinematic region of $q^2$.
The Boyd-Grinstein-Lebed (BGL) parameterization~\cite{Boyd:1994tt, Boyd:1997kz, Boyd:1995sq} provides such a framework that enables us to reliably extend our results beyond the large-recoil limit.
Our basic strategy is to determine the free parameters in the BGL parameterization using our LCSR predictions in combination with recent lattice estimates valid in the \emph{zero}-recoil-limit~\cite{MILC:2015uhg, Na:2015kha}.
The fitted BGL parameterization for the $B\to D$ form factors are then used to make predictions for the entire region of $q^2$ of our interest. 

It has not escaped our attention that alternative approaches to the numerical analysis of the $B\to D$ form factors are available, most noticeably the Caprini-Lellouch-Neubert (CLN)~\cite{Caprini:1997mu} and the Bourrely-Caprini-Lellouch (BCL)~\cite{Bourrely:2008za} parameterization.  
While the former exploits the symmetries in HQET to further constrain the unitarity bounds, the latter was proposed to cure the unphysical behaviors of a finite order truncation of the BGL parameterization in the kinematic region of $|q^2|\gtrapprox (m_B + m_D)^2$.
On the one hand, as the quality of the lattice data has become significantly more precise in recent years, higher order corrections in $\Lambda_{\rm QCD}/m_c$ to the CLN parameterization have to be included, 
which raises the possibility of it being inadequate in incorporating the precision achieved in modern lattice calculations~\cite{Bigi:2016mdz}.
Therefore, we will not consider the CLN parameterization in this work.
On the other hand, the BCL parameterization employs a more complicated form making it less feasible to apply to the \emph{strong} unitarity bounds.
Thus, in order to impose the strong unitarity bounds to our phenomenological studies without sacrificing the high precision results from lattice calculations, we will adopt the BGL parameterization hereafter for our numerical analysis with the premise of keeping $|q^2|$ well below the threshold.

The starting point of the BGL parameterization is to map the entire complex plane of $q^2$ onto a unit disk as follows,
\begin{equation}
z(q^2,t_0)=\frac{\sqrt{t_+-q^2}-\sqrt{t_+-t_0}}{\sqrt{t_+-q^2}+\sqrt{t_+-t_0}}\, ,\qquad
z(q^2)\equiv z(q^2,t_-) =\frac{\sqrt{t_+-q^2}-\sqrt{t_+-t_-}}{\sqrt{t_+-q^2}+\sqrt{t_+-t_-}}\, ,
\end{equation}
with $t_{\pm}=(m_B^{(*)}\pm m_D^{(*)})^2$. 
Here we have tacitly included all possible $B$ and $D$ resonance states with $(*)$. $t_-$ corresponds to the physical kinematic upper bound of $q^2$ in the $B\to D$ decays. The form factors develop cuts at $t_+$ after analytic continuation of $q^2$ to the entire complex plane. $t_0<t_+$ is a free parameter determining which point in the complex plane of $q^2$ is mapped onto the origin of the  complex $z$-plane. It is easy to convince oneself that $|z(q^2,t_0)|\leq1$ and $|z(q^2,t_-)|\leq1$ for arbitrary $q^2\in {\mathbb C}$. In the following we will only consider the case $t_0=t_-$ with losing generality. 

The two $B\to D$ decay form factors are then parametrized as follows
\begin{align}
f_{BD}^+(z)&=\frac{1}{P_+(z)\phi_+(z)}\sum_{n=0}^\infty a_n z^n\, ,\qqqquad  f_{BD}^0(z)=\frac{1}{P_0(z)\phi_0(z)}\sum_{n=0}^\infty b_n z^n\, ,
\label{eq:BGL-form}
\end{align}
where $P_\ell(z)$ and $\phi_\ell(z)$ are called Blaschke factors and outer functions, respectively. 
Assuming that $B\to D$ constitutes the only decay channel, namely neglecting all the excited hadronic states in the decay process, the coefficients $a_n$ and $b_n$ in eq.~\eqref{eq:BGL-form} are then subject to the \emph{weak} unitarity bound
\begin{align}\label{eq:BGL-weak}
\sum_{n=0}^\infty  a_n^2< 1\, , && \sum_{n=0}^\infty  b_n^2< 1\, .
\end{align}
This follows from dispersion relations, analyticity of the form factors, the crossing symmetry, as well as the quark-hadron duality. For further discussion of the unitarity bounds see~\cite{DiCarlo:2021dzg,Martinelli:2021onb}. The outer functions, which depend on how the helicity amplitudes enter the form factors, are given by 
\begin{align}
\phi_{+}(z)=&~
k_+
 \ \frac{(1+z)^2\sqrt{1-z}}{[(1+r)(1-z)+2 \, \sqrt{r} \, (1+z)]^5} ,\notag
&k_+\simeq &~12.43,~\text{\cite{Bigi:2016mdz}}
 \\
\phi_{0}(z)=&~  
k_0 \ \frac{(1\!-\!z^2)\sqrt{1-z}}{[(1+r)(1-z)+2 \, \sqrt{r} \, (1+z)]^4}, 
&k_0\simeq &~ 10.11 \, \phantom{,}~\text{\cite{Bigi:2016mdz}}
\label{eq:outerf}
\end{align}
with $r=m_D/m_B$.
$k_+$ and $k_0$ are calculable perturbatively at regions far away from the threshold, normally chosen at $q^2=0$. 
The Blaschke factors in eq.~\eqref{eq:BGL-form} take the form,
\begin{align}
&P_{+}(z)=\prod_{P_+=1}^{3}\frac{z-z_{P_+}}{1-zz_{P_+}}
\, ,
&&
P_{0}(z)=\prod_{P_0=1}^{2}\frac{z-z_{P_0}}{1-zz_{P_0}}\, ,
\label{eq:Blaschke}
\end{align}
where  $z_{P}$ is defined as
\begin{equation*}
z_{P}=\frac{\sqrt{t_{+}-m_{P}^2}-\sqrt{t_{+}-t_{0}}}{\sqrt{t_{+}-m_{P}^2}+\sqrt{t_{+}-t_{0}}},
\end{equation*}
with $m_{P}$ denoting the mass of the $P$-th $B_{\rm c}$ resonance below the threshold  for $BD$ pair production with the appropriate quantum number assignment of $1^-$ for $P_+(z)$ and $0^+$ for $P_0(z)$.
The role of the Blaschke factors is to remove any poles below the threshold from the form factors rendering the form factors to be analytic for all $q^2$ below the threshold.
The masses of all resonance states entering the Blaschke factors are collected in table \ref{Tab:resonance mass}.
\begin{table}[t]
\begin{center}
\begin{tabular}{|l|l|l|l|l|l|} 
\hline
$1^-$ state & Mass (GeV)  & ref.& 
$0^+$ state & Mass (GeV)  & ref. \\
\hline
$B_{\rm c}(1\, ^3\!S_1)$ & 6.332  & \cite{Dowdall:2012ab}  &
$B_{\rm c}(2\, ^3\!P_0)$ & 6.712  & \cite{Mathur:2018epb} \\
\hline
 $B_{\rm c}(2\, ^3\!S_1)$ & 6.925    &  \cite{Dowdall:2012ab,Sirunyan:2019osb}  
 & $B_{\rm c}(3\, ^3\!P_0)$ & 7.105  & \cite{Mathur:2018epb}  \\
\hline
$B_{\rm c}(3\, ^3\!D_1)$ & 7.007 &  \cite{Eichten:2019gig} 
& & & \\
\hline
$B_{\rm c}(3\, ^3\!S_1)$ & 7.280 &  \cite{Eichten:2019gig} 
& & &  \\
\hline
\end{tabular}
\end{center}
\caption{Relevant scalar ($0^+$) and vector ($1^-$)  $B_c$ masses. Note that the $B_{\rm c}(3\, ^3\!S_1)$ state is above the threshold and therefore does not enter the Blaschke factor for our process. }
\label{Tab:resonance mass}
\end{table}

The \emph{strong} unitarity bound~\cite{Boyd:1997kz} requires multiple decay channels with the correct quantum numbers to be considered. This also includes channels with higher multiplicity. 
The inclusion of additional channels reflects the physical picture more realistically, i.e., the intermediate process of the $B\to D$ decay should be thought of as a superposition of the two-particle $BD$-state with a collection of additional  resonances including $B_c$, which can pair-produce the $BD$ state, as well as a continuum of  states such as $BD\pi\pi$, and etc. 
The amplitudes of these extra channels then give rise to additional form factors via the dispersion relation based on their spin and parity as in the ground state case. 
For each form factor in consideration, one adopts the same parametrization as~\eqref{eq:BGL-form}. As an example relevant for our current study, we write, 
\begin{align}
&F_{+}^{(i)}=\frac{1}{P_{+}^{(i)}(z) \phi_{+}^{(i)}(z)}\sum_{n}a_{in}z^{n}\,,
&&F_{0}^{(i)}=\frac{1}{P_{0}^{(i)}(z) \phi_{0}^{(i)}(z)}\sum_{n}b_{in}z^{n}\,,
\end{align}
where $F_{+}^{(i)}$ denotes the vector form factor in general whereas $F_{0}^{(i)}$ corresponds to the scalar form factor.
Index $i$ labels the helicity of the form factor.
Applying the same procedures that produce eq.~\eqref{eq:BGL-weak} to each form factor then allows to write a series of bounds, 
 \begin{align}
& \sum_{i=0}^{h_i}\sum^\infty_{n=0} a^2_{in} \leq 1\,,
&& \sum_{j=0}^{h_j}\sum^\infty_{n=0} b^2_{jn} \leq 1\, ,
 \end{align}
 where $i,j$ counts all  the helicity states that enter the vector and scalar transition form factors, respectively. 
 The different form factors can further be related using heavy-quark symmetry with the possibility of incorporating the effect of heavy quark masses order by order~\cite{Bigi:2016mdz}.
 For our purpose, we take into account in total 7 vector and 3 scalar form factors describing the $B^{(*)} D^{(*)}$ states\footnote{Here both $B$ and $D$ can be labeled with or without $*$, hence giving four different combinations in total.}~\cite{Boyd:1997kz, Boyd:1995pq}. 
 These form factors are generated by $B_c$ resources below the physical threshold for $BD$ pair production and therefore indispensable in constraining the BGL parameterization for the $B\to D$ factor.
 In comparison to the ground state $B\to D$ decay, we have 6 (2) additional vector (scalar) form factors. The strong unitarity bound therefore becomes,
 \begin{align}
 & \sum_{i=0}^7\sum^\infty_{n=0} a^2_{in} \leq 1\,,
&& \sum_{j=0}^3\sum^\infty_{n=0} b^2_{jn} \leq 1\, .
\label{eq:unit-ab}
 \end{align}
 Following ref.~\cite{Bigi:2016mdz}, we truncate the infinite series at order ${\cal O}(z)$. 
 Adopting the relevant numerical values listed in table~\ref{Tab:resonance mass}, we find,
  \begin{align}\label{eq:ab-bound}
 &~ 34.95 a_{0}^2 + 33.25 a_{0} a_{1} + 16.74 a_{1}^2  \leq 1\,,\qquad
  ~ 3.76 b_{0}^2 + 2.53 b_{0} b_{1} + 2.07 b_{1}^2  \leq 1 \, ,
 \end{align}
 where relations between different form factors (labeled by $i,j$ for vector and scalar one in~\eqref{eq:unit-ab} respectively) due to heavy-quark symmetry including $1/m_b$ corrections have been applied, leaving only four free parameters $\{a_0,a_1,b_0,b_1\}$. 

We are now ready to extract the four parameters from the LCSR predictions of the $B\to D$ form factors. 
Employing the BGL parameterization to $f_{BD}^+(z)$ and $f_{BD}^0(z)$ with ${\cal O}(z)$ approximation, 
we have,
\begin{align}\label{eq:FF-fit}
f_{BD}^+(z)&=\frac{a_0+a_1 z}{P_+(z)\phi_+(z)}\, ,\qqqquad f_{BD}^0(z)=\frac{b_0+b_1 z}{P_0(z)\phi_0(z)}\, .
\end{align}
Applying eqs.~\eqref{eq:outerf} and \eqref{eq:Blaschke} to the boundary condition $f_{BD}^+(z=0)=f_{BD}^0(z=0)$ provides an exact relation among the four parameters,
\begin{align}\label{eq:b0=}
b_0 = 4.797 a_0+0.311 a_1- 0.065 b_1 \,.
\end{align}
Finally, we are able to fix the three parameters $\{a_0, a_1, b_1\}$ by fitting eq.~\eqref{eq:FF-fit} to the LCSR predictions for the form factors in combination with the data obtained from the lattice QCD calculations~\cite{MILC:2015uhg, Na:2015kha}. 
This allows us to access a much wider range of $q^2$. 
The relevant data points used for the fitting are given in tables~\ref{tab:lcsr} and~\ref{tab:lqcd}.
\begin{table}[t] 
\centering \setlength\tabcolsep{6pt} \def\arraystretch{1.4}
\begin{tabular}{|c|cccccc|} 
\hline 
$q^2~[\text{GeV}^2]$ & $-3.0$ &  $-2.0$  & $-1.0$  & $0.0$  & $1.0$ &  $2.0$ \\
\hline 
$f_{BD}^+$ & $0.479(191)$ & $0.502(199)$ & $0.526(208)$ & $0.552(216)$ & $0.581(226)$ & $0.611(235)$ \\
%\hline
$f_{BD}^0$ & $0.525(210)$ & $0.534(212)$ & $0.543(214)$ & $0.552(216)$ & $0.562(218)$ & $0.572(220)$ \\
\hline
\end{tabular} 
\caption{LCSR predictions of the $B\to D$ form factors obtained in this work used in the BGL fit.} 
\label{tab:lcsr}
\end{table}
In table~\ref{tab:lcsr}, we provide our LCSR predictions for the form factors in the range of $q^2\in [-3.0, 2.0]~ {\rm GeV^2}$ with error estimates. 
The relevant formulas can be found in eqs.~ \eqref{eq:NLL-master} and~\eqref{eq:FFs-general}.
\begin{table}[t] 
\centering \setlength\tabcolsep{6pt} \def\arraystretch{1.6}
\begin{tabular}{|c|ccc|ccc|} 
\hline 
& \multicolumn{3}{c|}{FNAL/MILC~\cite{MILC:2015uhg}} 
& \multicolumn{3}{c|}{HPQCD~\cite{Na:2015kha}} \\
\hline 
$q^2~[\text{GeV}^2]$ & $8.51$ & $10.08$ & $11.66$ & $9.30$ & $10.48$ & $11.46$ \\
\hline 
%\hline 
$f^+_{BD}$ & $1.005(12)$ & $1.094(10)$ & $1.199(10)$ & $1.038(39)$ & $1.105(42)$ & $1.167(45)$ \\
%\hline
$f^0_{BD}$ & $0.825(9)$ & $0.861(8)$ & $0.903(7)$ & $0.840(37)$ & $0.870(39)$ & $0.897(40)$\\
\hline
\end{tabular} 
\caption{Lattice QCD predictions of the form factors used in the BGL fit.} 
\label{tab:lqcd}
\end{table}
In table~\ref{tab:lqcd}, we collect the lattice QCD results for the form factors from two collaborations adopted in our BGL fit.
The combined data from both tables are then fed into eqs.~\eqref{eq:FF-fit} to extract the parameters $\{a_0,a_1,b_1\}$ using \eqref{eq:b0=} under the constraint of \eqref{eq:ab-bound}.
\begin{table}[t] 
\centering \setlength\tabcolsep{6pt} \def\arraystretch{1.6}
\begin{tabular}{|c|c|ccc|} 
\hline 
& & \multicolumn{3}{c|}{correlation~matrix} \\
\hline 
& value & $a_0$ & $a_1$ & $b_1$ \\
\hline 
$a_0$ & $0.0155(1)$ & $1$ & $0.170$ & $0.276$\\
$a_1$ & $-0.041(3)$ & & $1$ & $0.937$ \\
$b_1$ & $-0.206(15)$ & && $1$\\
\hline
\end{tabular}
\caption{Values of BGL parameters extracted from fitting the BGL parameterization against LCSR and lattice QCD predictions. Correlations between the three parameters are also provided.} 
\label{tab:zfit}
\end{table}

The final numerical results are listed in table~\ref{tab:zfit} together with the correlation matrix.
It is fair to say that the three parameters are highly correlated, in particular the correlation between $a_1,b_1$ is close to 100\% indicating that the data is highly restrictive, which can also be concluded from the small uncertainties of the parameters.
The values of $\{a_0,a_1,b_1\}$ are obtained using $\chi^2$-fitting with $\chi^2/{\rm d.o.f}=68.32/(24-3)=3.25$. 
Apparently,  the two unitariry bounds displayed in  (\ref{eq:ab-bound}) are not saturated in practice. One should not be surprised by this observation, since  we have not taken into account all the contributing exclusive channels (for instance $B_s \to D_s^{(\ast)}$, $B_c \to \eta_c$,  $B_c \to J/\psi$,  $\Lambda_b \to \Lambda_c^{(\ast)}$) in the construction of the strong unitarity bounds  (\ref{eq:unit-ab})  and  moreover  we have  truncated the infinite series appearing in  (\ref{eq:unit-ab}) at order ${\cal O}(z)$ in an attempt to derive the very constraints presented in (\ref{eq:ab-bound}).

 \begin{figure}[t] 
\begin{center} 
\includegraphics[width=0.48\textwidth]{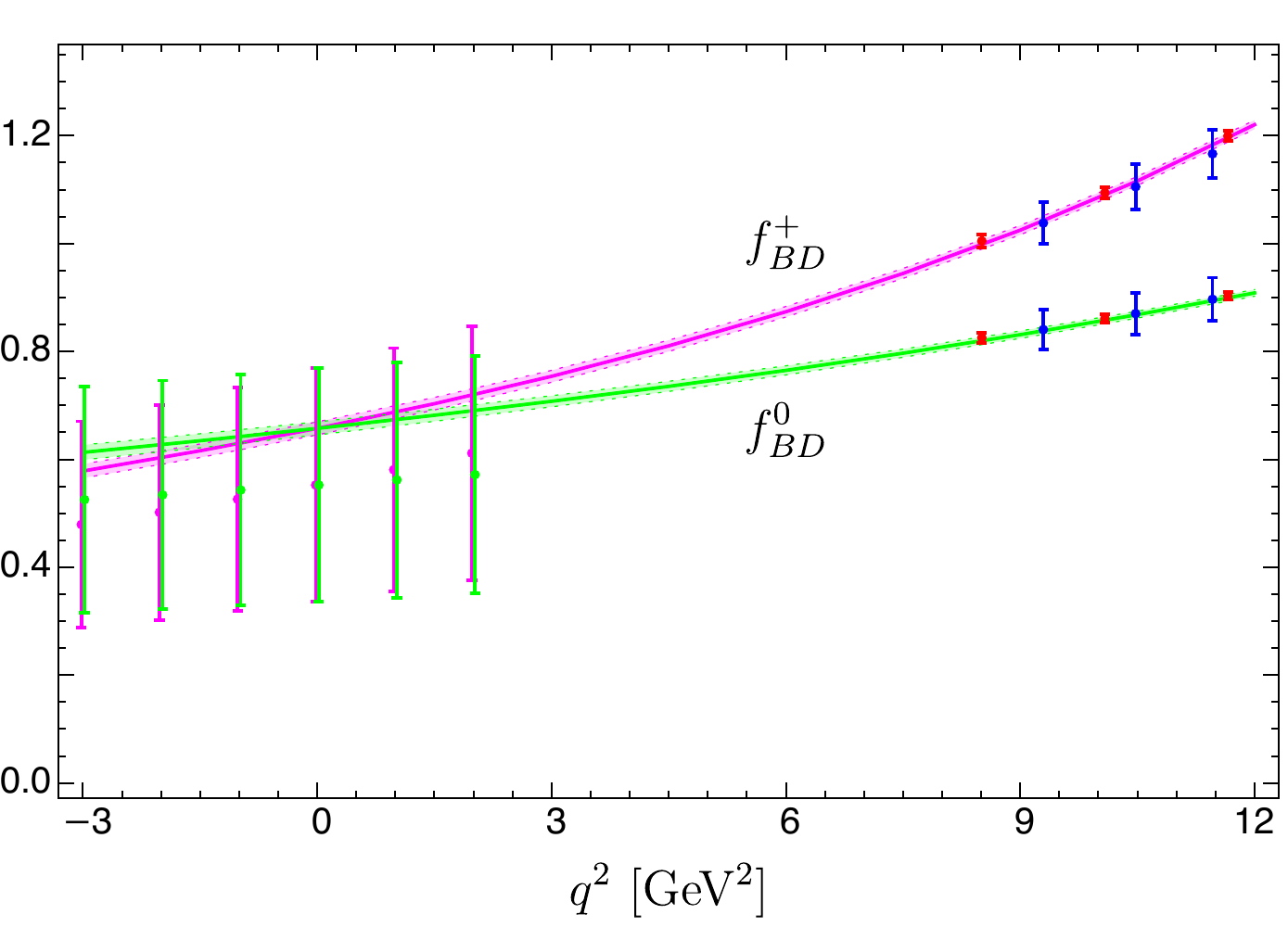}
\includegraphics[width=0.48\textwidth]{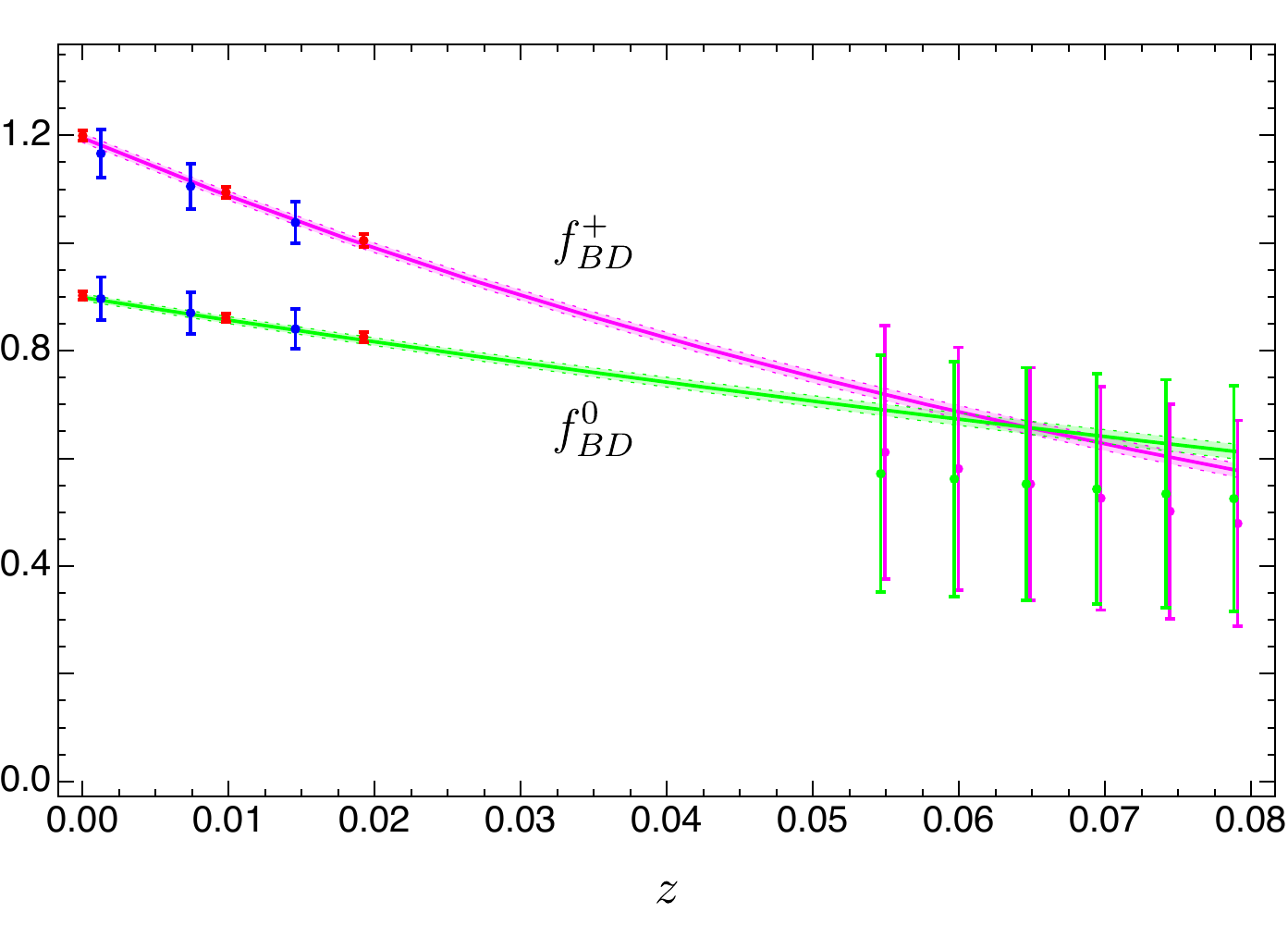} 
\end{center} 
\vspace{-0.5cm}
\caption{BGL fitted results of the form factors as functions of $q^2$ (left) and $z$ (right).
The red (blue) points with error bars represent the FNAL/MILC (HPQCD) data 
 whereas the magenta and green points are LCSR results from this work for $f_{BD}^+(q^2)$ and $f_{BD}^0(q^2)$, respectively.
The magenta (green) curves correspond to $f_{BD}^+(q^2)$ ($f_{BD}^0(q^2)$) obtained by fitting the combined data of LCSR and lattice predictions. The shaded regions are error bands due to uncertainties of the three parameters in table~\ref{tab:zfit}.}
\label{fig:zfit}
\end{figure}

The effectiveness of the BGL fitting is demonstrated in figure~\ref{fig:zfit} where the large error-bars of the LCSR predictions in the interval $q^2\in [-3,2]~{\rm GeV}^2$ are due to the large model uncertainties. 
The combined data, however, allows to generate predictions for the form factors across a large range of $q^2\in [-3,12]~{\rm GeV}^2$ with relatively small errors.

\subsection{Semileptonic decay rate and extraction of \texorpdfstring{$|V_{cb}|$}{Lg}}

One of the major uncertainties for our theory predictions comes from the first inverse moment $\lambda_B(1\rm{GeV})$ of the leading-twist LCDA, for which we have adopted the value obtained from double-radiative $B$-decay $B\to\gamma\gamma$~\cite{Shen:2020hfq} with $\lambda_B(1\rm{GeV})=(0.35\pm0.15)~\rm{GeV}$.
This value is comparable to the previous study based on $B$-meson leptonic radiative decay $B\to\gamma\ell\nu_\ell$~\cite{Beneke:2018wjp, Gelb:2018end}.

We are now ready to extract the value of $R(D)$ and the CKM matrix element $|V_{cb}|$, both of which play an utmost important role in probing the BSM physics. 
The former provides a handle on testing the assumption of lepton universality in the Standard Model, while the latter gives us hints on physics inaccessible directly to our current particle colliders.

The differential decay rate of  $B \to D \ell \nu_{\ell}$
in the rest frame of the $B$-meson can be computed as
\begin{align}
\frac{d \Gamma (B \to D \ell \nu_{\ell})}{d q^2}
=& \frac{\eta_{\rm EW}^2 \, G_F^2 \, |V_{cb}|^2}{24 \, \pi^3 \, m_B^2} \,
\left (1 - \frac{ m_\ell^2}{q^2} \right )^2  \, | \vec{p}_D | \,
\bigg [ \left ( 1 + \frac{m_\ell^2}{2 \, q^2} \right )\, m_B^2 \,
| \vec{p}_D |^2 \, \left | f^{+}_{B D}(q^2) \right |^2 \nonumber  \\
& + \frac{3 \, m_\ell^2}{8 \, q^2} \, (m_B^2 -m_D^2)^2 \,
\left | f^{0}_{B D}(q^2) \right |^2  \bigg ]  \,,
\end{align}
where $| \vec{p}_D | = \sqrt{\lambda \left (m_B^2, m_D^2, q^2 \right )}/ (2 m_B)$
with $\lambda(a, b, c)=a^2+b^2+c^2- 2 a b - 2 a c - 2 b c$ is the magnitude of the three-momentum
of the $D$-meson and the effective coefficient 
\begin{eqnarray}
\eta_{\rm EW} = 1 + \frac{\alpha_{\rm em}}{\pi }\, \ln \left ( \frac{m_Z}{m_B} \right )  \simeq 1.0066 
\end{eqnarray}
arises from the electroweak correction to the semileptonic $b \to c \ell \nu_{\ell}$ process \cite{Sirlin:1981ie} at one loop.

 It remains important to mention that the dedicated investigation of the low-energy QED correction to the electroweak penguin $\bar B \to \bar K \ell^{+} \ell^{-}$ has been carried out comprehensively in \cite{Isidori:2020acz} (including also interesting discussions on the electromagnetic correction to the semileptonic $B \to D \ell \nu_{\ell}$ decay) by employing the effective mesonic Lagrangian,  which surpasses the previous study \cite{Bordone:2016gaq} on the same topic in various aspects  (see Appendix A.2 of Ref. \cite{Isidori:2020acz} for more details).  In particular, the logarithmically enhanced QED corrections proportional to $\alpha/(4\pi)\log(m_b^2/m_\ell^2)$ to the inclusive $\bar B \to X_s \ell^{+} \ell^{-}$ decay distributions have been addressed intensively in \cite{Huber:2005ig,Huber:2007vv,Huber:2015sra,Huber:2020vup} in anticipation of the precision measurements at the Belle II experiment\footnote{see Appendix A.1.1 of Ref. \cite{Isidori:2020acz} for the further comparison with the exclusive  $\bar B \to \bar K \ell^{+} \ell^{-}$ decay.}. Monte Carlo studies on the actual size of the QED logarithms due to angular and energetic cuts were carried out in~\cite{Huber:2015sra,Huber:2020vup}, revealing that the cuts tame the logarithms in the electronic case toward the size of that of the muons. While a complete Monte Carlo study along the lines of~\cite{Huber:2015sra,Huber:2020vup} is beyond the scope of the present paper, we believe that due to the universality of the collinear bremsstrahlung effect this type of QED corrections to $R(D)$ will be small.
 
 Moreover, the general analysis presented in \cite{Isidori:2020acz} tends to indicate that the structure-independent QED correction to  $B \to D \ell \nu_{\ell}$ is not expected to  generate exceedingly large impact in practice.  Furthermore, the short-distance electromagnetic effects in the exclusive $B$-meson decays have been explored with the ${\rm QCD \times QED}$ factorization technique (or equivalently the soft-collinear effective theory framework) for $\bar B_q \to \mu^{+} \mu^{-}$ \cite{Beneke:2017vpq,Beneke:2019slt}, for the charmless hadronic two-body $B$-meson decays \cite{Beneke:2020vnb} (see  Section 6 of this article on the conceptual difference when compared with the earlier study of the soft photon effect \cite{Baracchini:2005wp}),
 and for the two-body hadronic and semileptonic $B$-meson decays with heavy-light final states \cite{Beneke:2021jhp}, demanding actually the introduction of light-cone hadronic  distribution amplitudes with qualitatively new features \cite{Beneke:2021pkl} in comparison with the conventional quantities in QCD-only.
 Additionally, the lattice QCD method has been developed to incorporate the electromagnetic effects in  the exclusive  semileptonic decays \cite{Carrasco:2015xwa,Sachrajda:2019uhh,Seng:2020jtz} and in the leptonic decay processes \cite{Giusti:2017dwk,DiCarlo:2019thl,Carlo:2022zkw}. A complete analysis of the QED correction to the semileptonic $B \to D \ell \nu_{\ell}$ decay is apparently an important task of its own (see a variety of open issues discussed in \cite{Isidori:2020acz,Beneke:2017vpq,Beneke:2019slt}) from both the theoretical and phenomenological aspects and goes well beyond the scope of our work mainly focusing on the intricate strong interaction dynamics encoded in the exclusive $B \to D$ decay form factors.

\begin{figure}[t] 
\begin{center} 
\includegraphics[width=0.8\textwidth]{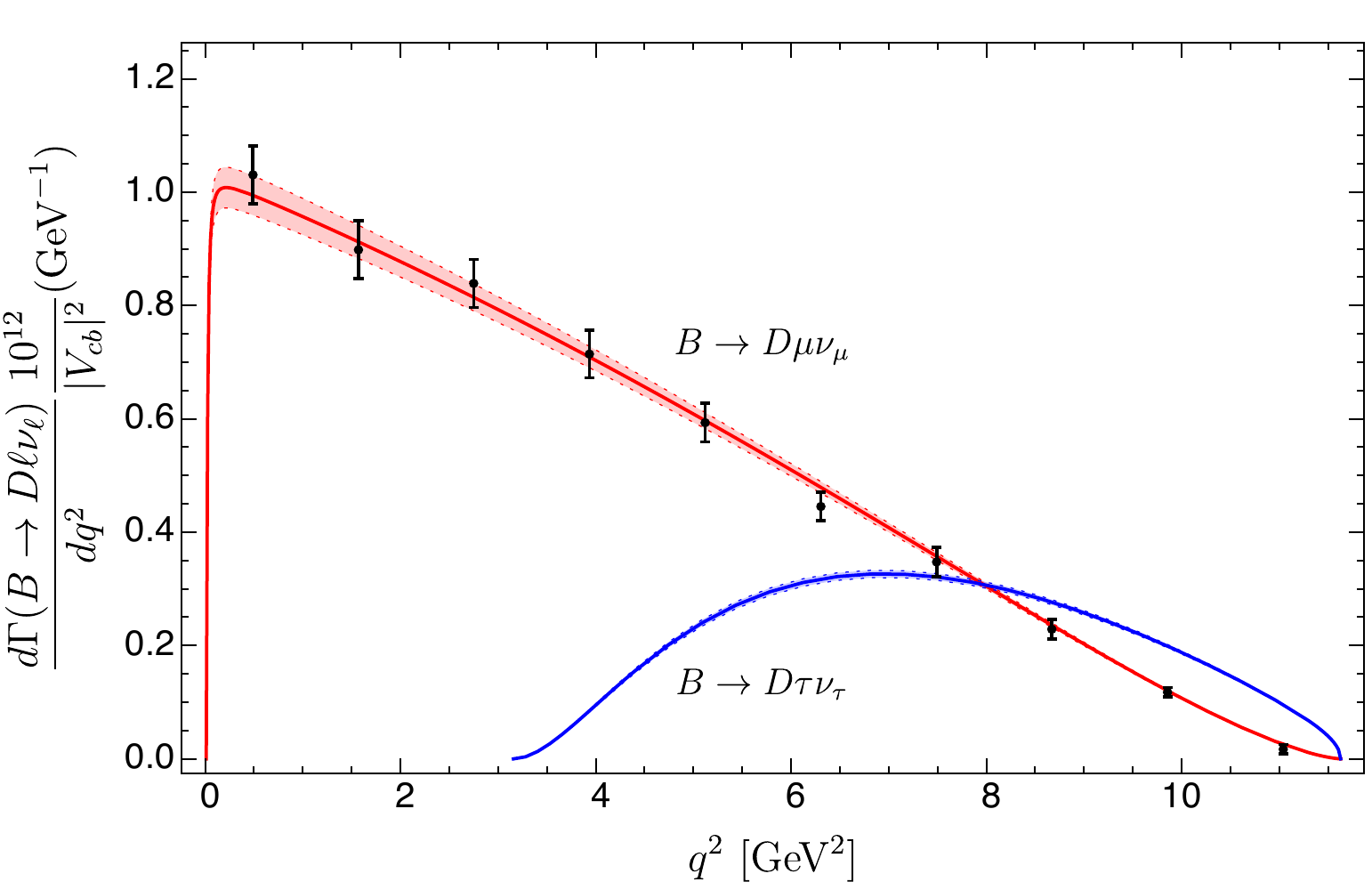} 
\end{center} 
\vspace{-0.5cm}
\caption{The differential decay rates for $B\to D\mu\nu_{\mu}$ (red curve) and $B\to D\tau\nu_{\tau}$ (blue curve) with error bands obtained from this work. The black points with error bars are the experimental data for $B\to D\mu\nu_{\mu}$ from the Belle collaboration \cite{Glattauer:2015teq}.}
\label{fig:BRdiff}
\end{figure}

In figure~\ref{fig:BRdiff}, we show that our prediction for the differential decay rate of  
$B \to D \mu \nu_{\mu}$ is in good agreement with the Belle results~\cite{Glattauer:2015teq}. In addition, we also provide predictions for the differential decay rate of $B \to D \tau \nu_{\tau}$.

\begin{table}[t] 
\centering \setlength\tabcolsep{6pt} \def\arraystretch{1.6}
\begin{tabular}{|c|cc|c|cc|} 
\hline
[$t_{1}$,$t_{2}$] & \multicolumn{2}{c|}{$\Delta \Gamma_{\mu} (t_{1},t_{2})~(10^{-12})$ GeV} 
& [$t_{1}$,$t_{2}$] & \multicolumn{2}{c|}{$\Delta \Gamma_{\mu} (t_{1},t_{2})~(10^{-12})$ GeV}  \\
(GeV$^{2}$) & this work & Belle \cite{Glattauer:2015teq} 
& (GeV$^{2}$) & this work & Belle \cite{Glattauer:2015teq} \\
\hline
$\text{[0.00,0.98]}$ & $\text{0.94$\pm $0.03}$ & $\text{1.01$\pm $0.05}$ & $\text{[5.71,6.90]}$ & $\text{0.57$\pm $0.01}$ & $\text{0.53$\pm $0.03}$ \\
$\text{[0.98,2.16]}$ & $\text{1.08$\pm $0.03}$ & $\text{1.06$\pm $0.06}$ & $\text{[6.90,8.08]}$ & $\text{0.42$\pm $0.01}$ & $\text{0.41$\pm $0.03}$ \\
$\text{[2.16,3.34]}$ & $\text{0.96$\pm $0.03}$ & $\text{0.99$\pm $0.05}$ & $\text{[8.08,9.26]}$ & $\text{0.28$\pm $0.00}$ & $\text{0.27$\pm $0.02}$ \\
$\text{[3.34,4.53]}$ & $\text{0.84$\pm $0.02}$ & $\text{0.85$\pm $0.05}$ & $\text{[9.26,10.45]}$ & $\text{0.14$\pm $0.00}$ & $\text{0.14$\pm $0.01}$ \\
$\text{[4.53,5.71]}$ & $\text{0.70$\pm $0.02}$ & $\text{0.70$\pm $0.04}$ & $\text{[10.45,11.63]}$ & $\text{0.03$\pm $0.00}$ & $\text{0.02$\pm $0.01}$ \\
\hline
\end{tabular} 
\caption{Theory predictions for $B\to D\mu\nu_{\mu}$ in different bins compared against the Belle data.} 
\label{tab:BRdiff}
\end{table}

The binned (normalized) partial decay rate of
$B \to D \ell \nu_{\ell}$ reads,
\begin{eqnarray}
\Delta \Gamma_{\ell}(t_1, t_2) = \int_{t_1}^{t_2} \, d q^2  \,\,
\frac{d \Gamma (B \to D \ell \nu_{\ell})}{d q^2}  \, \frac{1}{ |V_{cb}|^2} \,.
\end{eqnarray}
In table~\ref{tab:BRdiff}, we collect our predictions for the partial decay rate $\Delta \Gamma_{\mu}$ at different intervals of $(t_1,t_2)$ and compare them with the Belle results \cite{Glattauer:2015teq}.
We also predict the total normalized decay rate of $B\to D\mu\nu_\mu$ in the entire kinematic region of $q^2\in [m_\mu^2,(m_B-m_D)^2]=[0.01, 11.63]~ {\rm GeV}^2$ to be,
\begin{align}
\Delta \Gamma_{\mu}(0.01 {\rm GeV}^{2}, 11.63{\rm GeV}^{2})
=(5.97 \pm 0.16) \times 10^{-12}~{\rm GeV}\, .
\end{align}

By fitting our theoretical predictions of $\Delta\Gamma_\mu(t_1,t_2)$ to the BarBar and Belle data, we are able to extract the value of $|V_{cb}|$,
\begin{eqnarray}
|V_{cb}| = \left\{
\begin{array}{l}
\left( 40.2^{+0.6}_{-0.5} {\big |_{\rm th}}\,\, {}^{+1.4}_{-1.4} {\big |_{\rm exp}} \right)\times 10^{-3}\,,  \qquad
[{\rm BaBar} \,\,  2010 ~ \text{\cite{Aubert:2009ac}}] \vspace{0.4 cm} \\
\left( 40.9^{+0.6}_{-0.5} {\big |_{\rm th}}\,\, {}^{+1.0}_{-1.0} {\big |_{\rm exp}} \right)\times 10^{-3}\,,  \qquad
[{\rm Belle} \,\,  2016~ \text{\cite{Glattauer:2015teq}}]
\end{array}
 \hspace{0.5 cm} \right. \, ,
 \label{eq:Vcb}
\end{eqnarray}
where ``th'' and ``exp'' denote the theoretical and experimental uncertainties, respectively.

Now we consider ratios of differential decay widths integrated over different $q^2$ intervals,
\begin{eqnarray}
\Delta {\cal R}(t_1, t_2) =
\frac{\Delta \Gamma_{\tau}(t_1, t_2)}
{\Delta \Gamma_{\mu}(t_1, t_2)}    \,.
\end{eqnarray}

\begin{table}[t]
\centering \setlength\tabcolsep{6pt} \def\arraystretch{1.6}
\begin{tabular}{|c|c|cc|}
\hline
[$t_{1}$,$t_{2}$] & $\Delta \Gamma_{\tau} (t_{1},t_{2})~(10^{-12})$ GeV
& \multicolumn{2}{c|}{$\Delta {\cal R} (t_{1},t_{2})$}  \\
(GeV$^{2}$) & this work & this work &  \cite{Celis:2016azn} \\
\hline
$\text{[4.00,4.53]}$ & $\text{0.072$\pm $0.002}$ & $\text{0.199$\pm $0.000}$ & $\text{0.199$\pm $0.001}$ \\
$\text{[4.53,5.07]}$ & $\text{0.112$\pm $0.003}$ & $\text{0.330$\pm $0.001}$ & $\text{0.330$\pm $0.001}$ \\
$\text{[5.07,5.60]}$ & $\text{0.139$\pm $0.003}$ & $\text{0.455$\pm $0.001}$ & $\text{0.455$\pm $0.001}$ \\
$\text{[5.60,6.13]}$ & $\text{0.158$\pm $0.003}$ & $\text{0.571$\pm $0.001}$ & $\text{0.571$\pm $0.002}$ \\
$\text{[6.13,6.67]}$ & $\text{0.172$\pm $0.004}$ & $\text{0.680$\pm $0.002}$ & $\text{0.680$\pm $0.002}$ \\
$\text{[6.67,7.20]}$ & $\text{0.173$\pm $0.003}$ & $\text{0.786$\pm $0.002}$ & $\text{0.786$\pm $0.003}$ \\
$\text{[7.20,7.73]}$ & $\text{0.170$\pm $0.003}$ & $\text{0.892$\pm $0.003}$ & $\text{0.892$\pm $0.003}$ \\
$\text{[7.73,8.27]}$ & $\text{0.165$\pm $0.003}$ & $\text{1.006$\pm $0.004}$ & $\text{1.006$\pm $0.004}$ \\
$\text{[8.27,8.80]}$ & $\text{0.150$\pm $0.003}$ & $\text{1.135$\pm $0.004}$ & $\text{1.135$\pm $0.005}$ \\
$\text{[8.80,9.33]}$ & $\text{0.134$\pm $0.002}$ & $\text{1.293$\pm $0.006}$ & $\text{1.294$\pm $0.006}$ \\
$\text{[9.33,9.86]}$ & $\text{0.116$\pm $0.002}$ & $\text{1.508$\pm $0.007}$ & $\text{1.513$\pm $0.007}$ \\
$\text{[9.86,10.40]}$ & $\text{0.096$\pm $0.001}$ & $\text{1.851$\pm $0.010}$ & $\text{1.860$\pm $0.010}$ \\
$\text{[10.40,11.63]}$ & $\text{0.116$\pm $0.002}$ & $\text{3.150$\pm $0.022}$ & $-$ \\
\hline
\end{tabular} 
\caption{Theory predictions for the partial decay rates of $B \to D \tau \nu_{\tau}$
and for the binned distributions of $\Delta {\cal R}(t_1, t_2)$. Our predictions are comparable to \cite{Celis:2016azn} albeit with slightly larger error bars in general due to large uncertainties from LCDAs.} 
\label{tab:deltaR}
\end{table}

In table~\ref{tab:deltaR}, we collect the partial decay rates of $B \to D \tau \nu_{\tau}$ and the ratio $\Delta {\cal R}$ in different intervals of $q^2$. Our predictions for $\Delta {\cal R}$ are in good agreement with the previous one~\cite{Celis:2016azn}.

Summing up the total branching ratio of $B\to D\tau\nu_\tau$ and $B\to D\mu\nu_\mu$  in the entire kinematic region relevant for each channel, namely $q^2\in[m_\tau^2,(m_B-m_D)^2]=[3.16,11.63]~{\rm GeV}^2$ for $B\to D\tau\nu_\tau$ and $q^2\in[m_\mu^2, (m_B-m_D)^2] = [0.01,11.63]~{\rm GeV}^2$ for $B\to D\mu\nu_\mu$, we find the Standard Model prediction for the ratio $R(D)$ which has become the gold standard for testing lepton universality and hence probing the BSM physics,
\begin{align}
R(D) \equiv \frac{{\mathcal{B}}(B \to D \tau \nu_{\tau})}
{{\mathcal{B}}(B \to D \mu \nu_{\mu})}=
\frac{\Delta \Gamma_{\tau}(m_\tau^2,(m_B-m_D)^2)}
{\Delta \Gamma_{\mu}(m_\mu^2,(m_B-m_D)^2)}= 0.302\pm 0.003\, ,\label{eq:RD}
\end{align}
where the branching ratio $\cal B$ is the decay rate normalized to the total $B$-meson decay width, which is of course unimportant for computing $R(D)$.
Our prediction for $R(D)$ is in reasonable agreement with the experimental value $0.340(27)(13)$~\cite{Bozek:2010xy, Lees:2012xj, Lees:2013uzd, Huschle:2015rga, HFLAV:2019otj, Abdesselam:2019dgh}, the lattice average $0.2934(53)$~\cite{Aoki:2021kgd}, %previous theory predictions
as well as the extracted interval $0.298 (6)$~\cite{Cohen:2019zev}  from the model-independent global analysis of exclusive hadronic $b \to c \ell \nu_{\ell}$  processes within theory uncertainties.
In spite of the slightly smaller central values, the resulting predictions of $|V_{cb}|$ and $R(D)$ obtained in~\cite{Bordone:2019vic,Bordone:2019guc} are compatible with our results.
In table~\ref{tab:VRs} we give an overview of how our numbers compare to those of other recent studies following the techniques of LCSR, lattice QCD, and HQET.

\begin{table}[t] 
\centering \setlength\tabcolsep{6pt} \def\arraystretch{1.6}
\scalebox{0.92}{\begin{tabular}{|c|c|c|c|c|} 
\hline
 & LCSR\footnotemark & Lattice & HQET\footnotemark & This work\\
\hline
\multirow{3}{*}{$|V_{cb}|\times 10^3$} & $41.3(1.4)$\cite{Bigi:2017jbd}, $40.6^{+1.2}_{-1.3}$\cite{Bigi:2017jbd} & 39.36(68)\cite{Aoki:2021kgd}&  & $ 
40.2^{+0.6}_{-0.5} {\big |_{\rm th}}\,\, {}^{+1.4}_{-1.4} {\big |_{\rm BaBar}} $\\
& $39.2^{+3.5}_{-3.4}$\cite{Wang:2017jow}, 41.4(4.0)\cite{Wang:2017jow} & 38.40(0.7)\cite{FermilabLattice:2021cdg} & $39.3(1.0)$\cite{Bernlochner:2017jka} &  
\\
& 40.3(0.8)\cite{Bordone:2019vic} & 43.0(2.7)\cite{Harrison:2021tol} &  & $ 
 40.9^{+0.6}_{-0.5} {\big |_{\rm th}}\,\, {}^{+1.0}_{-1.0} {\big |_{\rm Belle}} $ \\
\hline
\multirow{3}{*}{$R(D)$} & $0.305^{+0.022}_{-0.025}$\cite{Wang:2017jow}  & 0.299(11)\cite{MILC:2015uhg} & &\\ 
& $0.296(6)$\cite{Gubernari:2018wyi} &  0.300(8)\cite{Na:2015kha} & 0.299(3)\cite{Bernlochner:2017jka} & 0.302(3)\\
& 0.297(3)\cite{Bordone:2019vic} & 0.301(6)\cite{Monahan:2017uby} & &\\
\hline
\end{tabular}}
\caption{Recent determinations of $|V_{cb}|$ and $R(D)$ from different techniques.} 
\label{tab:VRs}
\end{table}

%\newpage
%
%%%%%%%%%%%%%%%%%%%%%%%%%%%%%%%%%%%%%%%%%%%%%%%%%%%%%%%%%%%%%%%%%%%%%%%%%%%%
\section{Summary}
\label{sec:summary}
%%%%%%%%%%%%%%%%%%%%%%%%%%%%%%%%%%%%%%%%%%%%%%%%%%%%%%%%%%%%%%%%%%%%%%%%%%%%
%

In this work, we have presented a comprehensive study on the subleading power corrections to the $B\to D$ decay with the $B$-meson LCDAs providing the nonperturbative information of the hadronic state. 
The calculation relies on the LCSR technique to gain access to the $B\to D$ form factors in the large-recoil region by employing the power-counting scheme $n\cdot p\sim {\cal O}(m_b),~|\bar n\cdot p|\sim {\cal O}(\lambda^2 m_b)$, and $m_c\sim {\cal O}(\lambda m_b)$.
The  corrections that we consider are generated by the higher-twist $B$-meson LCDAs including both two- and three-particle contributions, by the NLP terms in expanding the charm-quark propagator, and by the subleading terms in the HQET representation of the bottom-quark field. The applied techniques that we used to derive the higher-power corrections comprise, amongst others, the systematic inclusion of higher-twist HQET LCDAs as dictated by power-counting, the classical EOM relations between the non-local operators, the diagrammatic factoriztion approach as well as the dispersion relations. These aspects, together with a non-negligible amount of bookkeeping, render our calculation non-trivial.
\addtocounter{footnote}{-1}
\footnotetext{Here we quote the LCSR predictions including the lattice inputs.}
\addtocounter{footnote}{+1}\footnotetext{Lattice and QCD sum rule inputs are used in combination with HQET constraints.}

We then proceeded to explore the phenomenological implications of the LCSR by constructing models for the $B$-meson LCDAs up to the twist-six level.
These models, inspired by previous studies~\cite{Beneke:2018wjp} with extensions to LCDAs beyond twist four, are consistent with all known constraints by construction.
The leading-power contribution was evaluated at the (partial) NLL level where it was demonstrated that the factorization scale dependence is negligible.
This is consistent with the previous study of the two-loop $B$-meson LCDA evolution.
We found that the subleading power corrections are minuscule compared to their leading-power counterparts in the kinematic region where LCSR is applicable.
Numerically, they account for approximately $\sim20\%$ in total to the theory prediction of the form factors.
This is encouraging as it indicates the presence of a well-established power hierarchy.
The LCSR predictions at the large hadronic recoil are then complemented by the lattice data, which is valid at the other end of the kinematic spectrum, allowing us to fix the coefficients in BGL parametrization under the strong unitarity bound with small uncertainties.
In this way, we are able to evaluate the $B\to D$ form factors in the entire kinematic region of $q^2$.

Subsequently, the form factors are used to predict the decay rate of $B\to D\ell\nu_\ell$, from which we have extracted the CKM matrix element $|V_{cb}|$ in eq.~\eqref{eq:Vcb} from two experimental data sets, as well as the ratio $R(D)=0.302\pm0.003$, crucial for testing lepton universality. 
In addition, we also provide theory predictions for the physical observables $\Delta\Gamma_\tau(t_1,t_2)$ and $\Delta{\cal R}(t_1,t_2)$ in $q^2$ bins.

Let us conclude by mentioning a couple of further directions to be addressed in future work. First, the perturbative corrections can be extended to two loops at the leading-power approximation and to one loop for the next-to-leading power contribution.
Second, since the two-loop evolution kernel for the leading-twist LCDA $\phi_B^+(\omega,\mu)$ has become available recently~\cite{Braun:2019wyx}, this allows for a complete next-to-leading-logarithmic resummation to be accomplished for certain decays (e.g., $B\to\gamma\ell\nu_\ell$) with the resulting two-loop RGE subsequently solved in the Mellin~\cite{Braun:2019zhp} and Laplace space~\cite{Galda:2020epp}. 
For future perspectives of the $B\to D$ decays in this regard, it is desirable to refine the Wandzura-Wilczek part, which is the dominating component of the two-particle twist-3 LCDA  $\phi_B^-(\omega,\mu)$, to the NLL level. 
Finally, an interesting future study will be to reformulate the subleading-power corrections in the SCET framework. While this task is technically challenging it certainly deserves a dedicated further investigation.

%%%%%%%%%%%%%%%%%%%%%%%%%%%%%%%%%%%%%%%%%%%%%%%%%%%%%%%%%%%%%%%%%%%%%%%
\section*{Acknowledgements}
J.G.\ is partially supported by the Deutscher Akademischer Austauschdienst (DAAD).
The research of T.H.\ and Y.J.\ was supported in part by the Deutsche Forschungsgemeinschaft
(DFG, German Research Foundation) under grant  396021762 - TRR 257. Y.J.\ also acknowledges the support of DFG grant SFB TR 110/2.
 C.W.\ is supported in part by the National Natural Science Foundation of China with Grant No. 12105112.
Y.M.W.\ acknowledges support from the National Youth Thousand Talents Program,
the Youth Hundred Academic Leaders Program of Nankai University,
the  National Natural Science Foundation of China  with Grant No. 11675082, 11735010 and 12075125,
and  the Natural Science Foundation of Tianjin with Grant No. 19JCJQJC61100. 
Y.B.W.\ is supported in part by the Alexander-von-Humboldt Stiftung.

%%%%%%%%%%%%%%%%%%%%%%%%%%%%%%%%%%%%%%%%%%%%%%%%%%%%%%%%%%%%%%%%%%%%%%%

\appendix
\section*{Appendix}
\label{sec:apps}
\addcontentsline{toc}{section}{Appendices}

\renewcommand{\theequation}{\Alph{section}.\arabic{equation}}
\renewcommand{\thetable}{\Alph{table}}
\setcounter{section}{0}
\setcounter{table}{0}

%
%%%%%%%%%%%%%%%%%%%%%%%%%%%%%%%%%%%%%%%%%%%%%%%%%%%%%%%%%%%%%%%%%%%%%%%%%%%

\section{Conventions}
\label{App:DAs}
%%%%%%%%%%%%%%%%%%%%%%%%%%%%%%%%%%%%%%%%%%%%%%%%%%%%%%%%%%%%%%%%%%%%%%%%%%%
%

Following the convention established in~\cite{Grozin:1996pq},  the $B$-meson LCDAs are defined 
in terms of the renormalized non-local operators composed of an effective heavy quark field $h_v(0)$ and light degrees
of freedom with a light-like separation, sandwiched between the vacuum and the $B$-meson state $| \bar B_v\rangle$ in the  HQET framework~\cite{Braun:2017liq}
\begin{subequations}
\begin{align}
& \langle 0| \,\bar q(x) \, \Gamma \, [x,0] \, h_v(0) \, |\bar B_v\rangle 
\nonumber \\
=&
-\frac{i}2  \widetilde{f}_B(\mu)m_B \,\bigg\{ \Tr\Big[\gamma_5 \Gamma P_+ \Big] \,
\big(\Phi_B^+ + x^2 G_B^+\big)
\nonumber\\[0.2cm]
&-\frac{1}{2}  \Tr\Big[\gamma_5 \Gamma P_+ 
\slashed{x} \Big] \frac{1}{v\cdot x}
\Big[\big(\Phi_B^+-\Phi_B^-\big) 
+ x^2 \big(G_B^+-G_B^-\big)\Big]\bigg\}(v\cdot x,\mu)\,,
\label{def:two} 
\\
\lefteqn{\langle 0| \, \bar q(z_1\bar n) \,[z_1\bar n,z_2\bar n]\, 
g_sG_{\mu\nu}(z_2\bar n) \,\Gamma \, [z_2\bar n,0] \, h_v(0) \,
|\bar B_v\rangle }
\nonumber\\[0.15cm]
=&~
\frac12 \widetilde{f}_B(\mu)m_B \,\Tr\biggl\{\gamma_5 \Gamma P_+
\biggl[ (v_\mu\gamma_\nu-v_\nu\gamma_\mu)  
\big[{\Psi}_A-{\Psi}_V \big]-i\sigma_{\mu\nu}{\Psi}_V
- (\bar n_\mu v_\nu-\bar n_\nu v_\mu){X}_A
\nonumber\\
& + \,(\bar n_\mu \gamma_\nu-\bar n_\nu \gamma_\mu)\big[W+{Y}_A\big]
- i\epsilon_{\mu\nu\alpha\beta} \bar n^\alpha v^\beta \gamma_5 \widetilde{X}_A
+ i\epsilon_{\mu\nu\alpha\beta} \bar n^\alpha \gamma^\beta\gamma_5 \widetilde{Y}_A
\nonumber\\[0.1cm]
& - \,(\bar n_\mu v_\nu-\bar n_\nu v_\mu)\slashed{\bar n}\,{W} + 
(\bar n_\mu \gamma_\nu-\bar n_\nu \gamma_\mu)\slashed{\bar n}\,{Z}
\biggr]\biggr\}(z_1,z_2;\mu)\,,
\label{def:three}
\end{align}
\end{subequations}
where $\gamma_5=i\gamma^0\gamma^1\gamma^2\gamma^3$, $P_+=\frac12(1+\slashed{v})$, $\Gamma$ represents an arbitrary Dirac structure, the  totally antisymmetric Levi-Civita tensor follows $\epsilon_{0123}=1$, $\widetilde f_B(\mu)$ denotes the $B$-meson decay constant in HQET,
and
\begin{equation}\label{WL}
{}[z\bar n,0] \equiv {\rm Pexp}\left[ig\int_0^1\!du\,\bar n_\mu A^\mu(uz\bar n)\right]
\end{equation}
is the Wilson line connecting the fundamental fields that ensures gauge invariance. 
Such factors are always implied if not shown explicitly. 
We have expanded the two-particle matrix element to ${\cal O}(x^2)$ 
to match our current accuracy where the two-particle LCDAs $\Phi^+_B$, $\Phi^-_B$, $G^+_B$ and $G^-_B$ are of the twist counting two, three, four and five, respectively.
The physical interpretation of such an expansion in the partonic picture is that the light quark 
carries nonvanishing transverse momenta along the light-cone.
The momentum space distributions are defined naturally as
\begin{subequations}
\begin{align}
\Phi_B^\pm(z,\mu)&=\int^\infty_0d\omega\,\e^{-i\omega z}\,\phi_B^\pm(\omega,\mu)\, ,
\label{def:LCDAFourier}
\\
 {\Psi}_A (z_1,z_2) &=
\int_0^\infty \!d\omega_1\!  \int_0^\infty \!\!d\omega_2\, 
e^{-i\omega_1 z_1-i\omega_2 z_2}\, {\psi}_A (\omega_1,\omega_2)\,,
\label{3pt-momspace}
\end{align}
\end{subequations}
and similarly for the other two- and three-particle LCDAs.
We adopt the convention to use upper (lower) case letters for coordinate (momentum) space distributions.

The eight invariant three-particle functions from Lorentz structure decomposition can be expanded in terms of LCDAs with definite collinear twist. 
One finds one LCDA of twist three
\begin{align}\label{eq:3pt3}
\Phi_3 &= {\Psi}_A-{\Psi}_V\,,
\end{align}
three twist-four LCDAs
\begin{align}\label{eq:3pt4}
  \Phi_4 = {\Psi}_A+{\Psi}_V\,, 
&&
  \Psi_4 = {\Psi}_A+{X}_A\,,
&&
  \widetilde{\Psi}_4 =  {\Psi}_V- \widetilde{X}_A\,,
\end{align}
three twist-five LCDAs 
\begin{align}\label{eq:3pt5}
\widetilde{\Phi}_5 & = \Psi_A +  \Psi_V + 2 Y_A - 2 \widetilde{Y}_A + 2 W\,,
\notag\\
  \Psi_5 &= -{\Psi}_A  +  {X}_A  -  2 Y_A\,,
\nonumber\\
  \widetilde{\Psi}_5 &= - {\Psi}_V  -  \widetilde{X}_A  + 2 \widetilde{Y}_A\,,
\end{align}
and one twist-six LCDA~\cite{Braun:2017liq}
\begin{align}\label{eq:3pt6}
 \Phi_6 &= {\Psi}_A- {\Psi}_V + 2{Y}_A + 2{W}+ 2 \widetilde{Y}_A - 4  Z\,.
\end{align}
It has been demonstrated that the one-loop RGEs for the three-particle LCDAs up to twist four are completely integrable in the large $N_c$ limit~\cite{Braun:2015pha, Braun:2017liq}
thanks to their relations to certain spin chain models~\cite{Braun:2018fiz}.

%
%%%%%%%%%%%%%%%%%%%%%%%%%%%%%%%%%%%%%%%%%%%%%%%%%%%%%%%%%%%%%%%%%%%%%%%%%%%
\section{NLO correction to the leading-power contribution}
\label{App:NLO}
%%%%%%%%%%%%%%%%%%%%%%%%%%%%%%%%%%%%%%%%%%%%%%%%%%%%%%%%%%%%%%%%%%%%%%%%%%%
%

The ``effective" DAs, which incorporate the contribution of the one-loop jet function, take the form~\cite{Wang:2017jow}
\begin{align}
\Phi_{+, n}^{\rm eff}(\omega^{\prime}, \mu) =& \frac{\alpha_s \, C_F}{4 \, \pi} \,\theta(\omega^{\prime} - \omega_c)\,
\bigg \{ r_c \, \Big[ (1-r_c)  \,
\ln \Big(\frac{1}{r_c}-1\Big) - 1  \Big] \,
 \phi_B^{+}(\omega^{\prime} - \omega_c)   \nonumber \\
& +  \int_0^{\infty}  \, d \omega \,
\bigg [ \frac{\omega}{ (\omega+\omega_c)^2 } \, \Big ({\cal P} \, 
\frac{\omega_c}{\omega^{\prime} -\omega -\omega_c}
- r_c  \Big ) 
+ \frac{r^2_c}{ \omega+\omega_c} 
 \nonumber \\
& 
- \frac{1-r_c}{ \omega} \,
\theta(\omega+\omega_c-\omega^{\prime})
+ \frac{r_c}{\omega-\omega^{\prime}}\,
\theta(\omega^{\prime} - \omega - \omega_c)
  \bigg ]   \, \phi_B^{+}(\omega) \, \bigg \} \,,\label{eq:Phi-eff-1}\\
\Phi_{+, \bar{n}}^{\rm eff}(\omega^{\prime}, \mu) =&~  \frac{\alpha_s \, C_F}{4 \, \pi} \,
 \theta(\omega^{\prime} - \omega_c) \,
\bigg \{ m_c \, \bigg [ \frac{1}{\omega'}\,\Big[ (1-r_c)\,
\ln \Big(\frac{1}{r_c}-1\Big)  - 1\Big]   \,
\phi_B^{+}(\omega^{\prime} - \omega_c)  \label{eq:Phi-eff-2}\\
&+ \int_{0}^{\infty} \, d \omega \,
\frac{1}{(\omega+\omega_c)^2}\,
\bigg ( 
{\cal P} \, \frac{\omega}{\omega^{\prime} -\omega -\omega_c}
 +  \Big(1+\frac{\omega_c}{\omega}\Big)\,r_c \,
(1-r_c)  +r_c
\bigg ) \, \phi_B^{+}(\omega)   \nonumber \\
& - \int_{0}^{\infty} \, d \omega \,  
\theta(\omega^{\prime} - \omega - \omega_c)\,
\frac{\omega^{\prime} - \omega - \omega_c}{(\omega^{\prime} - \omega)^2}\,  
\frac{\phi_B^{+}(\omega)}{ \omega}    \bigg ]
\nonumber \\
& + \, r \int_{0}^{\infty} \, d \omega \,\bigg [\,
r^2_c
-\theta(\omega + \omega_c - \omega^{\prime})
-\theta(\omega^{\prime}-\omega - \omega_c )   \,
\frac{\omega_c^2}{(\omega^{\prime}-\omega)^2}
\bigg ]
\frac{\phi_B^{+}(\omega)}{ \omega} 
 \bigg \}\notag
\,,\\
\Phi_{-, n}^{\rm eff}(\omega^{\prime}, \mu) =&  - \phi_B^{-}(\omega^{\prime} - \omega_c) \,
 \theta(\omega^{\prime} - \omega_c) \,, \label{eq:Phi-eff-3}\\
 \Phi_{-, \bar n}^{\rm eff}(\omega^{\prime}, \mu) =&  - \phi_B^{-}(\omega^{\prime} - \omega_c) \,
 \theta(\omega^{\prime} - \omega_c) +
 \frac{\alpha_s \, C_F}{4 \, \pi} \, \bigg \{   \phi_B^{-}(\omega^{\prime} - \omega_c) \,
\theta(\omega^{\prime} - \omega_c) \, \rho^{(1)}(\omega^{\prime})  \label{eq:Phi-eff-4}\\
& + \left [ \frac{d }{ d \omega^{\prime}} \phi_B^{-}(\omega^{\prime}-\omega_c) \right ] \,
\theta(\omega^{\prime}-\omega_c) \, \rho^{(2)}(\omega^{\prime})
+ \phi_B^{-}(0) \, \rho^{(3)}(\omega^{\prime})   
\nonumber \\
&+ \, \int_{0}^{\infty} \, d \omega \,  \left [ \rho^{(4)}(\omega, \omega^{\prime}) + \rho^{(5)}(\omega, \omega^{\prime}) \frac{d}{ d \omega} +  \frac{\rho^{(6)}(\omega, \omega^{\prime})}{\omega}\left(\frac{d}{d\omega}-\frac{1}{\omega}\right)\right ] \phi_B^{-}(\omega) \,
\bigg \}\, ,  \notag
  \end{align}
where ${\cal P}$ stands for the  principal value and
\begin{align}
\rho^{(1)}(\omega^{\prime}) =&~   
-\ln^2 \frac{\mu^2}{ n \cdot p \, \omega'} \,
+\ln(1-r_c) \,
\Big[2\,\ln \frac{\mu^2}{ n \cdot p \, (\omega'-\omega_c)} 
+4\,\ln\Big(\frac{1}{r_c}-1\Big) 
- (1- r_c )^2
\Big] 
\nonumber \\
& - (1+2\,r_c-r^2_c)\, \ln r_c 
+ 2 \, {\rm Li}_2 (1-r_c)  
- r_c
-\frac{ \pi^2}{2} \,,
\\
\rho^{(2)}(\omega^{\prime}) =&~  2 \, \omega_c \,
\left [ 3 \,  \ln \frac{\mu^2}{n \cdot p \, \omega_c}   +4 \right ] \,, 
\\
\rho^{(3)}(\omega^{\prime}) =&~  2\, \omega_c \, \delta(\omega_c-\omega^{\prime}) \,
\left [ 3 \,  \ln \frac{\mu^2}{ n \cdot p \, \omega_c}    +4 \right ]\notag\\
& +2 \, \theta(\omega^{\prime}-\omega_c )\, \theta(\omega^{\prime} )\,\Big[
\ln (1-r_c)
-\ln^2 (1-r_c)\Big]\,,
\\
\rho^{(4)}(\omega, \omega^{\prime}) =&~  \theta(\omega'-\omega_c) \,\bigg\{
{\cal P} \, 
\frac{1} { \omega^{\prime} - \omega - \omega_c} \,
\bigg[  -2 \,
\ln \frac{ \mu^2} { n \cdot p \,\omega^{\prime}}
+ 4  \,\ln(1-r_c)
-  \Big(\frac{\omega} { \omega+\omega_c} \Big)^2 
\bigg]
\nonumber \\
&
+4 \,
\bigg[\frac{\theta(\omega + \omega_c -\omega')} { \omega^{\prime} - \omega - \omega_c}\,
\ln\frac{\omega_c}{\omega + \omega_c -\omega'}
+\frac{\theta(\omega'-\omega - \omega_c)} { \omega^{\prime} - \omega - \omega_c} \, 
\ln\frac{\omega'-\omega}
{\omega'- \omega - \omega_c}\bigg]
\nonumber \\
& 
+  {\cal P} \, \frac{4\, \theta(\omega+\omega_c-\omega^{\prime})} { \omega - \omega^{\prime}} \,
\ln\frac{\omega_c}{\omega+\omega_c-\omega'}
+ \frac{4\, \theta(\omega^{\prime}- \omega-\omega_c)} { \omega - \omega^{\prime}} \,
\ln(1-r_c)
\nonumber \\
& + \Big( r_c
- \frac{\omega} { \omega+\omega_c}  -1 \Big) \,
\frac{r_c} { \omega+\omega_c} 
+\theta(\omega'-\omega-\omega_c)\,
\frac{\omega+\omega_c-\omega'}{(\omega'-\omega)^2}\bigg\}\,,
\\
\rho^{(5)}(\omega, \omega^{\prime}) =&~
2\,\theta(\omega'-\omega_c) \, 
\bigg\{
\theta(\omega'-\omega-\omega_c) \,
\bigg[
2\,\ln\frac{\omega'-\omega-\omega_c}{\omega'-\omega_c}\,
\ln\frac{\mu^2}{n\cdot p \, (\omega'-\omega)}\notag\\
&\quad
- \ln^2\frac{\omega'-\omega-\omega_c}{\omega'}
+ 
\ln\frac{\omega'-\omega-\omega_c}{\omega'}\,
\Big(2\,\ln\frac{\omega'-\omega}{\omega'} +1 \Big)
\bigg]\notag\\
&
-  2 \, \theta( \omega + \omega_c - \omega^{\prime})  \,
{\rm Li}_2 \left ( \frac{\omega^{\prime} - \omega} { \omega^{\prime} - \omega - \omega_c} \right )
\bigg\}\,,
\\
\rho^{(6)}(\omega, \omega^{\prime}) =&~  2 \,\theta( \omega + \omega_c - \omega^{\prime})  \,
\theta(\omega^{\prime} - \omega_c) \, \left (\omega_c - \omega^{\prime} \right ) \,
\ln\frac{\omega+\omega_c-\omega^{\prime}} { \omega^{\prime} - \omega_c} \,,
\end{align}
with $r_c=\omega_c/\omega'$.

The (renormalized) hard coefficient functions $C_{\pm,\bar n}(n\cdot p,\mu)$ and $C_{\pm,n}(n\cdot p,\mu)$ to one-loop accuracy 
were derived in ref.~\cite{Wang:2017jow} taking the form,
\begin{align}
C_{+,n}(n\cdot p,\mu)&=C_{+,\bar n}=1\, ,\qquad C_{-,n}=-\frac{\alpha_s\,C_F}{4\pi}\left[\frac{1}{r-1}\left(1+\frac{r}{1-r}\ln r\right)\right]\, ,\notag\\
C_{-,\bar n} (n\cdot p,\mu)&=1-\frac{\alpha_s}{4\pi}\left[2\ln^2\frac{\mu}{n\cdot p}+5\ln\frac{\mu}{m_b}-2\Li_2\left(1-\frac{1}{r}\right)-\ln^2 r +\frac{2-r}{r-1}\ln r\right.\notag\\
&\qquad\qquad\quad\left.+\frac{\pi^2}{12}+5\right]\, ,
\label{eq:hard-coeff}
\end{align}
where $r=n\cdot p/m_b$.

The evolution of hard coefficient functions $C_{+,(n,\bar n)}$ and $C_{-,n}$ are irrelevant at the one-loop matching and therefore
are not considered.
The RGEs governing the scale dependence of $C_{-,\bar n}$ and the HQET decay constant read~\cite{Wang:2017jow},
\begin{align}
\frac{d} { d \ln \mu}{C}_{-, \bar n}(n \cdot p,\mu) &=
\Big[-  \Gamma_{\rm cusp}(\alpha_s) \ln \frac{ \mu} { n
\cdot p} + \gamma(\alpha_s)\Big]\, {C}_{-,\bar n}(n \cdot p, \mu) \,,
\nonumber \\
\frac{d} { d \ln \mu} \,\widetilde{f}_B(\mu)
&= \widetilde{\gamma}(\alpha_s)\,\widetilde{f}_B(\mu)\,,
\end{align}
where the anomalous dimensions of the hard function 
\begin{align}
\gamma(\alpha_s)&=\sum_{n=0}^{\infty} \left(\frac{\alpha_s}{4\pi}\right)^{n+1} \gamma^{(n)}\, ,&& \widetilde \gamma(\alpha_s)=\sum_{n=0}^{\infty} \left(\frac{\alpha_s}{4\pi}\right)^{n+1} \widetilde\gamma^{(n)}
\end{align}
to the two-loop order are as follows~\cite{Beneke:2011nf}
\begin{align} 
\gamma^{(0)} &=-5 C_F\,,
&& 
\gamma^{(1)} = C_F\Big[-\frac{1585}{18}-\frac{5\pi^2}{6}+34\zeta(3)+ 
n_l\big(\frac{125}{27}+\frac{\pi^2}{3}\big)\Big]\,,\nonumber  \\
\widetilde{\gamma}^{(0)} &= 3 \,C_F\,, 
&&~ \widetilde{\gamma}^{(1)}=C_F\Big[ \frac{127} {6} + \frac{14\, \pi^2}{9} - \frac{5} {3}\, n_l\Big]
\end{align}
with $n_l$ being the number of light flavors and $\zeta(n)$ being the Riemann zeta function.
The cusp anomalous dimension
\begin{align}
\Gamma_{\rm cusp}&\equiv\sum_{n=0}^{\infty}\left(\frac{\alpha_s}{4\pi}\right)^{n+1}\Gamma^{(i)}_{\rm cusp}
\end{align}
to the three-loop order reads,
\begin{align}
\Gamma^{(0)}_{\rm cusp}=&~ 4 C_F,  
&&~ 
\Gamma^{(1)}_{\rm cusp}=
C_F\Big[\frac{268}{3}-4\pi^2-\frac{40}{9} n_l\Big], 
\label{eq:cusps} \\
\Gamma^{(2)}_{\rm cusp} =&~ 
C_F\bigg\{1470-\frac{536\pi^2}{3}+\frac{44\pi^4}{5}\,
\hspace{-0.6cm}&& +264\zeta_3
+n_l\Big[-\frac{1276}{9}+\frac{80\pi^2}{9}-\frac{208}{3}\zeta(3)\Big] 
-\frac{16}{27}n_l^2\bigg\}\,.\notag
\end{align}
The RGEs can then be solved analytically with the solution
\begin{align}
C_{-,\bar n}(n \cdot p,\mu)&=
U_1(n\cdot p,\mu_{h_1},\mu)\,C_{-,\bar n}(n \cdot p,\mu_{h_1})\,,
\nonumber \\
\widetilde{f}_B(\mu)&=U_2(\mu_{h_2},\mu)\,
\widetilde f_B(\mu_{h_2})
\end{align}
with the explicit expressions of  the evolution factors $U_1$ and $U_2$ given in~\cite{Wang:2016qii}.

%
%%%%%%%%%%%%%%%%%%%%%%%%%%%%%%%%%%%%%%%%%%%%%%%%%%%%%%%%%%%%%%%%%%%%%%%%%%%
\section{Next-to-leading power corrections}
\label{App:NLP}
%%%%%%%%%%%%%%%%%%%%%%%%%%%%%%%%%%%%%%%%%%%%%%%%%%%%%%%%%%%%%%%%%%%%%%%%%%%
%

The $B$-meson LCDAs contribute to the LCSR for the $B\to D$ form factors as shown in eq.~\eqref{master formula of the NLP LCSR}, via the following functionals  
\begin{align}
\mathcal{F}_{2,1}(\phi)
=& -\e^{-\omega_c/\omega_M} \int^{\omega_s-\omega_c}_0 d\omega\,\e^{-\omega/\omega_M}\,
\phi(\omega)\, ,
\label{eq:cal-F21}
\\
\mathcal{F}_{2,2}(\phi)=&~ 
\e^{-\omega_s/\omega_M}\,
\phi(\omega_s-\omega_c)
+\frac{\e^{-\omega_c/\omega_M}}{\omega_M}
\int^{\omega_s-\omega_c}_0\,d\omega\,\e^{-\omega/\omega_M}\,\phi(\omega)\,,
\label{eq:cal-F22} \\
\mathcal{F}_{3,2}(\phi)= &~
\e^{-\omega_s/\omega_M}
\int^{\omega_s-\omega_c}_0 d\omega \int^\infty_{\omega_s-\omega_c-\omega}
\frac{d\xi}{\xi}\,
\phi\left(\frac{\omega_s-\omega_c-\omega}{\xi},\omega,\xi\right)
\nonumber \\
&+
\frac{\e^{-\omega_c/\omega_M}}{\omega_M}\int^{\omega_s-\omega_c}_0\,
d\omega'
\int^{\omega'}_0 d\omega \int^\infty_{\omega'-\omega}\frac{d\xi}{\xi}\,
\,\e^{-\omega'/\omega_M}\,\phi\left(\frac{\omega'-\omega}{\xi},\omega,\xi\right)\, ,
\label{eq:calF-32}
\\
\mathcal{F}_{2,3}(\phi)= &~
-\frac{1}{2}\,\e^{-\omega_s/\omega_M}\,\Big[
\frac{d}{d\omega_s}\phi(\omega_s-\omega_c)
+\frac{1}{\omega_M}\,
\phi(\omega_s-\omega_c)\Big]
\nonumber \\
&-\frac{1}{2}\,\frac{\e^{-\omega_c/\omega_M}}{\omega^2_M}
\int^{\omega_s-\omega_c}_0\,d\omega\,\e^{-\omega/\omega_M}\,\phi(\omega)\,,
\label{eq:cal-F23}
\\
\mathcal{F}_{3,3}(\phi)
= &~
\frac{1}{2}\,\e^{-\omega_s/\omega_M}\,\bigg\{
\int^{\omega_s-\omega_c}_0 d\omega \int^\infty_{\omega_s-\omega_c-\omega}
\frac{d\xi}{\xi}\,\bigg[\Big(-\frac{1}{\xi}\frac{d}{du}-\frac{1}{\omega_M}\Big)\,
\phi(u,\omega,\xi)\bigg]
\bigg|_{u=\frac{\omega_s-\omega_c-\omega}{\xi}}
\nonumber \\
&-\int^\infty_0 \frac{d\xi}{\xi}\,\phi(0,\omega_s-\omega_c,\xi)
+\int^{\omega_s-\omega_c}_0 \frac{d\xi}{\xi}\,
\phi(1,\omega_s-\omega_c-\xi,\xi)\bigg\}
\nonumber \\
&-\frac{1}{2}\,
\frac{\e^{-\omega_c/\omega_M}}{\omega^2_M}\,\int^{\omega_s-\omega_c}_0\,
d\omega'
\int^{\omega'}_0 d\omega \int^\infty_{\omega'-\omega}\frac{d\xi}{\xi}\,
\,\e^{-\omega'/\omega_M}\,\phi\left(\frac{\omega'-\omega}{\xi},\omega,\xi\right)\, ,
\label{eq:cal-F33}
\end{align}
where $\phi$ represents a general LCDA with an appropriate particle content 
entering the functional ${\cal F}_{i,j}$.  
Here $i$ and $j$ respectively indicate the number of particles composing $\phi$ and the power in the denominator of the coefficient function accompanying $\phi$,  which together contribute to  $\Pi_\mu$  
(see eq.~\eqref{eq:2pLCSR}).

It is easy to show that taking $\omega_c\to0$ reproduces the corresponding expressions in ref.~\cite{Lu:2018cfc}.
The underlined term can potentially produce power-suppressed contribution depending on the specific 
LCDA from which $\phi(u,\omega,\xi)$ is constructed. 
More specifically, the power counting is determined by the asymptotic behavior of the three-particle LCDA at the small momenta 
region $\omega,\xi\sim 0$ dictated by the conformal spins of the fundamental fields building up the LCDA~\cite{Braun:2017liq}.

More generally, one obtains for the two-particle case,
\begin{align}
{\cal F}_{2,k}(\phi_2)\equiv&\int^{\omega_s}_0\,\frac{d\omega'}{\omega'-\bar n\cdot p}\,
\frac{1}{\pi}\,{\rm Im}_{\omega'}\int^\infty_0 d\omega\,
\frac{\phi_{2}(\omega)}{(\omega'-\omega-\omega_c+i\epsilon)^k}
\label{eq:2pLCSR}
\\
=&~ \frac{0!}{(k-1)!}\int^{\omega_s}_0\,\frac{d\omega'}{\omega'-\bar n\cdot p}\,
\int^\infty_0 d\omega\,\phi_{2}(\omega)\,
\frac{1}{\pi}\,{\rm Im}_{\omega'}
\Big[\frac{d^{k-1}}{d\omega^{k-1}}\frac{1}{\omega'-\omega-\omega_c+i\epsilon}\Big]
\nonumber \\
=&~ \frac{0!}{(k-1)!}\int^{\omega_s}_0\,\frac{d\omega'}{\omega'-\bar n\cdot p}\,
\int^\infty_0 d\omega\,\phi_{2}(\omega)\,
\Big[(-1)\,\frac{d^{k-1}}{d\omega^{k-1}}\,\delta(\omega'-\omega-\omega_c)\Big]
\nonumber \\
=&~\frac{(-1)^{k}}{(k-1)!}\,\sum^{k-1}_{l=1}\,
\frac{(l-1)!}{(\omega_s-\bar n\cdot p)^{l}}\,
\phi^{(k-l-1)}_{2}(\omega_s-\omega_c)
+\int^{\omega_s-\omega_c}_0\,d\omega\,
\frac{(-1)^{k}\phi_{2}(\omega)}{(\omega+\omega_c-\bar n\cdot p)^{k}}\, ,\notag
\end{align}
where ${\rm Im}_{z}f(z)\equiv {\rm Im}f(z+i0)$, and we have used integration-by-parts repeatedly in the last step and applied $\omega_s>\omega_c$.
The three-particle dispersion relation can be written in a similar fashion,
\begin{align}
{\cal F}_{3,k}(\phi_3)\equiv&\int^{\omega_s}_0\,\frac{d\omega'}{\omega'-\bar n\cdot p}\,
\frac{1}{\pi}\,{\rm Im}_{\omega'}\int^\infty_0 d\omega \,d\xi\int^1_0 du\,
\frac{f(u)\,\phi_{3}(\omega,\xi)}{(\omega'-\omega-u\xi-\omega_c+i\epsilon)^k}
\nonumber \\
=&\frac{(-1)^{1}\,0!}{(k-1)!}
\int^\infty_0 d\omega \,d\xi\,\phi_{3}(\omega,\xi)
\int^1_0 du\,f(u)\,\bigg[\frac{d^{k-1}}{d\omega^{k-1}_c}
\int^{\omega_s}_0\,d\omega'\,\frac{\delta(\omega'-\omega-u\xi-\omega_c)}{\omega'-\bar n\cdot p}
\bigg]
\nonumber \\
=&\sum^{k-1}_{l=1}\,\frac{(-1)^{l+1}}{(k-1)!}\,
\frac{(l-1)!}{(\omega_s-\bar n\cdot p)^l}\,
\frac{d^{k-l-1}}{d\omega_c^{k-l-1}}
\int^{\omega_{sc}}_0 d\omega \int^\infty_{\omega_{sc}-\omega}\frac{d\xi}{\xi}\,
\phi_{3}(\omega,\xi)\,f\left(\frac{\omega_{sc}-\omega}{\xi}\right)
\nonumber \\
&+
(-1)^{k}\int^{\omega_{sc}}_0\,
\frac{d\omega'}{(\omega'+\omega_c-\bar n\cdot p)^k}\,
\int^{\omega'}_0 d\omega \int^\infty_{\omega'-\omega}\,
\frac{d\xi}{\xi}\,\phi_{3}(\omega,\xi)\,f\left(\frac{\omega'-\omega}{\xi}\right)\, ,
\end{align}
where $\omega_{sc}=\omega_s-\omega_c$, and $f(u)$ is an arbitrary function regular in $u$ integration.

%
%%%%%%%%%%%%%%%%%%%%%%%%%%%%%%%%%%%%%%%%%%%%%%%%%%%%%%%%%%%%%%%%%%%%%%%%%%%
\section{Modeling the \texorpdfstring{$B$}{Lg}-meson LCDAs}
\label{App:Model}
%%%%%%%%%%%%%%%%%%%%%%%%%%%%%%%%%%%%%%%%%%%%%%%%%%%%%%%%%%%%%%%%%%%%%%%%%%%
%

The LCDAs are, however, not independent as they are constrained by the EOMs.
At tree-level in coordinate space, the following identities hold~\cite{Kawamura:2001jm,Braun:2017liq},
\begin{subequations}
\label{KKQT}
\begin{align}
\hspace*{-0.5cm}  \Big[z\frac{d}{dz}+1\Big]\Phi_B^-(z) &=  \Phi_B^+(z)  + 2 z^2  \int_0^1\! udu\,\Phi_3(z,uz)\,,
\label{KKQT1}
\\
2 z^2  G_B^+(z) & =
-  \Big[ z \frac{d}{dz} - \frac12  + i z \bar \Lambda \Big] \Phi_B^+(z)
-  \frac{1}{2}\Phi_B^-(z)
- z^2  \int_0^1\! \bar udu\,{\Psi}_4(z,uz)\,,
\label{KKQT2}
\\
 2 z^2 G_B^-(z)
&= -  \Big[ z \frac{d}{dz} - \frac12  + i z \bar \Lambda \Big] \Phi_B^-(z) - \frac12  \Phi_B^+(z)
- z^2  \int_0^1\! \bar udu\,{\Psi}_5(z,uz)\,,
\label{KKQT3}
\\
 \Phi_B^-(z)
&= \left(z \frac{d}{dz}+1 + 2i z \bar \Lambda  \right) \Phi_B^+(z) +
2 z^2 \int_0^1\! du\,  \Big[ u \Phi_4(z,uz) + {\Psi}_4(z,uz)\Big],
\label{KKQT4}
\end{align}
\end{subequations}
For later convenience, we introduce $\widehat{G}^{-}_{B}$ which plays a similar role as the WW-term in $\phi_B^-$
\begin{align}
G^{-}_{B}(z)= & ~\widehat G^{-}_{B}(z)
-\frac{1}{2}\int^{1}_{0} du\,\bar{u}\,\Psi_{5}(z,uz)\,, \label{def:G-}
 \end{align}
 satisfying the following condition,
 \begin{align}
2\,z^{2}\,\widehat G^{-}_{B}(z)= & 
-\Big[z\,\frac{\partial}{\partial z}-\frac{1}{2}+i\,z\,\bar{\Lambda}
\Big]\,\Phi^{-}_{B}(z)-\frac{1}{2}\,\Phi^{+}_{B}(z) \,.\label{eq:G-}
\end{align}
Further, neglecting systematically contributions of twist-four four-particle operators of the type $\bar qGGh_v$ and $\bar qq\bar qh_v$
leads to the following identity due to Lorentz-invariance,
 \begin{align}
&2 \frac{d}{dz_1}z_1 \Phi_4(z_1,z_2) = \left(\frac{d}{dz_2} z_2+1\right)
\left[\Psi_4 (z_1,z_2)+\widetilde\Psi_4 (z_1,z_2)\right]\,.\label{eq:tw4-constraint}
\end{align}

The $B$-meson LCDAs at the reference scale $\mu_0$ (commonly taken to be $1~{\rm GeV}$) can be systematically constructed~\cite{Beneke:2018wjp}
in such a way that both the constraints in eqs.~\eqref{KKQT}, \eqref{eq:tw4-constraint}, and the normalization conditions of the 
LCDAs~\cite{Braun:2017liq, Grozin:1996pq} are satisfied. 
\begin{align}
\phi_B^+(\omega) &= \omega \, \mathbb{F}(\omega;-1) \,, \qquad\quad
\phi_B^{-\rm WW}(\omega) = \mathbb{F}(\omega;0) \,,
\nonumber \\
\phi_B^{-\rm t3}(\omega) &= \frac{1}{6}\,\mathcal{N} \, 
(\lambda^2_E - \lambda^2_H ) \,
\Big[
-\omega^2\,\mathbb{F}(\omega;-2)
+ 4\,\omega\,\mathbb{F}(\omega;-1) -2\, \mathbb{F}(\omega;0)
\Big] \,,
\nonumber \\
\phi_3(\omega_1,\omega_2) &=  \frac{1}{2}\,\mathcal{N} \, 
(\lambda^2_E - \lambda^2_H ) \, 
\omega_1\,\omega^2_2\, \mathbb{F}(\omega_1+\omega_2;-2) \,,
\nonumber \\
\widehat{g}_B^+(\omega) &=
\frac{1}{4}\bigg[
2\,\omega\,(\omega-\bar{\Lambda})\,\mathbb{F}(\omega;0)
+(3\omega-2\bar{\Lambda})\,\mathbb{F}(\omega;1)
+3\,\mathbb{F}(\omega;2)
\nonumber \\
& -\frac{1}{6} \, \mathcal{N} \, (\lambda^2_E - \lambda^2_H ) \, 
\omega^2 \,\mathbb{F}(\omega;0)\bigg]\,,
\nonumber \\
\widehat{g}_B^-(\omega) &=
\frac{1}{4}\,\bigg\{ (3\omega-2\bar{\Lambda})\,\mathbb{F}(\omega;1)
+3\, \mathbb{F}(\omega;2)
\nonumber \\
& +\frac{1}{3} \, \mathcal{N} \, 
(\lambda^2_E - \lambda^2_H ) \, 
\omega\,\bigg[
\omega\, (\bar{\Lambda}-\omega) \, \mathbb{F}(\omega;-1)
-\Big(2\,\bar{\Lambda} -\frac{3}{2}\,\omega \Big) \,
\mathbb{F}(\omega;0) \bigg]\bigg\}\,,
\nonumber \\
\phi_4(\omega_1,\omega_2) &= \frac{1}{2}\,\mathcal{N} \, 
(\lambda^2_E + \lambda^2_H ) \, 
\omega^2_2\, \mathbb{F}(\omega_1+\omega_2;-1) \,,
\nonumber \\
\psi_4(\omega_1,\omega_2) 
&= \mathcal{N} \, 
\lambda^2_E  \, 
\omega_1\, \omega_2\,  \mathbb{F}(\omega_1+\omega_2;-1) \,,\qquad
\widetilde{\psi}_4(\omega_1,\omega_2) = \mathcal{N} \, 
\lambda^2_H  \, 
\omega_1\, \omega_2\,  \mathbb{F}(\omega_1+\omega_2;-1) \,,
\nonumber \\
\phi_5(\omega_1,\omega_2) &= \mathcal{N} \, 
(\lambda^2_E + \lambda^2_H ) \, 
\omega_1\, \mathbb{F}(\omega_1+\omega_2;0) \, , \hspace{2.7mm}
\psi_5(\omega_1,\omega_2) = - \mathcal{N} \, 
\lambda^2_E \, 
\omega_2\,  \mathbb{F}(\omega_1+\omega_2;0) \,,
\nonumber \\
\widetilde{\psi}_5(\omega_1,\omega_2) &= - \mathcal{N} \, 
\lambda^2_H \, 
\omega_2\,  \mathbb{F}(\omega_1+\omega_2;0) \,,
\nonumber \\
\phi_6(\omega_1,\omega_2) &= \mathcal{N} \, 
(\lambda^2_E - \lambda^2_H ) \, \mathbb{F}(\omega_1+\omega_2;1)\,,\label{model:ansatz}
\end{align}
where 
\begin{align}
\mathcal{N} &= \frac{1}{3}\,\frac{\beta\,(\beta+1)}{\alpha\,(\alpha+1)}\,
\frac{1}{\omega^2_0} \, ,\qquad\qquad\qquad
\bar{\Lambda} = \frac{3}{2}\, \frac{\alpha}{\beta} \, \omega_0\,,
\notag\\
\mathbb{F}(\omega;n) &\equiv 
\omega^{n-1}_0 \,  
U(\beta-\alpha,2-n-\alpha,\omega/\omega_0) \,
\frac{\Gamma(\beta)}{\Gamma(\alpha)}\,
e^{-\omega/\omega_0} \,, \label{model:bb-F}
\end{align}
with $U(a, b, z)$ being the hypergeometric $U$ function.
Here we have exploited the relationship between the first moment of the leading-twist LCDA and $\bar\Lambda$
\begin{align}
\int^\infty_0d\omega\,\omega\,\phi_B^+(\omega)=\frac{4}{3}\bar\Lambda\, .\label{eq:1stmom}
\end{align}
We follow the procedure of fixing the model parameters by a set of $\{\lambda_B(\mu_0), \widehat\sigma_1(\mu_0), \bar\Lambda(\mu_0)\}$ 
at the reference scale $\mu_0=1~\rm{GeV}$. 
The $\lambda_E$ and $\lambda_H$ are defined by the matrix element of local quark-gluon-quark operator,
\begin{align}
\langle 0| \bar q(0) g_sG_{\mu\nu}(0)\Gamma h_v(0)|\bar B(v)\rangle &=-\frac{i}{6} \widetilde{f}_B(\mu)m_B \lambda^2_H \Tr\Big[\gamma_5\Gamma P_+ \sigma_{\mu\nu}\Big]
\label{def:lambdaEH}\\
&-\vspace*{3mm}\frac{1}{6} \widetilde{f}_B(\mu)m_B\Big( \lambda^2_H- \lambda^2_E\Big)
  \Tr\Big[\gamma_5\Gamma P_+(v_\mu\gamma_\nu-v_\nu\gamma_\mu)\Big]\,.\notag
\end{align}
The matrix element can be evaluated using QCD sum rules yielding, 
\begin{align}
\lambda^2_E = 0.11\pm 0.06~\text{GeV}^2, && 
\lambda^2_H =  0.18\pm 0.07~\text{GeV}^2,&& \text{\cite{Grozin:1996pq}}
\\
\lambda^2_E = 0.03\pm 0.02~\text{GeV}^2, && 
\lambda^2_H = 0.06\pm 0.03~\text{GeV}^2, &&  \text{\cite{Nishikawa:2011qk}}
\\
\lambda^2_E = 0.01\pm 0.01~\text{GeV}^2, && 
\lambda^2_H = 0.15\pm 0.05~\text{GeV}^2, &&  \text{\cite{Rahimi:2020zzo}}
\label{QCDSR:lambdaEH}
\end{align}
where the last estimate, which has some overlap with the previous ones, has not been fully incorporated into our current study.

In principle, the model is expected to be valid only at the small momenta region $\omega, \omega_1,\omega_2\sim 0$ due to the development of a 
large momentum tail from the evolution of LCDAs.  
For inverse (logarithmic) moments, this does not pose as a major issue as the tail contribution is suppressed and the inverse (logarithmic) moments are
sensitive only to the small momentum behavior of DA adequately captured by the model.
It is, however, certainly not the case for $\bar\Lambda$. 
We, therefore, emphasize that $\bar\Lambda$ is only taken as an input parameter \emph{not} determined from the LCDA models but fixed by physical values. 
This is a consequence of the insufficient knowledge regarding the large momentum behaviors of the LCDAs at the moment.
Then the leading-twist LCDA is evolved to the factorization scale with no reference to the value of $\bar\Lambda$ any more.
In other words, the model satisfies eq.~\eqref{eq:1stmom} only at $\mu_0=1~\rm GeV$.

It is convenient to define the (logarithmic) inverse moments of the leading-twist $B$-meson LCDA as follows,
\begin{align}
\frac{1}{\lambda_B(\mu)} &=  
\int^\infty_0 \frac{d\omega}{\omega}\,\phi_B^+(\omega,\mu) \,,
\nonumber \\
\frac{\widehat{\sigma}_n(\mu)}{\lambda_B(\mu)} &= 
\int^\infty_0 \frac{d\omega}{\omega}\,
\ln^n\frac{e^{-\gamma_E}\lambda_B(\mu)}{\omega}\,
\phi_B^+(\omega,\mu) \, ,
\label{eq:log-moms}
\end{align}
which are relatable to the parameters of the model~\eqref{model:ansatz} via
\begin{align}
\lambda_B(\mu) &= \frac{\alpha-1}{\beta-1}\,\omega_0\,,
\nonumber \\
\widehat \sigma_1(\mu) &= 
\psi(\beta-1) -\psi(\alpha-1) + \ln\frac{\alpha-1}{\beta-1}\,,
\nonumber \\
\widehat \sigma_2(\mu)&= 
\widehat \sigma_1^2(\mu)+\psi'(\alpha-1)-\psi'(\beta-1)
+\frac{\pi^2}{6}
\label{eq:hatsigma}
\end{align}
with $\gamma_E$ and $\psi(x)$ being the Euler-Mascheroni constant and the digamma function, respectively.
The scale dependence of these moments at one-loop level read
\begin{align}
\frac{\lambda_B(\mu_0)}{\lambda_B(\mu)} 
&= 1+ \frac{\alpha_s(\mu_0)\,C_F}{4\pi}\ln\frac{\mu}{\mu_0}
\Big[2-2\ln\frac{\mu}{\mu_0} -4\sigma_1(\mu_0) \Big] \,,
\nonumber \\
\widehat{\sigma}_1(\mu) 
&= \widehat{\sigma}_1(\mu_0)
+ \frac{\alpha_s(\mu_0)\,C_F}{4\pi}\,4\ln\frac{\mu}{\mu_0}
\Big[\widehat{\sigma}^2_1(\mu_0)-\widehat{\sigma}_2(\mu_0) \Big] \,.
\end{align}
The evolution of the second logarithmic moment $\widehat{\sigma}_2$ 
is not considered in predicting the form factors due to its dependence on the
 higher $\geq3$ logarithmic moment and the expected small effect.

 Another major advantage of introducing the general ansatz in~\eqref{model:ansatz} is
 that the LL resummation (evolution) for the twist-2 and 3 two-particle DAs can be accomplished analytically. 
Explicitly, we find~\cite{Beneke:2018wjp},
\begin{align}
\phi_B^+(\omega,\mu)
=&~  U_\phi(\mu,\mu_0) \,  \frac{1}{\omega^{p+1}}\,
\frac{\Gamma(\beta)}{\Gamma(\alpha)} \,
\mathcal{G}(\omega;0,2,1)\,,
\nonumber \\
\phi^{-\rm WW}_B(\omega,\mu)
=&~   U_\phi(\mu,\mu_0) \,  \frac{1}{\omega^{p+1}}\,
\frac{\Gamma(\beta)}{\Gamma(\alpha)} \,
\mathcal{G}(\omega;0,1,1)\,,
\nonumber \\
\phi^{-\rm t3}_B(\omega,\mu)
=&  - \frac{1}{6}\, U^{\rm t3}_\phi(\mu,\mu_0) \,\mathcal{N} \, 
(\lambda^2_E - \lambda^2_H ) \,\frac{\omega^2_0}{\omega^{p+3}} \,
\frac{\Gamma(\beta)}{\Gamma(\alpha)}\, 
\bigg\{ \mathcal{G}(\omega;0,3,3)
\nonumber \\
&
+ (\beta-\alpha) \, \bigg[ \frac{\omega}{\omega_0}\,
\mathcal{G}(\omega;0,2,2)
-\beta \, \frac{\omega}{\omega_0}\, 
\mathcal{G}(\omega;1,2,2)
- \mathcal{G}(\omega;1,3,3)\bigg]
\bigg\} \,,
\label{eq:LL_lcda}
\end{align}
where $\displaystyle p=\frac{\Gamma^{(0)}_{\rm cusp}}{2\beta_0}\ln[\alpha_{s}(\mu)/\alpha_{s}(\mu_{0})]$, the twist-3 two-particle LCDA $\phi_B^-(\omega,\mu)
=\phi_B^{-\WW}(\omega,\mu)+\phi_B^{-\rm t3}(\omega,\mu)$ is a linear combination 
of the (twist-2) WW term and the genuine twist-3 term, and
\begin{align}
\mathcal{G}(\omega;l,m,n) \equiv 
G^{21}_{23}\Big(\frac{\omega}{\omega_0}\,
\Big|\,{}^{1,\beta+l}_{p+m,\alpha,p+n}\Big)\,,
\end{align}
denotes the MeijerG function. 
The evolution factor $U_\phi(\mu,\mu_0)$ and $U^{\rm t3}_\phi(\mu,\mu_0)$ reads explicitly at one-loop order~\cite{Braun:2015pha,Braun:2017liq}\footnote{Note that here we have taken the cusp anomalous dimension to the two-loop order for the LL accuracy (one order higher than the evolution kernel) to compensate for the numerically large logarithm in the RGEs.}, 
\begin{eqnarray}
U_\phi(\mu,\mu_0) &=&
\exp\,\biggl\{-\frac{\Gamma^{(0)}_{\rm cusp}}{4\beta_0^2}\bigg(
\frac{4\pi}{\alpha_s(\mu_0)}\left[\ln r-1+\frac1r\right]
\nonumber\\[0.2cm]
&& \hspace*{-1.5cm}
-\frac{\beta_1}{2\beta_0}\ln^2r+\left(\frac{\Gamma^{(1)}_{\rm cusp}}{\Gamma^{(0)}_{\rm cusp}}-\frac{\beta_1}{\beta_0}\right)[r-1-\ln r]\bigg)\biggr\}
\,\left(\e^{2 \gamma _E}\mu_0\right)^{\frac{\Gamma^{(0)}_{\rm cusp}}{2\beta_0}\ln r}\, r^{\frac{\gamma_{\rm t2}^{(0)}}{2\beta_0}}\,,\nonumber\\
U_\phi^{\rm t3}(\mu,\mu_0) &=& U_\phi(\mu,\mu_0)\bigg|_{\gamma_{\rm t2}^{(0)}\to \gamma_{\rm t2}^{(0)} + \gamma_{\rm t3}^{(0)}}\, ,
\end{eqnarray}
where $r = {\alpha_s(\mu)}/{\alpha_s(\mu_0)}$, $\Gamma^{(i)}_{\rm cusp}$ are the cusp anomalous dimensions at various orders~\eqref{eq:cusps}, and
\begin{align}
&\gamma^{(0)}_{\rm t2} = - 2 C_F\,,
&&
\gamma^{(0)}_{\rm t3} = 2N_c \, .
\end{align}
Both evolution factors satisfy the boundary condition at the reference scale $\mu_0$
\begin{align}
&U_\phi(\mu_0,\mu_0)=1\,,
&&
U^{\rm t3}_\phi(\mu_0,\mu_0)=1\,.
\end{align}
%

%%%%%%%%%%%%
\iffalse
\begin{align}
\frac{dU_\phi(\mu,\mu_0)}{d\ln\mu} &= 
-\Big( \Gamma_{\rm cusp}\, \ln\frac{\mu}{e^{-2\gamma_E}} 
+ \gamma_{\rm t2}^{(0)} 
\Big) \,U_\phi(\mu,\mu_0)\,,
\nonumber \\
\frac{dU^{\rm t3}_\phi(\mu,\mu_0)}{d\ln\mu} &= 
-\Big( \Gamma_{\rm cusp}\, \ln\frac{\mu}{e^{-2\gamma_E}} 
+ \gamma_{\rm t2}^{(0)} + \gamma_{\rm t3}^{(0)}
\Big) \,U^{\rm t3}_\phi(\mu,\mu_0)\,,
\end{align}
where
%
%%%%%%%
\fi

%%%%%%%%%%%%%%%%%%%%%%%%%%%%%%%%%%%%%%%%%%%%%%%%%%%%%%%%%%%%%%%%%%%%%%%%%

\bibliography{references}
\bibliographystyle{JHEP}

\end{document}